\documentclass[onecolumn]{emulateapj}

\usepackage{graphicx,color,bm}
\usepackage{epstopdf}
\usepackage{natbib}
\usepackage{bibentry}
\usepackage{amsmath}

\newcommand{\WMAP}{\textsl{WMAP}}

\newcommand{\wmap}{{\WMAP}}

\newcommand{\Bennett}{{C. L. Bennett}}

\newcommand{\Dunkley}{{J. Dunkley}}
\newcommand{\Gold}{{B. Gold}}

\newcommand{\Halpern}{{M. Halpern}}
\newcommand{\Hill}{{R. S. Hill}}
\newcommand{\Hinshaw}{{G. Hinshaw}}
\newcommand{\Jarosik}{{N. Jarosik}}
\newcommand{\Kogut}{{A. Kogut}}
\newcommand{\Komatsu}{{E. Komatsu}}
\newcommand{\Smith}{{K. M. Smith}}
\newcommand{\Larson}{{D. Larson}}
\newcommand{\Limon}{{M. Limon}}
\newcommand{\Meyer}{{S. S. Meyer}}
\newcommand{\Nolta}{{M. R. Nolta}}
\newcommand{\Odegard}{{N. Odegard}}
\newcommand{\Page}{{L. Page}}

\newcommand{\Spergel}{{D. N. Spergel}}
\newcommand{\Tucker}{{G. S. Tucker}}

\newcommand{\Weiland}{{J. L. Weiland}}
\newcommand{\Wollack}{{E. Wollack}}
\newcommand{\Wright}{{E. L. Wright}}

\newcommand{\HinshawEM}{{hinshaw@physics.ubc.ca}}
\newcommand{\Adnet}{{ADNET Systems, Inc., 
            7515 Mission Dr., Suite A100 Lanham, Maryland 20706}}
\newcommand{\Brown}{{Dept. of Physics, Brown University, 
            182 Hope St., Providence, RI 02912-1843}}
\newcommand{\Cita}{{Canadian Institute for Theoretical Astrophysics, 
            60 St. George St, University of Toronto, 
	    Toronto, ON  Canada M5S 3H8}}
\newcommand{\Columbia}{{Columbia Astrophysics Laboratory, 
            550 W. 120th St., Mail Code 5247, New York, NY  10027-6902}}
\newcommand{\Goddard}{{Code 665, NASA/Goddard Space Flight Center, 
            Greenbelt, MD 20771}}
\newcommand{\Hopkins}{{Dept. of Physics \& Astronomy, 
            The Johns Hopkins University, 3400 N. Charles St., 
	    Baltimore, MD  21218-2686}}
\newcommand{\Minn}{{University of Minnesota, School of Physics \& Astronomy, 
            116 Church Street S.E., Minneapolis, MN 55455}}
\newcommand{\Oxford}{{Oxford Astrophysics, Denys Wilkinson Building, 
            Keble Road, Oxford, OX1 3RH, UK}}
\newcommand{\Perimeter}{{Perimeter Institute for Theoretical Physics, Waterloo, ON N2L 2Y5, Canada}}
\newcommand{\PrincetonPhysics}{{Dept. of Physics, Jadwin Hall, 
            Princeton University, Princeton, NJ 08544-0708}}
\newcommand{\PrincetonAstro}{{Dept. of Astrophysical Sciences, 
            Peyton Hall, Princeton University, Princeton, NJ 08544-1001}}
\newcommand{\UBC}{{Dept. of Physics and Astronomy, University of 
            British Columbia, Vancouver, BC  Canada V6T 1Z1}}
\newcommand{\UChicago}{{Depts. of Astrophysics and Physics, KICP and EFI, 
            University of Chicago, Chicago, IL 60637}}
\newcommand{\UTexas}{{Texas Cosmology Center and Dept. of Astronomy,  
            Univ. of Texas, Austin, 2511 Speedway, RLM 15.306, Austin, TX 78712}}
\newcommand{\UCLA}{{UCLA Physics \& Astronomy, PO Box 951547, 
            Los Angeles, CA 90095--1547}}
\newcommand{\MPA}{{Max-Planck-Institut f\"{u}r Astrophysik, Karl-Schwarzschild Str. 1, 
            85741 Garching, Germany}}
\newcommand{\IPMU}{{Kavli Institute for the Physics and Mathematics of the Universe, 
            Todai Institutes for Advanced Study, the University of Tokyo, Kashiwa, 
            Japan 277-8583 (Kavli IPMU, WPI)}}

\newcommand{\ddeg}         {\mbox{${\rlap.}^\circ$}}

\newcommand{\lcdm}{\ensuremath{\Lambda\mathrm{CDM}}}

\newcommand{\be}{\begin{equation}}
\newcommand{\ee}{\end{equation}}
\newcommand{\bea}{\begin{eqnarray}}
\newcommand{\eea}{\end{eqnarray}}

\renewcommand{\ell}{\ensuremath{l}}
\newcommand{\act}{ACT}
\newcommand{\spt}{SPT}
\newcommand{\planck}{\textsl{Planck}}

\begin{document}

\title{Nine-Year Wilkinson Microwave Anisotropy Probe (\wmap) Observations: Cosmological Parameter Results}

\author{
{\Hinshaw}\altaffilmark{1},
{\Larson}\altaffilmark{2},
{\Komatsu}\altaffilmark{3,4,5},
{\Spergel}\altaffilmark{6,4},
{\Bennett}\altaffilmark{2},
{\Dunkley}\altaffilmark{7},
{\Nolta}\altaffilmark{8},
{\Halpern}\altaffilmark{1},
{\Hill}\altaffilmark{9},
{\Odegard}\altaffilmark{9},
{\Page}\altaffilmark{10},
{\Smith}\altaffilmark{6,11},
{\Weiland}\altaffilmark{2},
{\Gold}\altaffilmark{12},
{\Jarosik}\altaffilmark{10},
{\Kogut}\altaffilmark{13},
{\Limon}\altaffilmark{14}, 
{\Meyer}\altaffilmark{15}, 
{\Tucker}\altaffilmark{16}, 
{\Wollack}\altaffilmark{13}, 
{\Wright}\altaffilmark{17}
}

\altaffiltext{1}{\UBC}
\altaffiltext{2}{\Hopkins}
\altaffiltext{3}{\MPA}
\altaffiltext{4}{\IPMU}
\altaffiltext{5}{\UTexas}
\altaffiltext{6}{\PrincetonAstro}
\altaffiltext{7}{\Oxford}
\altaffiltext{8}{\Cita}
\altaffiltext{9}{\Adnet}
\altaffiltext{10}{\PrincetonPhysics}
\altaffiltext{11}{\Perimeter}
\altaffiltext{12}{\Minn}
\altaffiltext{13}{\Goddard}
\altaffiltext{14}{\Columbia}
\altaffiltext{15}{\UChicago}
\altaffiltext{16}{\Brown}
\altaffiltext{17}{\UCLA}

\email{\HinshawEM}

\begin{abstract}
We present cosmological parameter constraints based on the final nine-year \wmap\ data, in conjunction with a number of additional cosmological data sets.  The \wmap\ data alone, and in combination, continue to be remarkably well fit by a six-parameter \lcdm\ model.  When \wmap\ data are combined with measurements of the high-$l$ cosmic microwave background (CMB) anisotropy, the baryon acoustic oscillation (BAO) scale, and the Hubble constant, the matter and energy densities, \ensuremath{\Omega_bh^2}, \ensuremath{\Omega_ch^2}, and \ensuremath{\Omega_\Lambda}, are each determined to a precision of $\sim$1.5\%.  The amplitude of the primordial spectrum is measured to within 3\%, and there is now evidence for a tilt in the primordial spectrum at the 5$\sigma$ level, confirming the first detection of tilt based on the five-year \wmap\ data.  At the end of the \wmap\ mission, the nine-year data decrease the allowable volume of the six-dimensional \lcdm\ parameter space by a factor of 68,000 relative to pre-\wmap\ measurements.  We investigate a number of data combinations and show that their \lcdm\ parameter fits are consistent.  New limits on deviations from the six-parameter model are presented, for example: the fractional contribution of tensor modes is limited to \ensuremath{r < 0.13\ \mbox{(95\% CL)}}; the spatial curvature parameter is limited to \ensuremath{\Omega_k = -0.0027^{+ 0.0039}_{- 0.0038}}; the summed mass of neutrinos is limited to \ensuremath{\sum m_\nu < 0.44\ \mbox{eV}\ \mbox{(95\% CL)}}; and the number of relativistic species is found to lie within \ensuremath{N_{\rm eff} = 3.84\pm 0.40}, when the full data are analyzed.  The joint constraint on \ensuremath{N_{\rm eff}} and the primordial helium abundance, \ensuremath{Y_{\rm He}}, agrees with the prediction of standard Big Bang nucleosynthesis.  We compare recent \planck\ measurements of the Sunyaev--Zel'dovich effect with our seven-year measurements, and show their mutual agreement. Our analysis of the polarization pattern around temperature extrema is updated.  This confirms a fundamental prediction of the standard cosmological model and provides a striking illustration of acoustic oscillations and adiabatic initial conditions in the early universe.
\end{abstract}


\section{INTRODUCTION}
\label{sec:intro}

Measurements of temperature and polarization anisotropy in the cosmic microwave background (CMB) have played a major role in establishing and sharpening the standard ``\lcdm'' model of cosmology: a six-parameter model based on a flat universe, dominated by a cosmological constant, $\Lambda$, and  cold-dark-matter (CDM), with initial Gaussian, adiabatic fluctuations seeded by inflation.  This model continues to describe all existing CMB data, including the \textit{Wilkinson Microwave Anisotropy Probe} (\wmap) nine-year data presented in this paper and its companion paper \citep{bennett/etal:prep}, the small-scale temperature data \citep{das/etal:2011a,keisler/etal:2011,reichardt/etal:2012a,story/etal:2012}, and the small-scale polarization data \citep{brown/etal:2009,chiang/etal:2010,quiet:2011,quiet2:2012}.
 
Despite its notable success at describing {\em all} current cosmological data sets, the standard model raises many questions: what is the nature of dark matter and dark energy?  What is the physics of inflation?  Further, there are open questions about more immediate physical parameters: are there relativistic species present at the decoupling epoch, beyond the known photons and neutrinos?  What is the mass of the neutrinos?  Is the primordial helium abundance consistent with  Big Bang nucleosynthesis? Are the initial fluctuations adiabatic? Tightening the limits on these parameters is as important as measuring the standard ones.  Over the past decade \wmap\ has provided a wealth of cosmological information which can be used to address the above questions.  In this paper, we present the final, nine-year constraints on cosmological parameters from \wmap.

The paper is organized as follows. In Section~\ref{sec:methods}, we briefly describe the nine-year \wmap\ likelihood code, the external data sets used to complement \wmap\ data, and we update our parameter estimation methodology.  Section~\ref{sec:lcdm} presents nine-year constraints on the minimal six-parameter \lcdm\ model.  Section~\ref{sec:lcdm_ext} presents constraints on parameters beyond the standard model, such as the tensor-to-scalar ratio, the running spectral index, the amplitude of isocurvature modes, the number of relativistic species, the mass of neutrinos, spatial curvature, the equation of state parameters of dark energy, and cosmological birefringence. In Section~\ref{sec:matter}, we discuss constraints on the amplitude of matter fluctuations, $\sigma_8$, derived from other astrophysical data sets.  Section~\ref{sec:sz} compares \wmap's seven-year measurements of the Sunyaev--Zel'dovich effect with recent measurements by \planck.   In Section~\ref{sec:peaks}, we update our analysis of polarization patterns around temperature extrema, and we conclude in Section~\ref{sec:conclusion}. 

\section{METHODOLOGY UPDATE}
\label{sec:methods}

Before discussing cosmological parameter fits in the remaining part of the paper, we summarize changes in our parameter estimation methodology and our choice of input data sets.  In \S\ref{sec:wmap_like} we review changes to the \wmap\ likelihood code.  In \S\ref{sec:ext_data} we discuss our choice of external data sets used to complement \wmap\ data in various tests.  Most of these data sets are new since the seven-year data release.  We conclude with some updates on our implementation of Markov Chains.

\subsection{\wmap\ Likelihood Code}
\label{sec:wmap_like}

For the most part, the structure of the likelihood code remains as it was in the seven-year \wmap\ data release.  However, instead of using the Monte Carlo Apodised Spherical Transform EstimatoR (MASTER) estimate \citep{hivon/etal:2002} for the $l>32$ TT spectrum, we now use an optimally-estimated power spectrum and errors based on the quadratic estimator from \citet{tegmark/etal:1997}, as discussed in detail in \citet{bennett/etal:prep}.  This $l>32$ TT spectrum is based on the template-cleaned V- and W-band data, and the KQ85y9 sky mask (see \citet{bennett/etal:prep} for an update on the analysis masks).  The likelihood function for $l>32$ continues to use the Gaussian plus log-normal approximation described in \citet{bond/jaffe/knox:1998} and \citet{verde/etal:2003}. 

The $l \le 32$ TT spectrum uses the Blackwell-Rao estimator, as before.  This is based on Gibbs samples obtained from a nine-year one-region bias-corrected ILC map described in \citep{bennett/etal:prep} and sampled outside the KQ85y9 sky mask.  The map and mask were degraded to HEALPix r5\footnote{The map resolution levels refer to the HEALPix pixelization scheme \citep{gorski/etal:2005} where r4, r5, r9, and r10 refer to $N_{\rm side}$ values of 16, 32, 512, and 1024, respectively.}, and 2 $\mu$K of random noise was added to each pixel in the map.

The form of the polarization likelihood is unchanged.  The $l>23$ TE spectrum is based on a MASTER estimate and uses the template-cleaned Q-, V-, and W-band maps, evaluated outside the KQ85y9 temperature and polarization masks.  The $l \le 23$ TE, EE, and BB likelihood retains the pixel-space form described in Appendix D of \citet{page/etal:2007}.  The inputs are template-cleaned Ka-, Q-, and V-band maps and the HEALPix r3 polarization mask used previously.

As before, the likelihood code accounts for several important effects:  mode coupling due to sky masking and non-uniform pixel weighting (due to non-uniform noise); beam window function uncertainty, which is correlated across the entire spectrum; and residual point source subtraction uncertainty, which is also highly correlated.  The treatment of these effects is described in \citet{verde/etal:2003, nolta/etal:2009, dunkley/etal:2009}.

\subsection{External Data Sets}
\label{sec:ext_data}

\subsubsection{Small-scale CMB measurements}
\label{sec:cmb}

Since the time when the seven-year \wmap\ analyses were published, there have been new measurements of small-scale CMB fluctuations by the Atacama Cosmology Telescope (\act) \citep{fowler/etal:2010, das/etal:2011a} and the South Pole Telescope (\spt) \citep{keisler/etal:2011,reichardt/etal:2012a}. They have reported the angular power spectrum at 148 and 217~GHz for \act, and at 95, 150, and 220~GHz for \spt, to $1'$ resolution, over $\sim$1000 deg$^2$ of sky. At least seven acoustic peaks are observed in the angular power spectrum, and the results are in remarkable agreement with the model predicted by the \wmap\ seven-year data \citep{keisler/etal:2011}.

\begin{figure}[ht]
\begin{center}
\includegraphics[width=0.8\textwidth]{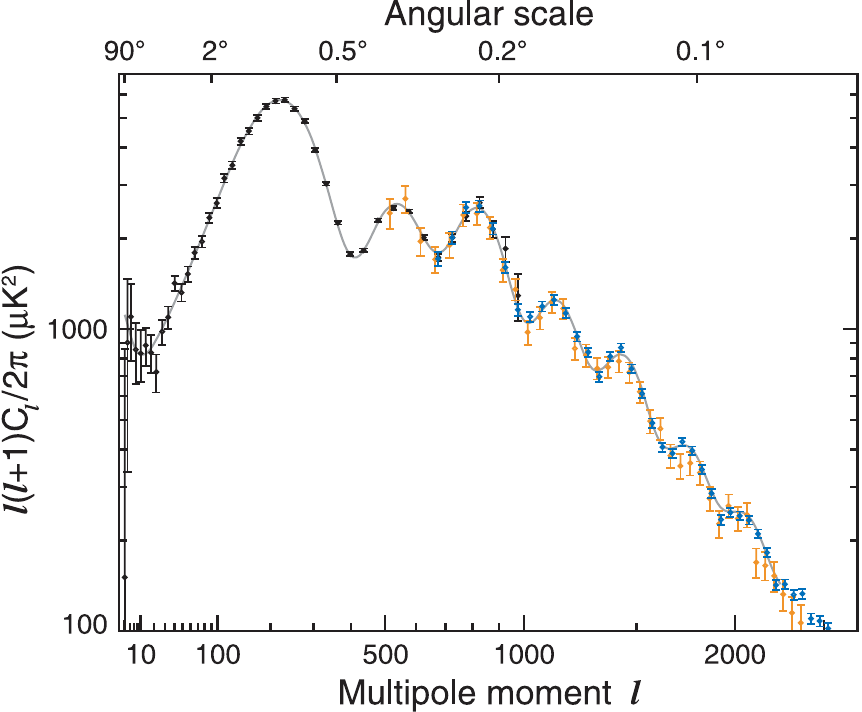}
\end{center}
\caption{A compilation of the CMB data used in the nine-year \wmap\ analysis.  The \wmap\ data are shown in black, the extended CMB data set -- denoted `eCMB' throughout -- includes \spt\ data in blue \citep{keisler/etal:2011}, and \act\ data in orange, \citep{das/etal:2011a}.  We also incorporate constraints from CMB lensing published by the \spt\ and \act\ groups (not shown).  The \lcdm\ model fit to the \wmap\ data alone (shown in grey) successfully predicts the higher-resolution data. \label{fig:cl_all}}
\end{figure}

Figure \ref{fig:cl_all} shows data from \act\ and \spt\ at 150~GHz, which constitutes the extended CMB data set used extensively in this paper (subsequently denoted `eCMB').  We incorporate the \spt\ data from \citet{keisler/etal:2011}, using 47 band-powers in the range $600 < l < 3000$. The likelihood is assumed to be Gaussian, and we use the published band-power window functions and covariance matrix, the latter of which accounts for noise, beam, and calibration uncertainty. Following the treatment of the \act\ and \spt\ teams, we account for residual extragalactic foregrounds by marginalizing over three parameters: the Poisson and clustered point source amplitudes, and the SZ amplitude \citep{keisler/etal:2011}. For \act\ we use the 148~GHz power spectrum from \citet{das/etal:2011a} in the multipole range $500< l <10000$, marginalizing over the same clustered point source and SZ amplitudes as in the \spt\ likelihood, but over a separate Poisson source amplitude.  See \S\ref{sec:wmap_chains} and \S\ref{sec:lcdm_cmb} for more details.

In addition to the temperature spectra, both \act\ and \spt\ have estimated the deflection spectra due to gravitational lensing \citep{das/etal:2011b, vanengelen/etal:2012}. These measurements are consistent with predictions of the $\Lambda$CDM model fit to \wmap.  When we incorporate \spt\ and \act\ data in the nine-year analysis, we {\em also} include the lensing likelihoods provided by each group\footnote{these codes are available at {\tt http://lambda.gsfc.nasa.gov}} to further constrain parameter fits. 

New observations of the CMB polarization power spectra have also been released by the QUIET experiment \citep{quiet:2011,quiet2:2012}; their TE and EE polarization spectra are in excellent agreement with predictions based primarily on \wmap\ temperature fluctuation measurements.  These data are the most recent in a series of polarization measurements at $l \ga 50$.  However, high-$l$ polarization observations do not (yet) substantially enhance the power of the full data to constrain parameters, so we do not include them in the nine-year analysis.

\subsubsection{Baryon Acoustic Oscillations}
\label{sec:bao}

The acoustic peak in the galaxy correlation function has now been detected over a range of redshifts from $z=0.1$ to $z=0.7$. This linear feature in the galaxy data provides a standard ruler with which to measure the distance ratio, $D_V/r_s$, the distance to objects at redshift $z$ in units of the sound horizon at recombination, independent of the local Hubble constant. In particular, the observed angular and radial BAO scales at redshift $z$ provide a geometric estimate of the effective distance,
\be
D_V(z) \equiv [(1+z)^2 \, D_A^2(z) \, cz \, /H(z)]^{1/3},
\label{eq:dvz}
\ee
where $D_A(z)$ is the angular diameter distance and $H(z)$ is the Hubble parameter. The measured ratio $D_V/r_s$, where $r_s$ is the co-moving sound horizon scale at the end of the drag era, can be compared to theoretical predictions. 

Since the release of the seven-year \wmap\ data, the acoustic scale has been more precisely measured by the Sloan Digital Sky Survey (SDSS) and SDSS-III Baryon Oscillation Spectroscopic Survey (BOSS) galaxy surveys, and by the WiggleZ and 6dFGS surveys. Previously, over half a million galaxies and LRGs from the SDSS-DR7 catalog had been combined with galaxies from 2dFGRS by \citet{percival/etal:2010} to measure the acoustic scale at $z=0.2$ and $z=0.35$. \citep[These data were used in the \wmap\ seven-year analysis, see also][]{kazin/etal:2010}.  Using the reconstruction method of \citet{eisenstein/etal:2007}, an improved estimate of the acoustic scale in the SDSS-DR7 data was made by \citet{padmanabhan/etal:2012}, giving $D_V(0.35)/r_s = 8.88 \pm 0.17$, and reducing the uncertainty from 3.5\% to 1.9\%.  More recently the SDSS-DR9 data from the BOSS survey has been used to estimate the BAO scale of the CMASS sample. They report $D_V(0.57)/r_s = 13.67 \pm 0.22$ for galaxies in the range $0.43<z<0.7$ (at an effective redshift $z=0.57$) \citep{anderson/etal:2012}.  This result is used to constrain cosmological models in \citet{sanchez/etal:2012}. 

\begin{deluxetable}{lccc}
\tablecaption{BAO Data Used in the Nine-year Analysis\label{tab:bao}}
\tablewidth{0pt}
\tablehead{\colhead{Redshift} & \colhead{Data Set} & \colhead{$r_s/D_V(z)$} & \colhead{Ref.}}
\startdata
0.1   &  6dFGS		& $0.336\pm0.015$		&  \cite{beutler/etal:2011} \\
\\
0.35 &  SDSS-DR7-rec	& $0.113\pm0.002$\tablenotemark{a} & \cite{padmanabhan/etal:2012} \\
0.57 &  SDSS-DR9-rec	& $0.073\pm0.001$\tablenotemark{a} & \cite{anderson/etal:2012} \\
\\
0.44 &  WiggleZ	& $0.0916\pm0.0071$		&  \cite{blake/etal:2012} \\
0.60 &  WiggleZ	& $0.0726\pm0.0034$		&  \cite{blake/etal:2012} \\
0.73 &  WiggleZ	& $0.0592\pm0.0032$		&  \cite{blake/etal:2012} \\
\enddata
\tablenotetext{a}{For uniformity, the SDSS values given here have been inverted from the published values: $D_V(0.35)/r_s = 8.88 \pm 0.17$, and $D_V(0.57)/r_s = 13.67 \pm 0.22$.}
\end{deluxetable}

The acoustic scale has also been measured at higher redshift using the WiggleZ galaxy survey. \citet{blake/etal:2012} report distances in three correlated redshift bins between 0.44 and 0.73. At lower redshift, $z=0.1$, a detection of the BAO scale has been made using the 6dFGS survey \citep{beutler/etal:2011}.  These measurements are summarized in Table~\ref{tab:bao}, and plotted as a function of redshift in Figure 19 of \citet{anderson/etal:2012}, together with the best-fit \lcdm\ model prediction from the \wmap\ seven-year analysis \citep{komatsu/etal:2011}. The BAO data are consistent with the CMB-based prediction over the measured redshift range. 

For the nine-year analysis, we incorporate these data into a likelihood of the form
\be
-2\ln L = ({\bf x-d})^T{\bf C}^{-1}({\bf x-d}),
\ee
where
\bea
{\bf x-d} &=& [r_s/D_V(0.1) - 0.336, \, D_V(0.35)/r_s - 8.88, \, D_V(0.57)/r_s - 13.67, \nonumber \\
&& r_s/D_V(0.44) - 0.0916, \, r_s/D_V(0.60) - 0.0726, \, r_s/D_V(0.73) - 0.0592]
\eea
and
\be
{\bf C}^{-1} = \left(\begin{array}{rrrrrr}
4444.4 & 0 & 0 & 0 & 0 & 0 \\
0 & 34.602 & 0 & 0 & 0 & 0 \\
0 & 0 & 20.661157 & 0 & 0 & 0 \\
0 & 0 & 0 & 24532.1  & -25137.7 & 12099.1 \\
0 & 0 & 0 & -25137.7 & 134598.4 & -64783.9 \\
0 & 0 & 0 & 12099.1 & -64783.9 & 128837.6 \\
\end{array}
\right).
\ee
The model distances are derived from the \lcdm\ parameters using the same scheme we used in the \wmap\ seven-year analysis \citep{komatsu/etal:2011}.

\subsubsection{Hubble Parameter}
\label{sec:h0}

It is instructive to combine \wmap\ measurements with measurements of the current expansion rate of the universe.  Recent advances in the determination of the Hubble constant have been made since the two teams using HST/WFPC2 observations reported their results \citep{freedman/etal:2001,sandage/etal:2006}.  Re-anchoring the HST Key Project distance ladder technique, \citet{freedman/etal:2012} report a significantly improved result of $H_0=74.3 \pm 1.5$ (statistical) $\pm 2.1$ (systematic) km s$^{-1}$ Mpc$^{-1}$.  The overall 3.5\% uncertainty must be taken with some caution however, since the uncertainties in all rungs are not fully propagated.

In a parallel approach, \citet{riess/etal:2009} redesigned the distance ladder and its observations to control the systematic errors that dominated the measurements. These steps include: the elimination of zero-point uncertainties by use of the same photometric system across the ladder; observations of Cepheids in the near-infrared to reduce extinction and sensitivity to differences in chemical abundance (the so-called ``metallicity effect''); the use of geometric distance measurements to provide a reliable absolute calibration; and the replacement of old Type Ia supernovae observations with recent ones that use the same photometric systems that define the Hubble flow.  This approach has led to a measurement of $H_0=73.8 \pm 2.4$ km s$^{-1}$ Mpc$^{-1}$ \citep{riess/etal:2011} with a fully propagated uncertainty of 3.3\%.  Since this uncertainty is smaller, we adopt it in our analysis.

\subsubsection{Type Ia Supernovae}
\label{sec:sn}

The first direct evidence for acceleration in the expansion of the universe came from measurements of luminosity distance as a function of redshift using Type Ia supernovae as standard candles \citep{riess/etal:1998, schmidt/etal:1998, perlmutter/etal:1999}. Numerous follow-up observations have been made, extending these early measurements to higher redshift. After the seven-year \wmap\ analysis was published, the Supernova Legacy Survey analyzed their three-year sample (`SNLS3') of high redshift supernovae \citep{guy/etal:2010, conley/etal:2011, sullivan/etal:2011}. They measured 242 Type Ia supernovae in the redshift range $0.08<z<1.06$, three times more than their first-year sample \citep{astier/etal:2006}. The SNLS team combined the 242 SNLS3 supernovae with 123 SNe at low redshift \citep{hamuy/etal:1996, riess/etal:1999, jha/etal:2006, hicken/etal:2009, contreras/etal:2010}, 93 SNe from the SDSS supernovae search \citep{holtzman/etal:2008}, and 14 SNe at $z>1$ from HST measurements by \citet{riess/etal:2007} to form a sample of 472 SNe. All of these supernovae were re-analyzed using both the {\tt SALT2} and {\tt SiFTO} light curve fitters, which give an estimate of the SN peak rest-frame B-band apparent magnitude at the epoch of maximum light in that filter.

\citet{sullivan/etal:2011} carry out a cosmological analysis of this combined data, accounting for systematic uncertainties including common photometric zero-point errors and selection effects.  They adopt a likelihood of the form
\be
-2\ln L = ({\bf m}_B - {\bf m}_{B}^{\rm th})^T {\bf C}^{-1} ({\bf m}_B - {\bf m}_{B}^{\rm th}),
\ee
where ${\bf m}_B$ is the peak-light apparent magnitude in B-band for each supernova, ${\bf m}_{B}^{\rm th}$ is the corresponding magnitude predicted by the model, and ${\bf C}$ is the covariance matrix of the data.  Their analysis assigns three terms to the covariance matrix, ${\bf C} = {\bf D}_{\rm stat} + {\bf C}_{\rm stat} + {\bf C}_{\rm sys}$, where ${\bf D}_{\rm stat}$ contains the independent (diagonal) statistical errors for each supernova, ${\bf C}_{\rm stat}$ includes the statistical errors that are correlated by the light-curve fitting, and ${\bf C}_{\rm sys}$ has eight terms to track systematic uncertainties, including calibration errors, Milky Way extinction, and redshift evolution.  The theoretical magnitude for each supernova is modeled as
\be
m_B^{\rm th} = 5\log[d_L(z)/\rm Mpc] - \alpha(s-1) + \beta C + \cal M,
\ee
where ${\cal M}$ is the empirical intercept of the $m_B^{\rm th} - z$ relation. The parameters $\alpha$ and $\beta$ quantify the stretch-luminosity and color-luminosity relationships, and the statistical error, ${\bf C}_{\rm stat}$, is coupled to both parameters.  Assuming a constant $w$ model, \citet{sullivan/etal:2011} measure $\alpha = 1.37 \pm 0.09$, and $\beta = 3.2 \pm 0.1$. Including the term ${\bf C}_{\rm sys}$ has a significant effect: it increases the error on the dark energy equation of state, $\sigma_w$ from $0.05$ to $0.08$ in a flat universe. 

The 472 Type Ia supernovae used in the SNLS3 analysis are consistent with the \lcdm\ model predicted by \wmap\ \citep{sullivan/etal:2011}, thus we can justify including these data in the present analysis.  However, the extensive study presented by the SNLS team shows that a significant level of systematic error still exists in current supernova observations.  Hence we restrict our use of supernova data in this paper to the subset of models that examine the dark energy equation of state.  When SNe data are included, we marginalize over the three parameters $\alpha$, $\beta$, and $M$. $\alpha$ and $\beta$ are sampled in the Markov Chain Monte Carlo (MCMC) chains, while $M$ is marginalized analytically \citep{lewis/bridle:2002}.   

\subsection{Markov Chain Methodology}
\label{sec:wmap_chains}

As with previous \wmap\ analyses, we use MCMC methods to evaluate the likelihood of cosmological parameters.  Aside from incorporating new likelihood codes for the external data sets described above, the main methodological update for the nine-year analysis centers on how we marginalize over SZ and point source amplitudes when analyzing multiple CMB data sets (i.e., ``\wmap+eCMB'').  We have also incorporated updates to the Code for Anisotropies in the Microwave Background \citep[{\tt CAMB},][]{lewis/challinor/lasenby:2000}, as described in \S\ref{sec:CAMB}.

{\bf SZ amplitude} - When combining data from multiple CMB experiments (\wmap, \act, \spt) we sample and marginalize over a single SZ amplitude, $A_{\rm SZ}$, that parameterizes the SZ contribution to all three data sets.  To do so, we adopt a common SZ power spectrum template, and scale it to each experiment as follows.  \citet{battaglia/etal:2011} compute a nominal SZ power spectrum at 150 GHz for the \spt\ experiment (their Figure 5, left panel, blue curve).  We adopt this curve as a spectral template and scale it by a factor of 1.05 and 3.6 to describe the relative SZ contribution at 148 GHz (for \act) and 61 GHz (for \wmap), respectively.  The nuisance parameter $A_{\rm SZ}$ then multiplies all three SZ spectra simultaneously.

The above frequency scaling assumes a thermal SZ spectrum.  For \wmap\ we assume an effective frequency of 61 GHz, even though the \wmap\ power spectrum includes 94 GHz data.  We ignore this error because \wmap\ data provide negligible constraints on the SZ amplitude when analyzed on their own.  In the \spt\ and \act\ frequency range, the thermal SZ spectrum is very similar to the kinetic SZ spectrum, so our procedure effectively accounts for that contribution as well.

{\bf Clustered point sources} - We adopt a common parameterization for the clustered point source contribution to both the \act\ and \spt\ data, namely $l(l+1)C_l/2\pi = A_{\rm cps} \, l^{0.8}$ \citep{addison/etal:2012}.  Both the \act\ and \spt\ teams use this form in their separate analyses at high $l$.  (At low $l$, the \spt\ group adopts a constant spectrum, but this makes a negligible difference to our analysis.)  By using a common amplitude for both experiments, we introduce one additional nuisance parameter.

{\bf Poisson point sources} - For unclustered residual point sources we adopt the standard power spectrum $C_l$ = const.  Since the \act\ and \spt\ groups use different algorithms for identifying and removing bright point sources, we allow the templates describing the residual power to have different amplitudes for the two experiments. This adds two additional nuisance parameters to our chains.

\subsubsection{CAMB}
\label{sec:CAMB}

Model power spectra are computed using the Code for Anisotropies in the Microwave Background \citep[{\tt CAMB},][]{lewis/challinor/lasenby:2000}, which is based on the earlier code {\tt CMBFAST} \citep{seljak/zaldarriaga:1996}.  We use the January 2012 version of CAMB throughout the nine-year analysis except when evaluating the ($w_0$, $w_a$) model, where we adopted the October 2012 version.  We adopt the default version of {\tt recfast} that is included with CAMB instead of other available options.  As in the seven-year analysis, we fix the reionization width to be $\Delta z = 0.5$. Since the \wmap\ likelihood code only incorporates low $l$ BB data (with low sensitivity), we set the {\tt accurate\_BB} flag to {\tt FALSE} and run the code with $C_l^{BB} = 0$.  We set the {\tt high\_accuracy\_default} flag to {\tt TRUE}.  When calling the \act\ and \spt\ likelihoods, we set {\tt k\_eta\_max\_scalar} = 15000 and {\tt l\_max\_scalar} = 6000.  The \act\ likelihood extends to $l=10000$, but foregrounds dominate beyond $l \approx 3000$ \citep{dunkley/etal:2011}, so this choice of $l_{\rm max}$ is conservative.  Except when exploring neutrino models, we adopt zero massive neutrinos and the nominal effective number of massless neutrino species.  The CMB temperature is set to 2.72548 K \citep{fixsen:2009}. 

\section{THE SIX-PARAMETER \lcdm\ MODEL}
\label{sec:lcdm}

In this section we discuss the determination of the standard \lcdm\ parameters, first using only the nine-year \wmap\ data, then, in turn, combined with the additional data sets discussed in \S\ref{sec:ext_data}.  Our analysis employs the same Monte Carlo Markov Chain (MCMC) formalism used in previous analyses \citep{spergel/etal:2003, verde/etal:2003, spergel/etal:2007, dunkley/etal:2009, komatsu/etal:2009, larson/etal:2011, komatsu/etal:2011}.  This formalism naturally produces parameter likelihoods that are marginalized over all other fit parameters in the model. Throughout this paper, we quote best-fit values as the mean of the marginalized likelihood, unless otherwise stated (e.g., mode or upper limits).  Lower and upper error limits correspond to the 16\% and 84\% points in the marginalized cumulative distribution, unless otherwise stated.  

The six parameters of the basic \lcdm\ model are: the physical baryon density, \ensuremath{\Omega_bh^2}; the physical cold dark matter density, \ensuremath{\Omega_ch^2}; the dark energy density, in units of the critical density, \ensuremath{\Omega_\Lambda}; the amplitude of primordial scalar curvature perturbations, $\Delta_{\cal R}^2$ at $k=0.002$ Mpc$^{-1}$; the power-law spectral index of primordial density (scalar) perturbations, \ensuremath{n_s}; and the reionization optical depth, \ensuremath{\tau}.  In this model, the Hubble constant, $H_0 = 100h$ km/s/Mpc, is implicitly determined by the flatness constraint, $\Omega_b + \Omega_c + \Omega_{\Lambda}=1$.  A handful of parameters in this model take assumed values that we further test in \S\ref{sec:lcdm_ext}; other parameters may be derived from the fit, as in Table~\ref{tab:lcdm_def}.  Throughout this paper we assume the initial fluctuations are adiabatic and Gaussian distributed (see \citet{bennett/etal:prep} for limits on non-Gaussian fluctuations from the nine-year \wmap\ data) except in \S\ref{sec:iso} where we allow the initial fluctuations to include an isocurvature component.

To assess \wmap\ data consistency, we begin with a comparison of the nine-year and seven-year results \citep{komatsu/etal:2011}; we then study the \lcdm\ constraints imposed by the nine-year \wmap\ data, in conjunction with the most recent external data sets available.

\begin{deluxetable}{llcc}
\tablecaption{Maximum Likelihood \lcdm\ Parameters\tablenotemark{a}\label{tab:lcdm_def}}
\tablewidth{0pt}
\tabletypesize{\footnotesize}
\tablehead{\colhead{Parameter} & \colhead{Symbol} & \colhead{\wmap\ data} & \colhead{Combined data\tablenotemark{b}}}
\startdata
\multicolumn{4}{c}{\bf Fit \lcdm\ parameters} \\[1mm]
Physical baryon density  
& \quad \ensuremath{\Omega_bh^2}
& \ensuremath{0.02256}
& \ensuremath{0.02240} \\ 
Physical cold dark matter density
& \quad \ensuremath{\Omega_ch^2} 
& \ensuremath{0.1142}
& \ensuremath{0.1146} \\
Dark energy density ($w=-1$)
& \quad \ensuremath{\Omega_\Lambda} 
& \ensuremath{0.7185}
& \ensuremath{0.7181} \\
Curvature perturbations, $k_0=0.002$ Mpc$^{-1}$
& \quad \ensuremath{10^9 \Delta_{\cal R}^2} 
& \ensuremath{2.40}
& \ensuremath{2.43} \\
Scalar spectral index
& \quad \ensuremath{n_s} 
& \ensuremath{0.9710}
& \ensuremath{0.9646} \\
Reionization optical depth
& \quad \ensuremath{\tau} 
& \ensuremath{0.0851}
& \ensuremath{0.0800} \\
\multicolumn{4}{c}{\bf Derived parameters} \\[1mm]
Age of the universe (Gyr)
& \quad \ensuremath{t_0}
& \ensuremath{13.76}
& \ensuremath{13.75} \\
Hubble parameter, $H_0 = 100h$ km/s/Mpc
& \quad \ensuremath{H_0}
& \ensuremath{69.7}
& \ensuremath{69.7} \\
Density fluctuations @ 8$h^{-1}$ Mpc
& \quad \ensuremath{\sigma_8} 
& \ensuremath{0.820}
& \ensuremath{0.817} \\
Baryon density/critical density
& \quad \ensuremath{\Omega_b} 
& \ensuremath{0.0464}
& \ensuremath{0.0461} \\
Cold dark matter density/critical density
& \quad \ensuremath{\Omega_c} 
& \ensuremath{0.235}
& \ensuremath{0.236} \\
Redshift of matter-radiation equality
& \quad \ensuremath{z_{\rm eq}} 
& \ensuremath{3273}
& \ensuremath{3280} \\
Redshift of reionization
& \quad \ensuremath{z_{\rm reion}} 
& \ensuremath{10.36}
& \ensuremath{9.97}
\enddata
\tablenotetext{a}{The maximum-likelihood \lcdm\ parameters for use in simulations.  Mean parameter values, with marginalized uncertainties, are reported in Table~\ref{tab:lcdm_wmap_ext}.}
\tablenotetext{b}{``Combined data'' refers to \wmap+eCMB+BAO+$H_0$.}
\end{deluxetable}

\subsection{Comparison With Seven-year Fits}
\label{sec:lcdm_7yr}

Table~\ref{tab:lcdm_wmap_79} gives the best-fit \lcdm\ parameters (mean and standard deviation, marginalized over all other parameters) for selected nine-year and seven-year data combinations.  In the case where only \wmap\ data are used, we evaluate parameters using both the $C^{-1}$-weighted spectrum and the MASTER-based one. For the case where we include BAO and $H_0$ priors, we use only the $C^{-1}$-weighted spectrum for the nine-year \wmap\ data, and we update the priors, as per \S\ref{sec:ext_data}.  The seven-year results are taken from Table~1 of \citet{komatsu/etal:2011}.

\subsubsection{\wmap\ Data Alone}
\label{sec:lcdm_wmap_cinv}

We first compare seven-year and nine-year results based on the MASTER spectra.  Table~\ref{tab:lcdm_wmap_79} shows that the nine-year \lcdm\ parameters are all within 0.5$\sigma$ of each other, with \ensuremath{\Omega_ch^2} having the largest difference.  We note that the combination $\ensuremath{\Omega_mh^2}+\ensuremath{\Omega_\Lambda}$ is approximately constant between the two models, reflecting the fact that this combination is well constrained by primary CMB fluctuations, whereas $\ensuremath{\Omega_m}-\ensuremath{\Omega_\Lambda}$ is less so due to the geometric degeneracy.  Turning to the $C^{-1}$-weighted spectrum, we note that the nine-year \lcdm\ parameters based on this spectrum are all within $\sim0.3\sigma$ of the seven-year values.  Thus we conclude that the nine-year model fits are consistent with the seven-year fit.  

\begin{figure}
\begin{center}
\includegraphics[width=0.8\textwidth]{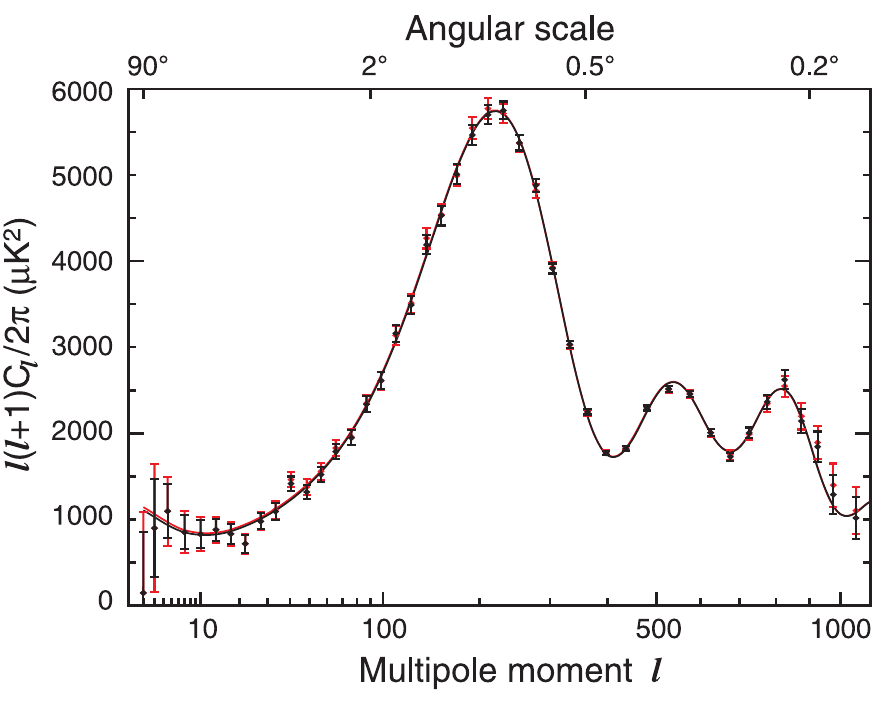}
\caption{Two estimates of the \wmap\ nine-year power spectrum along with the best-fit model spectra obtained from each; {\it black} - the $C^{-1}$-weighted spectrum and best fit model; {\it red} - the same for the MASTER spectrum and model.  The two spectrum estimates differ by up to 5\% in the vicinity of $l \sim 50$ which mostly affects the determination of the spectral index, \ensuremath{n_s}, as shown in Table~\ref{tab:lcdm_wmap_79}.  We adopt the $C^{-1}$-weighted spectrum throughout the remainder of this paper. \label{fig:Master_Cinv}}
\end{center}
\end{figure}

Next, we examine the consistency of the two \lcdm\ model fits, derived from the two nine-year spectrum estimates.  As seen in Table~\ref{tab:lcdm_wmap_79}, the six parameters agree reasonably well, but we note that the estimates for \ensuremath{n_s} differ by $0.75\sigma$, which we discuss below.  To help visualize the fits, we plot both spectra ($C^{-1}$-weighted and MASTER), and both models in Figure~\ref{fig:Master_Cinv}.  As noted in \citet{bennett/etal:prep}, the difference between the two spectrum estimates is most noticeable in the range $l \sim 30-60$ where the $C^{-1}$-weighted spectrum is lower than the MASTER spectrum, by up to 4\% in one bin.  However, the \lcdm\ model fits only differ noticeably for $l \la 10$ where the fit is relatively weakly constrained due to cosmic variance.

\begin{figure}
\begin{center}
\includegraphics[width=0.8\textwidth]{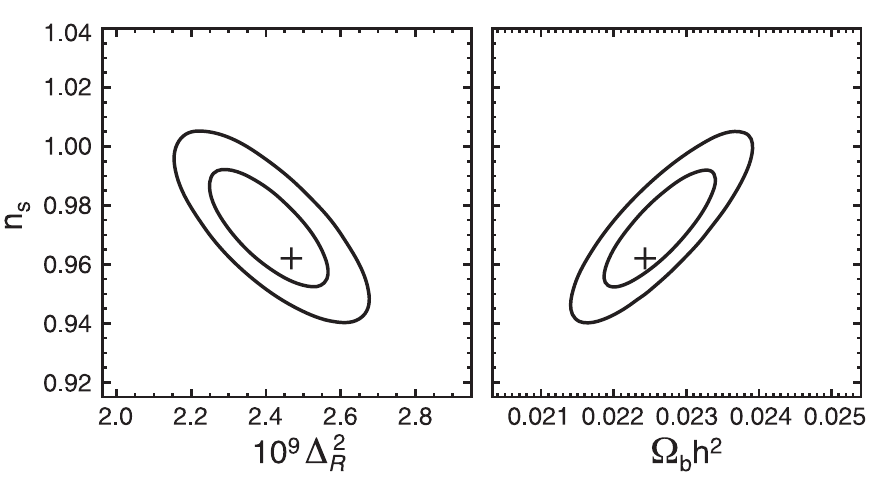}
\caption{68\% and 95\% CL regions for the \lcdm\ parameters \ensuremath{n_s}, \ensuremath{10^9 \Delta_{\cal R}^2}, and \ensuremath{\Omega_bh^2}.  There is a modest degeneracy between these three parameters in the six-parameter \lcdm\ model, when fit to the nine-year \wmap\ data.  The contours are derived from fits to the $C^{-1}$-weighted power spectrum, while the plus signs indicate the maximum likelihood point for the fit to the MASTER power spectrum.  As shown in Figure~\ref{fig:Master_Cinv}, the two model produce nearly identical spectra. \label{fig:ns_a_ob}}
\end{center}
\end{figure}

To understand why these two model spectra are so similar, we examine parameter degeneracies  between the six \lcdm\ parameters when fit to the nine-year \wmap\ data.  In Figure~\ref{fig:ns_a_ob} we show the two largest degeneracies that affect the spectral index \ensuremath{n_s}, namely \ensuremath{10^9 \Delta_{\cal R}^2} and \ensuremath{\Omega_bh^2}.  The contours show the 68\% and 95\% CL regions for the fits to the $C^{-1}$-weighted spectrum while the plus signs show the maximum likelihood points from the MASTER fit.  Note that the $C^{-1}$-weighted fits favor lower \ensuremath{10^9 \Delta_{\cal R}^2} and higher \ensuremath{\Omega_bh^2}, both of which push the $C^{-1}$-weighted fit towards higher \ensuremath{n_s}.  Given the consistency of the fit model spectra, we conclude that the underlying data are quite robust and in subsequent subsections, we look to external data to help break any degeneracies that remain in the nine-year data.

We conclude this subsection with a summary of some additional tests we carried out on simulations to assess the robustness of the $C^{-1}$-weighted spectrum estimate in general, and the \ensuremath{n_s} fits in particular.  The simulation data used were the 500 ``parameter recovery'' simulations developed for our seven-year analysis, described in detail in \citet{larson/etal:2011}.  These data include yearly sky maps for each differencing assembly, where the maps include simulated \lcdm\ signal (convolved with the appropriate beam) using the parameters given in Appendix A of \citet{larson/etal:2011}, and a model of correlated instrument noise appropriate to each differencing assembly.  For each realization in the simulation, we computed both the $C^{-1}$-weighted spectrum and the MASTER spectrum using the same prescription as was used for the flight data.  We found that both spectrum estimators were unbiased to within the standard spectrum errors divided by $\sqrt{500}$, i.e., to within the sensitivity of the test.

We next evaluated a number of difference statistics, but the one that was deemed most pertinent to understanding the \ensuremath{n_s} fit was the average power difference between $l=32-64$ (this is admittedly a posterior choice of $l$ range).  When the parameter recovery simulations were analyzed with the conservative KQ75y9 mask \citep{bennett/etal:prep}, more than one-third of the simulated spectrum pairs had a larger power difference ($C^{-1} -$MASTER, in the $l=32-64$ bin) than did the flight data.  This result indicates nothing unusual for that choice of mask.  However, when the same analysis was performed with the smaller KQ85y9 mask, only 2 out of 500 simulation realizations had a larger difference than did the flight data, which was 4\% in this bin.  Since the flight data appear to be unusual at the 0.4\% level (2/500) with the smaller mask, we suspected that (significant) residual foreground contamination might be present in the flight data, and that the two spectrum estimators might be responding to this differently.

To test this, we amended the CMB-only parameter recovery simulations with model foreground signals that we deemed to be representative of both the raw foreground signal outside the KQ85y9 mask, and an estimate of the residual contamination after template cleaning.  The simulated foreground signals were based on the modeling studies described in \citet{bennett/etal:prep}; in particular, the full-strength signal was based on the ``Model 9'' foreground model in \citet{bennett/etal:prep}, while the residual signal after cleaning was estimated from the $rms$ among the multiple foreground models studied.  With these foreground-contaminated simulations, we repeated the comparison of the two spectrum estimates considered above.  Both estimates showed slightly elevated power in the $l=32-64$ bin (a few percent), with the MASTER estimate being slightly more elevated.  However, the distribution of spectrum {\em differences} was not significantly different than with the CMB-only simulations: only 1\% of the simulated, foreground-contaminated difference spectra exceeded the difference seen in the flight data.  In the end, we attribute the spectrum differences to statistical fluctuations and we adopt the $C^{-1}$-weighted spectrum for our final analysis because it has lower uncertainties \citep{bennett/etal:prep} and because it was more stable to the introduction of foreground contamination in our simulations.   Nonetheless, we report \lcdm\ parameter fits for both spectrum estimates in Table~\ref{tab:lcdm_wmap_79} and Figure~\ref{fig:ns_a_ob} to give a sense of the potential systematic uncertainty in these parameters.

To conclude the seven-year/nine-year comparison, we note that the remaining 5 \lcdm\ parameters changed by less than $0.3 \sigma$ indicating very good consistency.  The overall effect of the nine-year \wmap\ data is to improve the average parameter uncertainty by about 10\%, with \ensuremath{\Omega_ch^2} and \ensuremath{\Omega_\Lambda} each improving by nearly 20\%.  The latter improvement is a result of higher precision in the third acoustic peak measurement \citep{bennett/etal:prep} which gives a better determination of \ensuremath{\Omega_ch^2}. This, in turn, improves \ensuremath{\Omega_\Lambda}, which is constrained by flatness (or in non-flat models, by the geometric degeneracy discussed in \S\ref{sec:olcdm}).  The overall volume reduction in the allowed 6-dimensional \lcdm\ parameter space in the switch from seven-year to nine-year data is a factor of 2, the majority of which derives from switching to the $C^{-1}$-weighted spectrum estimate.

\subsubsection{\wmap\ Data With BAO and $H_0$}
\label{sec:lcdm_7yr_ext}

To complete our comparison with seven-year results, we examine \lcdm\ fits that include the BAO and $H_0$ priors.  In \citet{komatsu/etal:2011} we argued that these two priors (then based on earlier data) provided the most robust and complementary parameter constraints, when used to supplement \wmap\ data.  In Table~\ref{tab:lcdm_wmap_79} we give results for the updated version of this data combination, which includes the nine-year $C^{-1}$-weighted spectrum for \wmap\ and the BAO and $H_0$ priors noted in \S\ref{sec:bao} and \S\ref{sec:h0}, respectively.  For comparison, we reproduce seven-year numbers from Table~1 of \citet{komatsu/etal:2011}.

As a measure of data consistency, we note that 4 of the 6 \lcdm\ parameters changed by less than $0.25 \sigma$ (in units of the seven-year $\sigma$) except for \ensuremath{\Omega_ch^2} and \ensuremath{\Omega_\Lambda} which changed by $\pm 0.8 \sigma$, respectively.  As noted above, these latter two parameters were more stable when fit to \wmap\ data alone (both in absolute value and in units of $\sigma$), so we conclude that this small change is primarily driven by the updated BAO and $H_0$ priors.  In particular, CMB data provide relatively weak constraints along the geometric degeneracy line (which corresponds to a line of nearly constant $\Omega_m + \Omega_{\Lambda}$ when spatial curvature is allowed), so external data are able to force limited anti-correlated changes in ($\Omega_m, \Omega_{\Lambda}$) with relatively little penalty from the \wmap\ likelihood.  In subsequent sections we explore the nine-year \lcdm\ fits more fully by adding external data sets to the \wmap\ data one at a time.

The combined effect of the nine-year \wmap\ data and updated the BAO and $H_0$ priors is to improve the average parameter uncertainty by nearly 25\%, with \ensuremath{\Omega_ch^2} and \ensuremath{\Omega_\Lambda} each improving by 37\%, due, in part, to improved constraints along the geometric degeneracy line.  The overall volume reduction in the allowed 6-dimensional \lcdm\ parameter space is a factor of 5, nearly half of which (a factor of 2) comes from the nine-year \wmap\ data alone.

\begin{deluxetable}{lcccccc}
\tablecaption{\wmap\ Seven-year to Nine-year Comparison of the Six-Parameter \lcdm\ Model\tablenotemark{a} \label{tab:lcdm_wmap_79}}
\tablewidth{0pt}
\tablehead{\colhead{\phn} & \multicolumn{3}{c}{\wmap-only\tablenotemark{b}} & \colhead{\phm{space}} & \multicolumn{2}{c}{\wmap+BAO+$H_0$\tablenotemark{b}} \\
\colhead{Parameter} & \colhead{Nine-year} & \colhead{Nine-year (MASTER)\tablenotemark{c}} & \colhead{Seven-year} & \colhead{\phm{space}} &
\colhead{Nine-year} & \colhead{Seven-year}}
\startdata
\multicolumn{4}{l}{Fit parameters} \\[1mm]
\quad \ensuremath{\Omega_bh^2}
& \ensuremath{0.02264\pm 0.00050}
& \ensuremath{0.02243\pm 0.00055}
& $0.02249^{+0.00056}_{-0.00057}$
& \phm{space}
& \ensuremath{0.02266\pm 0.00043}
& $0.02255 \pm 0.00054$ \\ 
\quad \ensuremath{\Omega_ch^2} 
& \ensuremath{0.1138\pm 0.0045}
& \ensuremath{0.1147\pm 0.0051}
& $0.1120 \pm 0.0056$
& \phm{space}
& \ensuremath{0.1157\pm 0.0023}
& $0.1126 \pm 0.0036$ \\
\quad \ensuremath{\Omega_\Lambda} 
& \ensuremath{0.721\pm 0.025}
& \ensuremath{0.716\pm 0.028}
& $0.727^{+0.030}_{-0.029}$
& \phm{space}
& \ensuremath{0.712\pm 0.010}
& $0.725 \pm 0.016$ \\
\quad \ensuremath{10^9 \Delta_{\cal R}^2} 
& \ensuremath{2.41\pm 0.10}
& \ensuremath{2.47\pm 0.11}
& $2.43 \pm 0.11$
& \phm{space}
& \ensuremath{2.427^{+ 0.078}_{- 0.079}}
& $2.430 \pm 0.091$ \\
\quad \ensuremath{n_s} 
& \ensuremath{0.972\pm 0.013}
& \ensuremath{0.962\pm 0.014}
& $0.967 \pm 0.014$
& \phm{space}
& \ensuremath{0.971\pm 0.010}
& $0.968 \pm 0.012$ \\
\quad \ensuremath{\tau} 
& \ensuremath{0.089\pm 0.014}
& \ensuremath{0.087\pm 0.014}
& $0.088 \pm 0.015$
& \phm{space}
& \ensuremath{0.088\pm 0.013}
& $0.088 \pm 0.014$ \\[1mm]
\multicolumn{4}{l}{Derived parameters} \\[1mm]
\quad \ensuremath{t_0} (Gyr)
& \ensuremath{13.74\pm 0.11}
& \ensuremath{13.75\pm 0.12}
& $13.77 \pm 0.13$
& \phm{space}
& \ensuremath{13.750\pm 0.085}
& $13.76 \pm 0.11$ \\
\quad \ensuremath{H_0} (km/s/Mpc)
& \ensuremath{70.0\pm 2.2}
& \ensuremath{69.7\pm 2.4}
& $70.4 \pm 2.5$
& \phm{space}
& \ensuremath{69.33\pm 0.88}
& $70.2 \pm 1.4$ \\
\quad \ensuremath{\sigma_8} 
& \ensuremath{0.821\pm 0.023}
& \ensuremath{0.818\pm 0.026}
& $0.811^{+0.030}_{-0.031}$
& \phm{space}
& \ensuremath{0.830\pm 0.018}
& $0.816 \pm 0.024$ \\
\quad \ensuremath{\Omega_b}
& \ensuremath{0.0463\pm 0.0024}
& \ensuremath{0.0462\pm 0.0026}
& $0.0455 \pm 0.0028$
& \phm{space}
& \ensuremath{0.0472\pm 0.0010}
& $0.0458 \pm 0.0016$ \\
\quad \ensuremath{\Omega_c} 
& \ensuremath{0.233\pm 0.023}
& \ensuremath{0.237\pm 0.026}
& $0.228 \pm 0.027$
& \phm{space}
& \ensuremath{0.2408^{+ 0.0093}_{- 0.0092}}
& $0.229 \pm 0.015$ \\
\quad \ensuremath{z_{\rm reion}} 
& \ensuremath{10.6\pm 1.1}
& \ensuremath{10.5\pm 1.1}
& $10.6 \pm 1.2$
& \phm{space}
& \ensuremath{10.5\pm 1.1}
& $10.6 \pm 1.2$
\enddata
\tablenotetext{a}{Comparison of 6-parameter \lcdm\ fits with seven-year and nine-year \wmap\ data, with and without BAO and $H_0$ priors.}
\tablenotetext{b}{The first three data columns give results from fitting to \wmap\ data only.  The last two columns give results when BAO and $H_0$ priors are added.  As discussed in \S\ref{sec:ext_data}, these priors have been updated for the nine-year analysis.  The seven-year results are taken directly from Table~1 of \citet{komatsu/etal:2011}.}
\tablenotetext{c}{Unless otherwise noted, the nine-year \wmap\ likelihood uses the $C^{-1}$-weighted power spectrum whereas the seven-year likelihood used the MASTER-based spectrum.  The column labeled `Nine-year (MASTER)' is a special run for comparing to the seven-year results.}
\end{deluxetable}

\subsection{\lcdm\ Constraints From CMB Data}
\label{sec:lcdm_cmb}

From the standpoint of astrophysics, primary CMB fluctuations, combined with CMB lensing, arguably provide the cleanest probe of cosmology because the fluctuations dominate Galactic foreground emission over most of the sky, and they can (so far) be understood in terms of linear perturbation theory and Gaussian statistics.  Thus we next consider parameter constraints that can be obtained when adding additional CMB data to the nine-year \wmap\ data.  Specifically, we examine the effects of adding \spt\ and \act\ data (see \S\ref{sec:cmb}): the best-fit parameters are given in the ``+eCMB'' column of Table~\ref{tab:lcdm_wmap_ext}.  

With the addition of the high-$l$ CMB data, the constraints on the energy density parameters \ensuremath{\Omega_bh^2}, \ensuremath{\Omega_ch^2}, and \ensuremath{\Omega_\Lambda} all improve by 25\% over the precision from \wmap\ data alone.  The improvement in the baryon density measurement is due to more precise measurements of the Silk damping tail in the power spectrum at $l \gtrsim 1000$; the improvements in \ensuremath{\Omega_ch^2} and \ensuremath{\Omega_\Lambda} are due in part to improvements in the high-$l$ TT data, but also to the detection of CMB lensing in the \spt\ and \act\ data \citep{das/etal:2011b, vanengelen/etal:2012}, which helps to constrain $\Omega_m$ by fixing the growth rate of structure between $z=1100$ and $z=1-2$ (the peak in the lensing kernel).  Taken together, CMB data available at the end of the \wmap\ mission produce a 1.6\% measurement of \ensuremath{\Omega_bh^2} and a 3.0\% measurement of \ensuremath{\Omega_ch^2}.

The increased $k$-space lever arm provided by the high-$l$ CMB data improves the uncertainty on the scalar spectral index by 25\%, giving \ensuremath{n_s} = \ensuremath{0.9646\pm 0.0098}, which implies a non-zero tilt in the primordial spectrum (i.e., $n_s < 1$) at 3.6$\sigma$.  We examine the implications of this measurement for inflation models in \S\ref{sec:primordial}.

If we assume a flat universe, which breaks the CMB's geometric degeneracy, then CMB data alone now provide a 2.3\% measurement of the Hubble parameter, \ensuremath{H_0} = \ensuremath{70.5\pm 1.6\ \mbox{km/s/Mpc}}, independent of the cosmic distance ladder.  As discussed in \S\ref{sec:lcdm_cmb_H0}, this is consistent with the recent determination of the Hubble parameter using the cosmic distance ladder: $H_0 = 73.8 \pm 2.4$ km/s/Mpc \citep{riess/etal:2011}; we explore the effect of adding this prior in \S\ref{sec:lcdm_cmb_H0}.  We relax the assumption of flatness in \S\ref{sec:olcdm}.

We conclude by comparing our results for the \act\ and \spt\ foreground ``nuisance'' parameters to those found by the \act\ and \spt\ teams.  For example, we find $\ensuremath{A_{\rm Poisson}^{\rm ACT}} = \ensuremath{14.8^{+ 2.3}_{- 2.4}}$ while the \act\ team finds $\ensuremath{A_{\rm Poisson}^{\rm ACT}} = 12.0 \pm 1.9$.  (Note that we do not expect perfect agreement because we use nine-year \wmap\ data and we fit the clustered source amplitude jointly with \spt\ data, unlike the \act\ team's treatment.)  The \act\ team conclude that ``The \lcdm\ cosmological model (fit to) the 148 GHz spectrum (and the seven-year \wmap\ data), marginalized over SZ and source power is a good fit to the data'' \citep{dunkley/etal:2011}.  The complete set of foreground parameters fit to the \act\ and \spt\ data may be found at {\tt http://lambda.gsfc.nasa.gov/} for all the models reported in this paper.

\subsection{Adding BAO Data}
\label{sec:lcdm_cmb_bao}

Acoustic structure in the large scale distribution of galaxies is manifest on a co-moving scale of 152 Mpc, where the evolution of matter fluctuations is largely within the linear regime.  A number of authors have studied the degree to which the acoustic structure could be perturbed by nonlinear evolution \citep[e.g.,][]{seo/eisenstein:2005,seo/eisenstein:2007,jeong/komatsu:2006,jeong/komatsu:2009,crocce/scoccimarro:2008,matsubara:2008,taruya/hiramatsu:2008,padmanabhan/white:2009}, and the effects are well below the current measurement uncertainties.  Because it is based on the same well-understood physics that governs the CMB anisotropy, we consider measurements of the BAO scale to be the next-most robust cosmological probe after CMB fluctuations.  The \lcdm\ parameters fit to CMB and BAO data are given in the ``+eCMB+BAO'' column of Table~\ref{tab:lcdm_wmap_ext}.

Measurements of the tangential and radial BAO scale at redshift $z$ measure the effective distance $D_V(z)$, given in equation~(\ref{eq:dvz}), in units of the sound horizon \ensuremath{r_s(z_d)}.  This quantity is primarily sensitive to the total matter and dark energy densities, and to the current Hubble parameter.  Since the BAO scale is relatively insensitive to the baryon density, \ensuremath{\Omega_bh^2}, this parameter does not improve significantly with the addition of the BAO prior.  However, the low-redshift distance information imposes complementary constraints on the matter density and Hubble parameter, improving the precision on \ensuremath{\Omega_ch^2} from 3.0\% to 1.6\%, and on \ensuremath{H_0} from 2.3\% to 1.2\%.  In the context of standard \lcdm\, these improvements lead to a measurement of the age of the universe with 0.4\% precision: \ensuremath{t_0} = \ensuremath{13.800\pm 0.061\ \mbox{Gyr}}.

\begin{figure}[ht]
\begin{center}
\includegraphics[width=.9\textwidth]{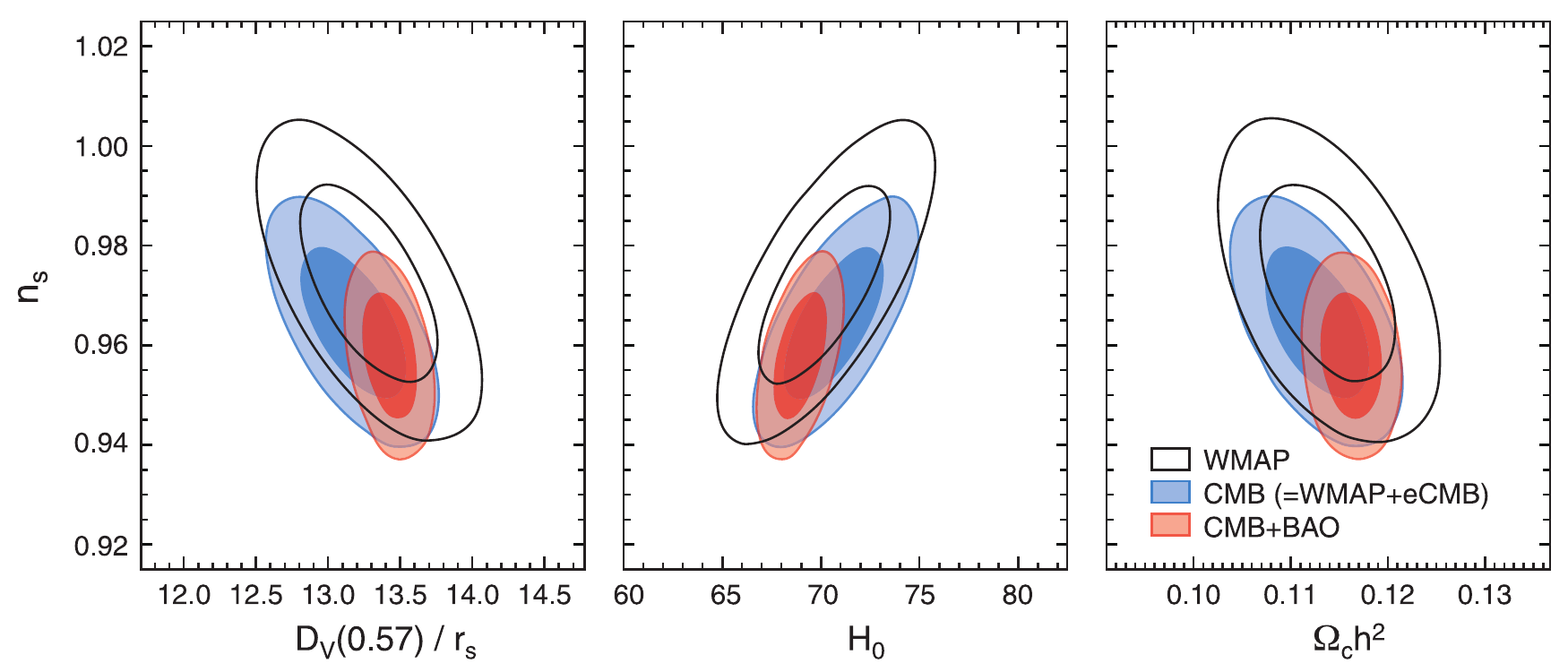}
\caption{Measurements of the scalar spectral index with CMB and BAO data. {\it Left to right} - contours of ($D_V(0.57)/r_s$,\ensuremath{n_s}), (\ensuremath{H_0},\ensuremath{n_s}), (\ensuremath{\Omega_ch^2},\ensuremath{n_s}).  Black contours show constraints using \wmap\ nine-year data alone; blue contours include \spt\ and \act\ data (\WMAP+eCMB); red contours add the BAO prior(\WMAP+eCMB+BAO).  The BAO prior provides an independent measurement of the low-redshift distance, $D_v(z)/r_s,$ which maps to constraints on \ensuremath{\Omega_ch^2} and \ensuremath{H_0}.  When combined with CMB data, the joint constraints require a tilt in the primordial spectral index (\ensuremath{n_s}$<1$) at the 5$\sigma$ level. \label{fig:ns_bao}}
\end{center}
\end{figure}

The addition of the BAO prior helps to break some residual degeneracy between the primordial spectral index, \ensuremath{n_s}, on the one hand, and \ensuremath{\Omega_ch^2} and \ensuremath{H_0} on the other.  Figure~\ref{fig:ns_bao} shows the 2-dimensional parameter likelihoods for (\ensuremath{n_s},\ensuremath{\Omega_ch^2}) and (\ensuremath{n_s},\ensuremath{H_0}) for the three data combinations considered to this point. With only CMB data (black and blue contours) there remains a weak degeneracy between \ensuremath{n_s} and the other two.  When the BAO prior is added (red), it pushes \ensuremath{\Omega_ch^2} towards the upper end of the range allowed by the CMB, and vice versa for \ensuremath{H_0}.  Both of these results push $n_s$ towards the lower end of its CMB-allowable range; consequently, with the BAO prior included, the marginalized measurement of the primordial spectral index is \ensuremath{n_s} = \ensuremath{0.9579^{+ 0.0081}_{- 0.0082}} {\em which constitutes a 5$\sigma$ measurement of tilt} ($\ensuremath{n_s} < 1$) {\em in the primordial spectrum.}  We discuss the implications of this measurement for inflation models in \S\ref{sec:primordial}.

\begin{deluxetable}{lcccccc}
\tablecaption{Six-Parameter \lcdm\ Fit; \wmap\ plus External Data\tablenotemark{a}\label{tab:lcdm_wmap_ext}}
\tablewidth{0pt}
\tabletypesize{\footnotesize}
\tablehead{\colhead{Parameter} & \colhead{\wmap} & \colhead{+eCMB}& \colhead{+eCMB+BAO}  & \colhead{+eCMB+$H_0$} & \colhead{+eCMB+BAO+$H_0$}}
\startdata
\multicolumn{3}{l}{Fit parameters} \\[1mm]
\quad \ensuremath{\Omega_bh^2}
& \ensuremath{0.02264\pm 0.00050}
& \ensuremath{0.02229\pm 0.00037}
& \ensuremath{0.02211\pm 0.00034}
& \ensuremath{0.02244\pm 0.00035}
& \ensuremath{0.02223\pm 0.00033} \\
\quad \ensuremath{\Omega_ch^2} 
& \ensuremath{0.1138\pm 0.0045}
& \ensuremath{0.1126\pm 0.0035}
& \ensuremath{0.1162\pm 0.0020}
& \ensuremath{0.1106\pm 0.0030}
& \ensuremath{0.1153\pm 0.0019} \\
\quad \ensuremath{\Omega_\Lambda} 
& \ensuremath{0.721\pm 0.025}
& \ensuremath{0.728\pm 0.019}
& \ensuremath{0.707\pm 0.010}
& \ensuremath{0.740\pm 0.015}
& \ensuremath{0.7135^{+ 0.0095}_{- 0.0096}} \\
\quad \ensuremath{10^9 \Delta_{\cal R}^2} 
& \ensuremath{2.41\pm 0.10}
& \ensuremath{2.430\pm 0.084}
& \ensuremath{2.484^{+ 0.073}_{- 0.072}}
& \ensuremath{2.396^{+ 0.079}_{- 0.078}}
& \ensuremath{2.464\pm 0.072} \\
\quad \ensuremath{n_s} 
& \ensuremath{0.972\pm 0.013}
& \ensuremath{0.9646\pm 0.0098}
& \ensuremath{0.9579^{+ 0.0081}_{- 0.0082}}
& \ensuremath{0.9690^{+ 0.0091}_{- 0.0090}}
& \ensuremath{0.9608\pm 0.0080} \\
\quad \ensuremath{\tau} 
& \ensuremath{0.089\pm 0.014}
& \ensuremath{0.084\pm 0.013}
& \ensuremath{0.079^{+ 0.011}_{- 0.012}}
& \ensuremath{0.087\pm 0.013}
& \ensuremath{0.081\pm 0.012} \\[1mm]
\multicolumn{3}{l}{Derived parameters} \\[1mm]
\quad \ensuremath{t_0} (Gyr)
& \ensuremath{13.74\pm 0.11}
& \ensuremath{13.742\pm 0.077}
& \ensuremath{13.800\pm 0.061}
& \ensuremath{13.702\pm 0.069}
& \ensuremath{13.772\pm 0.059} \\
\quad \ensuremath{H_0} (km/s/Mpc)
& \ensuremath{70.0\pm 2.2}
& \ensuremath{70.5\pm 1.6}
& \ensuremath{68.76\pm 0.84}
& \ensuremath{71.6\pm 1.4}
& \ensuremath{69.32\pm 0.80} \\
\quad \ensuremath{\sigma_8} 
& \ensuremath{0.821\pm 0.023}
& \ensuremath{0.810\pm 0.017}
& \ensuremath{0.822^{+ 0.013}_{- 0.014}}
& \ensuremath{0.803\pm 0.016}
& \ensuremath{0.820^{+ 0.013}_{- 0.014}} \\
\quad \ensuremath{\Omega_b} 
& \ensuremath{0.0463\pm 0.0024}
& \ensuremath{0.0449\pm 0.0018}
& \ensuremath{0.04678\pm 0.00098}
& \ensuremath{0.0438\pm 0.0015}
& \ensuremath{0.04628\pm 0.00093} \\
\quad \ensuremath{\Omega_c} 
& \ensuremath{0.233\pm 0.023}
& \ensuremath{0.227\pm 0.017}
& \ensuremath{0.2460\pm 0.0094}
& \ensuremath{0.216\pm 0.014}
& \ensuremath{0.2402^{+ 0.0088}_{- 0.0087}} \\
\quad \ensuremath{z_{\rm eq}} 
& \ensuremath{3265^{+ 106}_{- 105}}
& \ensuremath{3230\pm 81}
& \ensuremath{3312\pm 48}
& \ensuremath{3184\pm 70}
& \ensuremath{3293\pm 47} \\
\quad \ensuremath{z_{\rm reion}} 
& \ensuremath{10.6\pm 1.1}
& \ensuremath{10.3\pm 1.1}
& \ensuremath{10.0\pm 1.0}
& \ensuremath{10.5\pm 1.1}
& \ensuremath{10.1\pm 1.0}
\enddata
\tablenotetext{a}{\lcdm\ model fit to \wmap\ nine-year data combined with a progression of external data sets.  A complete list of parameter values for this model, with additional data combinations, may be found at {\tt http://lambda.gsfc.nasa.gov/}.}
\end{deluxetable}

\subsection{Adding $H_0$ Data}
\label{sec:lcdm_cmb_H0}

Measurements of the Hubble parameter using the cosmic distance ladder have a long history, and are subject to a variety of different systematic errors that have been steadily reduced over time.  However, an accurate, direct measurement of the current expansion rate is vital for testing the validity of the \lcdm\ model because the value derived from the CMB and BAO data is model-dependent.  Measurements of $H_0$ provide an excellent complement to CMB and BAO measurements.  The $H_0$ prior considered here has a precision that approaches the \lcdm-based value given above.  Consequently, we next consider the addition of the \ensuremath{H_0} prior discussed in \S\ref{sec:h0}, {\em without} the inclusion of the BAO prior.  The \lcdm\ parameters fit to CMB and \ensuremath{H_0} data are given in ``+eCMB+$H_0$'' column of Table~\ref{tab:lcdm_wmap_ext}.

\begin{figure}[ht]
\begin{center}
\includegraphics[width=.50\textwidth]{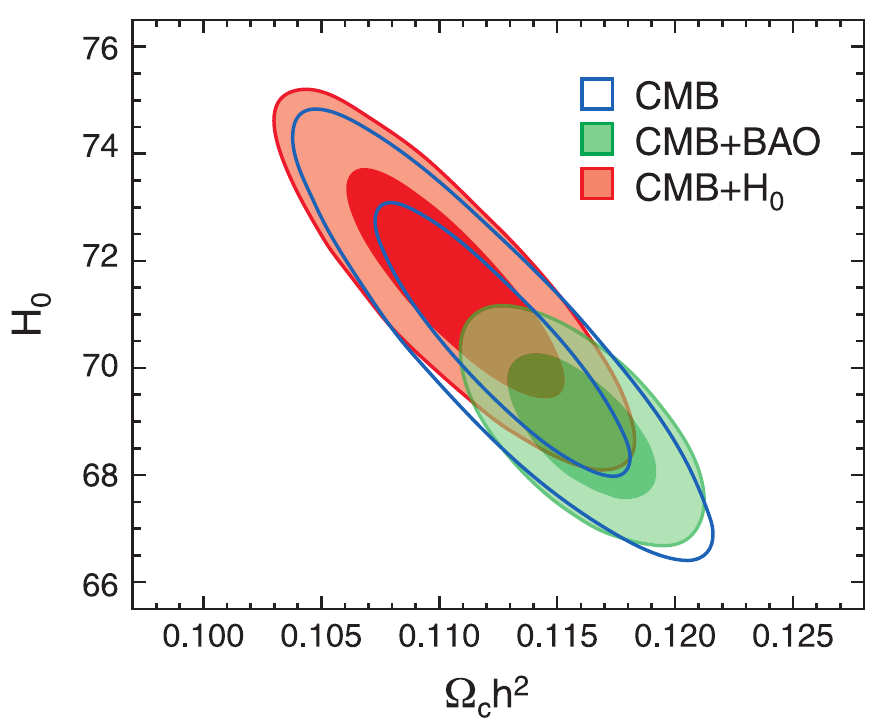}
\caption{Measurements of \ensuremath{\Omega_ch^2} and \ensuremath{H_0} from CMB data only (blue contours, \wmap+eCMB), from CMB and BAO data (green contours, \wmap+eCMB+BAO), and from CMB and $H_0$ data (red contours, \wmap+eCMB+$H_0$).  The two non-CMB priors push the constraints towards opposite ends of the range allowed by the CMB alone, but they are not inconsistent. \label{fig:bao_h0}}
\end{center}
\end{figure}

Two cosmological quantities that significantly shape the observed CMB spectrum are the epoch of matter radiation equality, \ensuremath{z_{\rm eq}}, which depends on \ensuremath{\Omega_ch^2}, and the angular diameter distance to the last scattering surface, \ensuremath{d_A(z_{*})}, which depends primarily on \ensuremath{H_0}.  As illustrated in Figure~\ref{fig:bao_h0} (see also \S\ref{sec:neff_spec}), the CMB data still admit a weak degeneracy between \ensuremath{\Omega_ch^2} and \ensuremath{H_0} that the BAO and $H_0$ priors help to break.  The black contours in Figure~\ref{fig:bao_h0} show the constraints from CMB data (\wmap+eCMB), the red from CMB and BAO data, and the blue from CMB with the $H_0$ prior.  While these measurements are all consistent, it is interesting to note that the BAO and $H_0$ priors are pushing towards opposite ends of the range allowed by the CMB data for this pair of parameters.  Given this minor tension, it is worth examining independent sets of constraints that do not share common CMB data.  A simple test is to compare the marginalized constraints on the Hubble parameter from the CMB+BAO data (\ensuremath{H_0} =  \ensuremath{68.76\pm 0.84\ \mbox{km/s/Mpc}}), to the direct, and independent, measurement from the distance ladder ($H_0 = 73.8 \pm 2.4$ km/s/Mpc).  In our Markov Chain that samples the \lcdm\ model with the \WMAP+eCMB+BAO data, we found that only 0.1\% of the $H_0$ values in the chain fell within the 1$\sigma$ range of the Hubble prior, but that 45\% fell within the 2$\sigma$ range of $73.8 \pm 4.8$ km/s/Mpc.  Based on this, we conclude that these measurements do not disagree, and that they may be combined to form more stringent constraints on the \lcdm\ parameters.

We conclude this subsection by noting that measurements of the remaining \lcdm\ parameters  are modestly improved by the addition of the $H_0$ prior to the CMB data, with \ensuremath{\Omega_ch^2} improving the most due to the effect discussed above and illustrated in Figure~\ref{fig:bao_h0}.

\subsection{\lcdm\ Fits to the Combined Data}
\label{sec:lcdm_cmb_bao_h0}

Given the consistency of the data sets considered above, we conclude with a summary of the \lcdm\ fits derived from the union of these data.  The marginalized results are given in the ``+eCMB+BAO+$H_0$'' column of Table~\ref{tab:lcdm_wmap_ext}.  The matter and energy densities, \ensuremath{\Omega_bh^2}, \ensuremath{\Omega_ch^2}, and \ensuremath{\Omega_\Lambda} are all now determined to $\sim$1.5\% precision with the current data.  The amplitude of the primordial spectrum is measured to within 3\%, and there is now evidence for tilt in the primordial spectrum at the 5$\sigma$ level.  

\begin{figure}
\begin{center}
\includegraphics[width=0.9\textwidth]{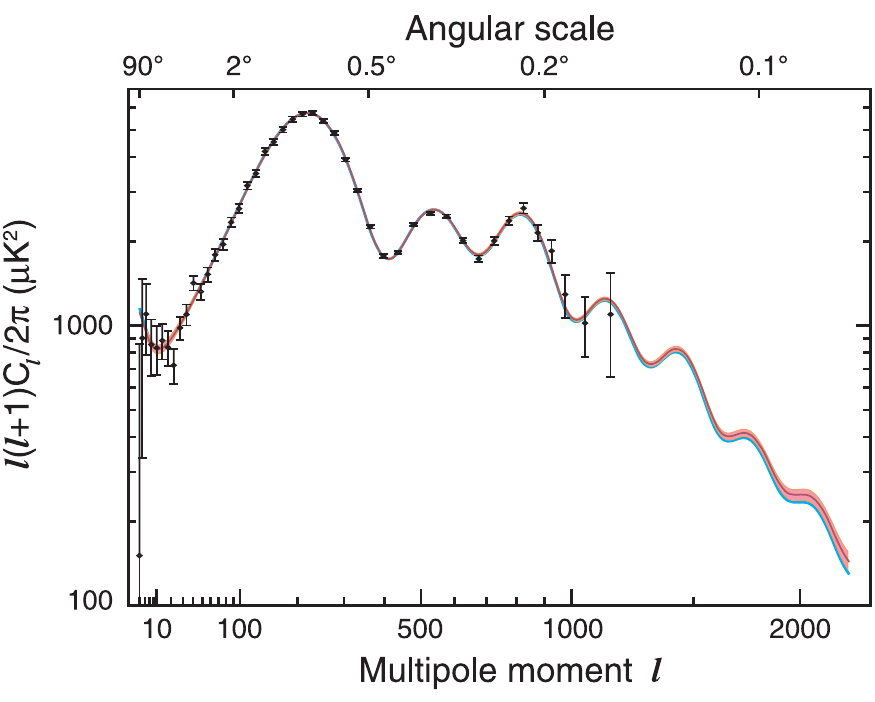}
\caption{The nine-year \wmap\ data (in black) are shown with the 1$\sigma$ locus of six-parameter \lcdm\ models allowed by the nine-year \wmap\ data.  The error band is derived from the Markov Chain of six-parameter model fits to the \wmap\ data alone.  The blue curve indicates the mean of the \lcdm\ model fit to the \wmap+eCMB data combination.  At high-$l$ this curve sits about 1$\sigma$ below the model fit to \wmap\ data alone.  The marginalized parameter constraints that define these models are given in the \wmap\ and \wmap+eCMB columns of Table~\ref{tab:lcdm_wmap_ext}. \label{fig:lcdm_range}}
\end{center}
\end{figure}

At the end of the \wmap\ mission, the nine-year data produced a factor of 68,000 decease in the allowable volume of the six-dimensional \lcdm\ parameter space, relative to the pre-\wmap\ measurements \citep{bennett/etal:prep}.  Specifically, the allowable volume is measured by the square root of the determinant of the $6 \times 6$ parameter covariance matrix, as discussed in \citet{larson/etal:2011}.  The pre-\wmap\ volume is determined from chains run with the data compiled by \citet{wang/etal:2002}.  When these data are combined with the eCMB+BAO+$H_0$ priors, we obtain an additional factor of 27 over the \wmap-only constraints.  As an illustration of the predictive power of the current data, Figure~\ref{fig:lcdm_range} shows the 1$\sigma$ range of high-l power spectra allowed by the six-parameter fits to the nine-year \wmap\ data.  As shown in Figure~\ref{fig:cl_all}, this model has already predicted the {\em current} small-scale measurements.  If future measurements of the spectrum, for example by \planck, lie outside this range, then either there is a problem with the six-parameter model, or a problem with the data.

Remarkably, despite this dramatic increase in precision, the six-parameter \lcdm\ model {\em still} produces an acceptable fit to all of the available data.  \citet{bennett/etal:prep} present a detailed breakdown of the goodness of fit to the nine-year \wmap\ data.  In the next section we place limits on parameters beyond the six {\em required} to describe our universe.

\section{BEYOND SIX-PARAMETER \lcdm}
\label{sec:lcdm_ext}

In this section we discuss constraints that can be placed on cosmological parameters beyond the standard model using the nine-year \wmap\ data combined with the external data sets discussed in \S\ref{sec:ext_data}.  In the following subsections, we consider limits that can be placed on additional parameters one or two at a time, beginning with constraints on initial conditions and proceeding through to the late-time effects of dark energy.

\subsection{Primordial Spectrum and Gravitational Waves}
\label{sec:primordial}

As noted in \S\ref{sec:lcdm_cmb_bao_h0}, the nine-year \wmap\ data, when combined with eCMB, BAO and $H_0$ priors, exclude a scale-invariant primordial power spectrum at 5$\sigma$ significance. For a power-law spectrum of primordial curvature perturbations,
\be
\Delta^2_{\cal R}(k) = \Delta^2_{\cal R}(k_0) \left(\frac{k}{k_0}\right)^{n_s-1},
\ee
with $k_0 = 0.002$ Mpc$^{-1}$, we find \ensuremath{n_s} = \ensuremath{0.9608\pm 0.0080}.  This result assumes that tensor modes (gravitational waves) contribute insignificantly to the CMB anisotropy.

At this time, the most sensitive limits on tensor modes are still obtained from the shape of the temperature power spectrum, in conjunction with additional data. For example, \citet{story/etal:2012} report $r < 0.18$ (95\% CL), where
\be
r \equiv \frac{\Delta^2_h(k_0)}{\Delta^2_{\cal R}(k_0)} = \frac{P_h(k_0)}{P_{\cal R}(k_0)}.
\ee
Due to confusion from density fluctuations, the lowest tensor amplitude that can be reliably detected from temperature data is \ensuremath{r} $\lesssim 0.13$ \citep{knox:1995}.  Several recent experiments are beginning to establish comparable limits from non-detection of B-mode polarization anisotropy, e.g., \citet{chiang/etal:2010} report $r<0.7$ (95\% CL) from BICEP and the \citet{quiet2:2012} reports $r<2.8$ (95\% CL).  A host of forthcoming experiments are targeting B-mode measurements that have the potential to detect or limit tensor modes at significantly lower levels than can be achieved with temperature data alone.  (Note that E-mode polarization, like temperature anisotropy, is dominated by scalar fluctuations, and since the E-mode signal is more than an order of magnitude weaker than the temperature signal, it contributes negligibly to constraints on tensor fluctuations.)

\begin{figure}[ht]
\begin{center}
\includegraphics{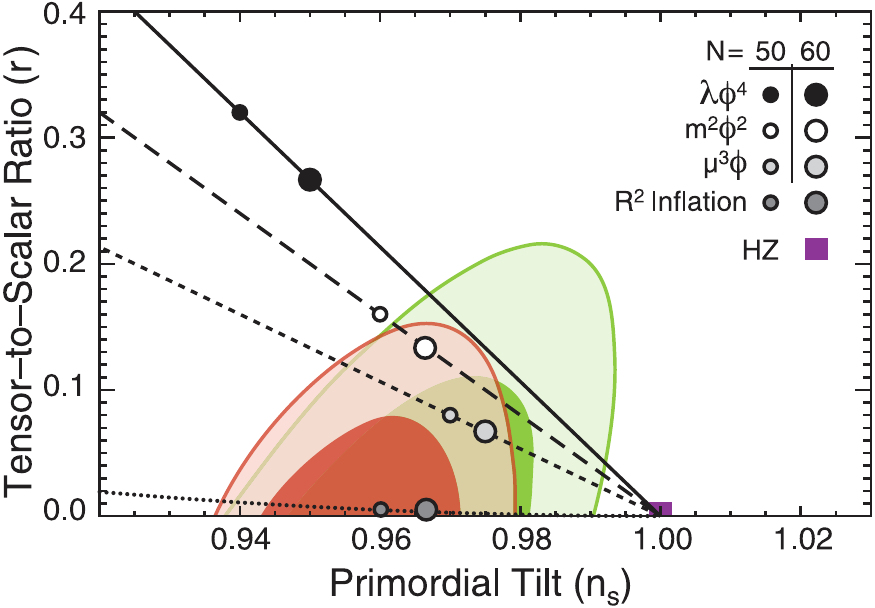}
\caption{Two-dimensional marginalized constraints (68\% and 95\% CL) on the primordial tilt, \ensuremath{n_s}, and the tensor-to-scalar ratio, \ensuremath{r}, derived with the nine-year \wmap\ in conjunction with: eCMB (green) and eCMB+BAO+$H_0$ (red). The symbols and lines show predictions from single-field inflation models whose potential is given by $V(\phi) \propto \phi^\alpha$ \citep{linde:1983}, with $\alpha=4$ (solid), $\alpha=2$ (long-dashed), and $\alpha=1$ \citep[short-dashed;][]{mcallister/silverstein/westphal:2010}. Also shown are those from the first inflation model, which is based on an $R^2$ term in the gravitational Lagrangian \citep[dotted;][]{starobinsky:1980}. Starobinsky's model gives $n_s=1-2/N$ and $r=12/N^2$ where $N$ is the number of $e$-folds between the end of inflation and the epoch at which the scale $k=0.002~{\rm Mpc}^{-1}$ left the horizon during inflation. These predictions are the same as those of inflation models with a $\xi\phi^2R$ term in the gravitational Lagrangian with a $\lambda\phi^4$ potential \citep{komatsu/futamase:1999}. See  Appendix~\ref{sec:starobinsky} for details. \label{fig:ns-r}} 
\end{center}
\end{figure}

In Table~\ref{tab:pri_spec}, we report limits on \ensuremath{r} from the nine-year \wmap\ data, analyzed alone and jointly with external data; the tightest constraint is
$$
\ensuremath{r < 0.13\ \mbox{(95\% CL)}} \qquad \mbox{\wmap+eCMB+BAO+}H_0.
$$
This is effectively at the limit one can reach without B-mode polarization measurements.  The joint constraints on \ensuremath{n_s} and \ensuremath{r} are shown in Figure~\ref{fig:ns-r}, along with selected model predictions derived from single-field inflation models.  Taken together, the current data strongly disfavor a pure Harrison-Zel'dovich (HZ) spectrum, even if tensor modes are allowed in the model fits.

\subsubsection{Running Spectral Index}
\label{sec:running}

Some inflation models predict a scale dependence or ``running'' in the (nearly) power-law spectrum of scalar perturbations.  This is conveniently parameterized by the logarithmic derivative of the spectral index, \ensuremath{dn_s/d\ln{k}}, which gives rise to a spectrum of the form \citep{kosowsky/turner:1995}
\be
\Delta^2_{\cal R}(k) = \Delta^2_{\cal R}(k_0) \left(\frac{k}{k_0}\right)^{n_s(k_0) - 1 + \frac12 \ln(k/k_0) \ensuremath{dn_s/d\ln{k}}}.
\ee
We do not detect a statistically significant deviation from a pure power-law spectrum with the nine-year \wmap\ data.  The allowed range of \ensuremath{dn_s/d\ln{k}} is both closer to zero and has a smaller confidence range with the nine-year data, \ensuremath{dn_s/d\ln{k} = -0.019\pm 0.025}.  However, with the inclusion of the high-$l$ CMB data, the full CMB data prefer a slightly more negative value, with a smaller uncertainty, \ensuremath{dn_s/d\ln{k} = -0.022^{+ 0.012}_{- 0.011}}.  While not significant, this result might indicate a trend as the $l$-range of the data expand.  The inclusion of BAO and $H_0$ data does not affect these results.

If we allow {\it both} tensors and running as additional primordial degrees of freedom, the data prefer a slight negative running, but still at less than $3\sigma$ significance, and only with the inclusion of the high-$l$ CMB data.  Complete results are given in Table~\ref{tab:pri_spec}.

\begin{deluxetable}{lcccc}
\tablecaption{Primordial spectrum: tensors \& running scalar index\tablenotemark{a}\label{tab:pri_spec}}
\tablewidth{0pt}
\tabletypesize{\footnotesize}
\tablehead{\colhead{Parameter} & \colhead{\WMAP} & \colhead{+eCMB} & \colhead{+eCMB+BAO} & \colhead{+eCMB+BAO+$H_0$}}
\startdata
\multicolumn{5}{l}{Tensor mode amplitude\tablenotemark{b}} \\[1mm]
\quad \ensuremath{r} 
& \ensuremath{< 0.38\ \mbox{(95\% CL)}}
& \ensuremath{< 0.17\ \mbox{(95\% CL)}}
& \ensuremath{< 0.12\ \mbox{(95\% CL)}}
& \ensuremath{< 0.13\ \mbox{(95\% CL)}} \\[1mm]
\quad \ensuremath{n_s} 
& \ensuremath{0.992\pm 0.019}
& \ensuremath{0.970\pm 0.011}
& \ensuremath{0.9606\pm 0.0084}
& \ensuremath{0.9636\pm 0.0084} \\[1mm]
\multicolumn{5}{l}{Running scalar index\tablenotemark{b}} \\[1mm]
\quad \ensuremath{dn_s/d\ln{k}} 
& \ensuremath{-0.019\pm 0.025}
& \ensuremath{-0.022^{+ 0.012}_{- 0.011}}
& \ensuremath{-0.024\pm 0.011}
& \ensuremath{-0.023\pm 0.011} \\[1mm]
\quad \ensuremath{n_s} 
& \ensuremath{1.009\pm 0.049}
& \ensuremath{1.018\pm 0.029}
& \ensuremath{1.020\pm 0.029}
& \ensuremath{1.020\pm 0.029} \\[1mm]
\multicolumn{5}{l}{Tensors and running, jointly\tablenotemark{b}} \\[1mm]
\quad \ensuremath{r} 
& \ensuremath{< 0.50\ \mbox{(95\% CL)}}
& \ensuremath{< 0.53\ \mbox{(95\% CL)}}
& \ensuremath{< 0.43\ \mbox{(95\% CL)}}
& \ensuremath{< 0.47\ \mbox{(95\% CL)}} \\[1mm]
\quad \ensuremath{dn_s/d\ln{k}} 
& \ensuremath{-0.032\pm 0.028}
& \ensuremath{-0.039\pm 0.016}
& \ensuremath{-0.039\pm 0.015}
& \ensuremath{-0.040\pm 0.016} \\[1mm]
\quad \ensuremath{n_s} 
& \ensuremath{1.058\pm 0.063}
& \ensuremath{1.076\pm 0.048}
& \ensuremath{1.068^{+ 0.045}_{- 0.044}}
& \ensuremath{1.075\pm 0.046}
\enddata
\tablenotetext{a}{A complete list of parameter values for these models, with additional data combinations, may be found at {\tt http://lambda.gsfc.nasa.gov/}.}
\tablenotetext{b}{The tensor mode amplitude and scalar running index parameter are each fit singly, and then jointly.  In models with running, the nominal scalar index is quoted at $k_0=0.002$ Mpc$^{-1}$.}
\end{deluxetable}

\subsection{Isocurvature Modes}
\label{sec:iso}

In addition to adiabatic fluctuations, where all species fluctuate in phase and therefore produce curvature fluctuations, it is possible to have isocurvature perturbations: an over-density in one species compensates for an under-density in another, producing no net curvature.  These entropy, or isocurvature perturbations have a measurable effect on the CMB by shifting the acoustic peaks in the power spectrum.  For cold dark matter and photons, we define the entropy perturbation field
\be
\mathcal{S}_{c,\gamma} \equiv \frac{\delta \rho_c}{\rho_c} - \frac{3 \delta \rho_\gamma}{4 \rho_\gamma}
\ee
\citep{bean/dunkley/pierpaoli:2006, komatsu/etal:2009}.  The relative amplitude of its power spectrum is parameterized by $\alpha$,
\be
\frac{\alpha}{1 - \alpha} \equiv \frac{P_\mathcal{S}(k_0)}{P_\mathcal{R}(k_0)},
\ee
with $k_0=0.002$ Mpc$^{-1}$.

We consider two types of isocurvature modes: those which are completely uncorrelated with the curvature modes (with amplitude \ensuremath{\alpha_{0}}), motivated by the axion model, and those which are anti-correlated with the curvature modes (with amplitude \ensuremath{\alpha_{-1}}), motivated by the curvaton model.  For the latter, we adopt the convention in which anti-correlation  increases the power at low multipoles \citep{komatsu/etal:2009}.

The constraints on both types of isocurvature modes are given in Table~\ref{tab:iso}.  We do not detect a significant contribution from either type of perturbation in the nine-year data, whether or not additional data are included in the fit.  With \wmap\ data alone, the limits are slightly improved over the seven-year results \citep{larson/etal:2011}, but the addition of the new eCMB data improves limits by a further factor of $\sim$2.  Adding the BAO data (\S\ref{sec:bao}) and $H_0$ data (\S\ref{sec:h0}) further improves the limits, to
$$
\begin{array}{llr}
\ensuremath{\alpha_{-1}} & \ensuremath{< 0.0039\ \mbox{(95\% CL)}} & \\[2mm]
\ensuremath{\alpha_{0}} & \ensuremath{< 0.047\ \mbox{(95\% CL)}} & \end{array} \qquad \mbox{\wmap+eCMB+BAO+}H_0,
$$
due to the fact that these data help to break a modest degeneracy in the CMB anisotropy between the isocurvature modes and the \lcdm\ parameters given in Table~\ref{tab:iso}.

\begin{deluxetable}{lcccc}
\tablecaption{Isocurvature modes\tablenotemark{a}\label{tab:iso}}
\tablewidth{0pt}
\tabletypesize{\footnotesize}
\tablehead{\colhead{Parameter} & \colhead{\WMAP} & \colhead{+eCMB} & \colhead{+eCMB+BAO} & \colhead{+eCMB+BAO+$H_0$}}
\startdata
\multicolumn{5}{l}{Anti-correlated modes\tablenotemark{b}} \\[1mm]
\quad \ensuremath{\alpha_{-1}}
& \ensuremath{< 0.012\ \mbox{(95\% CL)}}
& \ensuremath{< 0.0076\ \mbox{(95\% CL)}}
& \ensuremath{< 0.0035\ \mbox{(95\% CL)}}
& \ensuremath{< 0.0039\ \mbox{(95\% CL)}} \\[1mm]
\quad \ensuremath{\Omega_ch^2} 
& \ensuremath{0.1088\pm 0.0050}
& \ensuremath{0.1097\pm 0.0037}
& \ensuremath{0.1160\pm 0.0020}
& \ensuremath{0.1151\pm 0.0019} \\[1mm]
\quad \ensuremath{n_s} 
& \ensuremath{0.994\pm 0.017}
& \ensuremath{0.977\pm 0.011}
& \ensuremath{0.9631^{+ 0.0087}_{- 0.0088}}
& \ensuremath{0.9662^{+ 0.0085}_{- 0.0087}} \\[1mm]
\quad \ensuremath{\sigma_8} 
& \ensuremath{0.807^{+ 0.025}_{- 0.024}}
& \ensuremath{0.802\pm 0.018}
& \ensuremath{0.823^{+ 0.014}_{- 0.013}}
& \ensuremath{0.821^{+ 0.014}_{- 0.013}} \\[1mm]
\multicolumn{5}{l}{Uncorrelated modes\tablenotemark{c}} \\[1mm]
\quad \ensuremath{\alpha_{0}}
& \ensuremath{< 0.15\ \mbox{(95\% CL)}}
& \ensuremath{< 0.061\ \mbox{(95\% CL)}}
& \ensuremath{< 0.043\ \mbox{(95\% CL)}}
& \ensuremath{< 0.047\ \mbox{(95\% CL)}} \\[1mm]
\quad \ensuremath{\Omega_ch^2} 
& \ensuremath{0.1093\pm 0.0056}
& \ensuremath{0.1115\pm 0.0036}
& \ensuremath{0.1161\pm 0.0020}
& \ensuremath{0.1152\pm 0.0019} \\[1mm]
\quad \ensuremath{n_s} 
& \ensuremath{0.994\pm 0.021}
& \ensuremath{0.970\pm 0.011}
& \ensuremath{0.9608^{+ 0.0086}_{- 0.0085}}
& \ensuremath{0.9639^{+ 0.0085}_{- 0.0084}} \\[1mm]
\quad \ensuremath{\sigma_8} 
& \ensuremath{0.805\pm 0.027}
& \ensuremath{0.805\pm 0.018}
& \ensuremath{0.821\pm 0.014}
& \ensuremath{0.819\pm 0.014}
\enddata
\tablenotetext{a}{A complete list of parameter values for these models, with additional data combinations, may be found at {\tt http://lambda.gsfc.nasa.gov/}.}
\tablenotetext{b}{The anti-correlated isocurvature amplitude comprises one additional parameter in the \lcdm\ fit.  The remaining parameters in this table section are given for trending.}
\tablenotetext{c}{The uncorrelated isocurvature amplitude comprises one additional parameter in the \lcdm\ fit.  The remaining parameters in this table section are given for trending.}
\end{deluxetable}

\subsection{Number of Relativistic Species}
\label{sec:neff}

\subsubsection{The number of relativistic species and the CMB power spectrum}
\label{sec:neff_spec}

Let us write the energy density of relativistic particles near the epoch of photon decoupling, $z \approx 1090$, as
\be
\rho_r \equiv \rho_\gamma + \rho_\nu + \rho_{\rm er},
\ee
where, in natural units, $\rho_\gamma=\frac{\pi^2}{15}T_\gamma^4$ is the photon energy density, $\rho_\nu=\frac78\frac{\pi^2}{15}N_\nu T_\nu^4$ is the neutrino energy density, and $\rho_{\rm er}$ denotes the energy density of ``extra radiation species.'' (The factor of 7/8 in the neutrino density arises from the Fermi-Dirac distribution.)  In the standard model of particle physics, $N_\nu=3.046$ \citep{dicus/etal:1982,mangano/etal:2002}, while in the standard thermal history of the universe, $T_\nu=(4/11)^{1/3}\,T_\gamma$ \citep[e.g.,][]{weinberg:GAC}.

Since we don't know the nature of an extra radiation species, we cannot specify its energy density or temperature uniquely.  For example, $\rho_{\rm er}$ could be comprised of bosons or fermions.  Nevertheless, it is customary to parameterize the number of extra radiation species as if they were neutrinos, and write
\be
\rho_\nu + \rho_{\rm er} \equiv \frac{7\pi^2}{120} N_{\rm eff} \, T_\nu^4,
\ee
where $N_{\rm eff}$ is the {\em effective} number of neutrino species, which does not need to be an integer.  With this parameterization, the total radiation energy density is
\be
\rho_r = \rho_\gamma \left[1 + \frac78\left(\frac{4}{11}\right)^{4/3} N_{\rm eff}\right] 
\simeq \rho_\gamma (1+0.2271N_{\rm eff}).
\ee

While photons interact with baryons efficiently at $z \gtrsim 1090$, neutrinos do not interact much at all for $z \ll 10^{10}$.  As a result, one can treat neutrinos as free-streaming particles.  Here, we also treat extra radiation species as free-streaming. With this assumption, one can use the measured $C_\ell^{TT}$ spectrum to constrain $N_{\rm eff}$ \citep{hu/etal:1995,hu/etal:1999,bowen/etal:2002,bashinsky/seljak:2004}.  Section~6.2 of \cite{komatsu/etal:2009} and \S4.7 of \cite{komatsu/etal:2011} discuss previous attempts to constrain $N_{\rm eff}$ from the CMB and provide references.  More recently, \cite{dunkley/etal:2011} and \cite{keisler/etal:2011} constrain $N_{\rm eff}$ using the seven-year \wmap\ data combined with \act\ and \spt\ data, respectively.  In this paper, we assume the sound speed and anisotropic stress of any extra radiation species are the same as for neutrinos. See \citet{archidiacono/calabrese/melchiorri:2011,smith/das/zahn:2012,archidiacono/etal:2012} for constraints on other cases.

\begin{figure}[ht]
\begin{center}
\includegraphics[width=.96\textwidth]{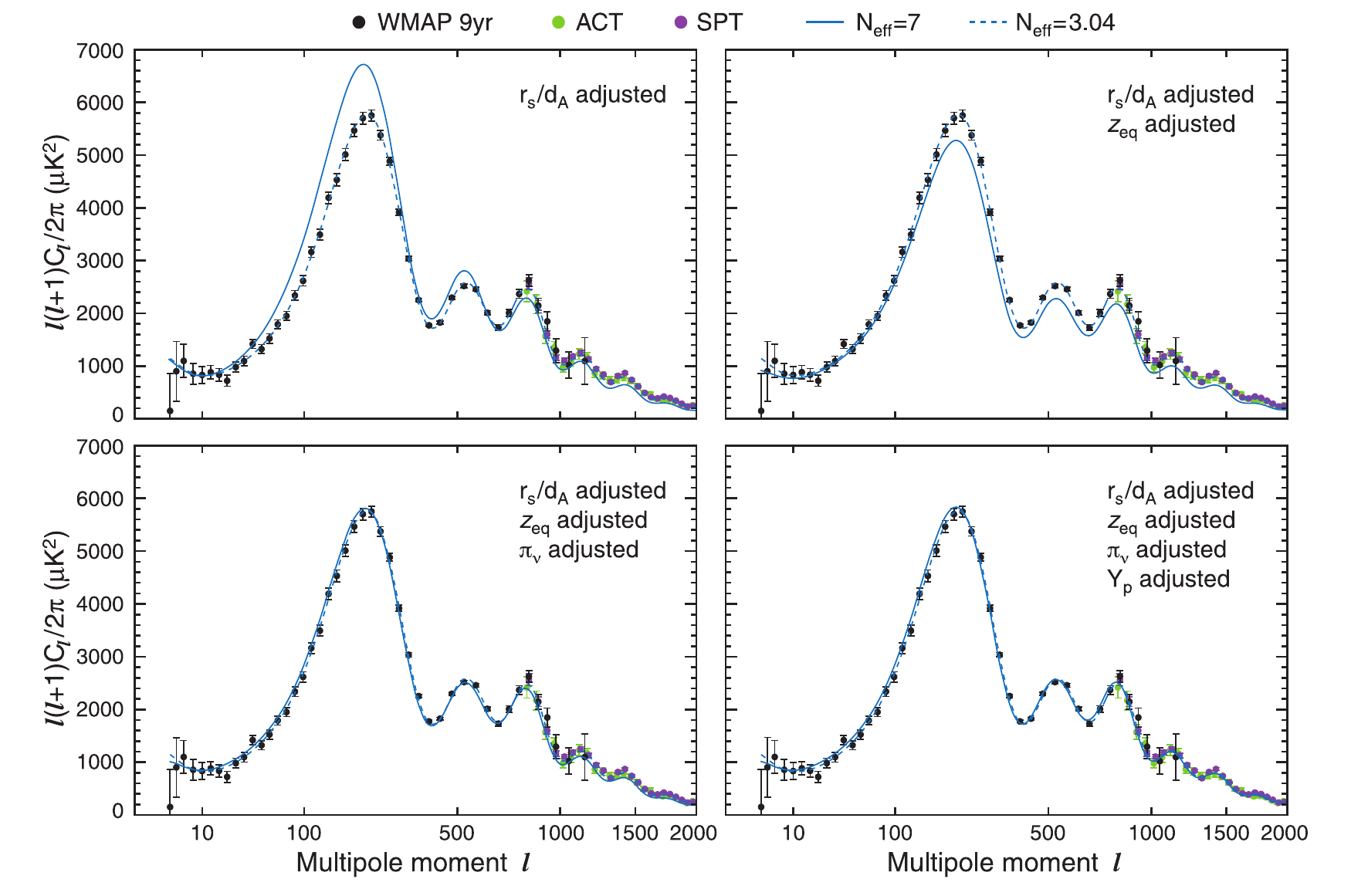}
\caption{An illustration of four effects in the CMB anisotropy that can compensate for a change in the total radiation density, $\rho_r$, parameterized here by an effective number of neutrino species, $N_{\rm eff}$.  The filled circles with errors show the nine-year \wmap\ data (in black), the \act\ data \citep[in green,][]{das/etal:2011a}, and the \spt\ data \citep[in violet,][]{keisler/etal:2011}.  The dashed lines show the best-fit model with $N_{\rm eff}=3.046$, while the solid lines show models with $N_{\rm eff}=7$ with selected adjustments applied. (The other parameters in the dashed model are $\Omega_bh^2=0.02270$, $\Omega_ch^2=0.1107$, $H_0=71.38$~km/s/Mpc, $n_s= 0.969$, $\Delta_{\cal R}^2=2.384\times 10^{-9}$, and $\tau=0.0856$.) {\it Top-left}: the $l$-axis for the $N_{\rm eff}=7$ model has been scaled so that both models have the same angular diameter distance, $d_A$, to the surface of last scattering. {\it Top-right}: the cold dark matter density, $\Omega_c h^2$, has been adjusted in the $N_{\rm eff}=7$ model so that both models have the same redshift of matter-radiation equality, $z_{\rm eq}$. {\it Bottom-left}: the amplitude of the $N_{\rm eff}=7$ model has been re- scaled to counteract the suppression of power that arises when the neutrino's anisotropic stress alters the metric perturbation. {\it Bottom-right}: the helium abundance, $Y_P$, in the $N_{\rm eff}=7$ model has been adjusted so that both models have the same diffusion damping scale. \label{fig:neff}}
\end{center}
\end{figure}

Neutrinos (and $\rho_{\rm er}$) affect the power spectrum, $C_\ell^{TT}$, in four ways.  To illustrate and explain each of these effects, Figure~\ref{fig:neff} compares models with $N_{\rm eff}=3.046$ and $N_{\rm eff}=7$, adjusted in stages to match the two spectra as closely as possible.

\begin{enumerate}
\item {\bf Peak locations} - The extra radiation density increases the early expansion rate via the Friedmann equation, $H^2=\frac{8\pi G}3(\rho_m+\rho_r)$. As a result, increasing $N_{\rm eff}$ from 3.046 to 7 reduces the comoving sound horizon, $r_s$, at the decoupling epoch, from 146.8~Mpc to 130.2~Mpc. The expansion rate after matter-radiation equality is less affected, so the angular diameter distance to the decoupling epoch, $d_A$, is only slightly reduced (by 2.5\%). Therefore, increasing $N_{\rm eff}$ reduces the angular size of the acoustic scale, $\theta_\ast \equiv r_s/d_A$, which determines the peak positions. A change in $\theta_\ast$ can be absorbed by rescaling $l$ by a constant factor. In the top-left panel of Figure~\ref{fig:neff}, we have rescaled $l$ for the $N_{\rm eff}=7$ model by a factor of 0.890, the ratio of $\theta_\ast$ for these two models ($\theta_\ast=0\ddeg5961$, $0\ddeg5306$ for $N_{\rm eff}=3.046$, 7, respectively). This rescaling brings the peak positions of these models into agreement, except for a small additive shift in peak positions; see \citet{bashinsky/seljak:2004}.  

\item {\bf Early Integrated-Sachs-Wolfe effect} - Extra radiation density delays the epoch of matter-radiation equality and thus enhances the first and second peaks via the Early Integrated-Sachs-Wolfe (ISW) effect \citep{hu/sugiyama:1995}.  This effect can be compensated by increasing the cold dark matter density in the $N_{\rm eff}=7$ model from $\Omega_ch^2=0.1107$ to 0.1817, which brings the matter-radiation equality epoch back into agreement.  (We do not change $\Omega_bh^2$, as that changes the first-to-second peak ratio.)  The top-right panel of Figure~\ref{fig:neff} shows the spectra after making this adjustment.  Note that changing $\Omega_ch^2$ also changes $\theta_\ast$, so the $l$ axis is rescaled by 0.957 for the $N_{\rm eff}=7$ model in this panel.

\item {\bf Anisotropic stress} - Relativistic species that do not interact effectively with themselves or with other species cannot be described as a (perfect) fluid. As a result, the distribution function, $f(\mathbf{x},\mathbf{p},t)$, of free-streaming particles has a non-negligible anisotropic stress,
\be
\pi_{ij} \equiv \int \frac{d^3p}{(2\pi)^3}~p\left(\hat{p}_i\hat{p}_j-\frac13\delta_{ij}\right)f(\mathbf{x},\mathbf{p},t),
\ee
as well as higher-order moments.  The energy density, pressure, and momentum are obtained from the distribution function by $\rho = (2\pi)^{-3} \int d^3p \, p \, f$, $P = (2\pi)^{-3} \int d^3p \,\frac{p}{3} \, f$, and $u^i = (2\pi)^{-3} \int d^3p \, p^i \, f$, respectively.  This term alters metric perturbations during the radiation era (via Einstein's field equations) and thus temperature fluctuations on scales $l \gtrsim 130$, since those scales enter the horizon during the radiation era.  On larger scales, fluctuations enter the horizon during the matter era and are less affected by this term.  Temperature fluctuations on these scales are given by the Sachs-Wolfe formula, $\delta T/T = -{\cal R}/5$, while those on smaller scales (ignoring the effect of baryons) are given by $\delta T/T = -\left(1+4f_\nu/15\right)^{-1}{\cal R}\cos(kr_s)$ \citep{hu/sugiyama:1996}, where $f_\nu$ is the fraction of the radiation density that is free-streaming,
\be
f_\nu(N_{\rm eff}) \equiv \frac{0.2271 N_{\rm eff}}{1+0.2271 N_{\rm eff}}.
\ee
The small-scale anisotropy is enhanced by a factor of $5(1+4f_\nu/15)^{-1}$ due to the decay of the gravitational potential at the horizon crossing during the radiation era.  Since the anisotropic stress alters the gravitational potential (via the field equations), it also alters the degree to which the small-scale anisotropy is enhanced relative to the large-scale anisotropy. Therefore, the effect of anisotropic stress can be removed by multiplying $C_\ell^{TT}(l \gtrsim 130)$ by $(1+4f_\nu/15)^2$. In the bottom-left panel of Figure~\ref{fig:neff}, we have multiplied $C_\ell^{TT}$ {\em at all $l$} by $[1+4 f_\nu(7)/15]^2/[1+4f_\nu(3.046)/15]^2$, where $f_\nu(7)=0.6139$ and $f_\nu(3.046) =0.4084$.  The two models now agree well, but the $N_{\rm eff}=7$ model is greater than the standard model at $l\lesssim 130$ because the anisotropic stress term does not affect these multipoles.

\item {\bf Enhanced damping tail} - While the increased expansion rate reduces the sound horizon, $r_s$, it also reduces the diffusion length, $r_d$, that photons travel by random walk.  The mean free path of a photon is $\lambda_C = 1/(\sigma_T n_e)$. Over the age of the universe, $t$, photons diffuse a distance $r_d \approx \sqrt{3ct/\lambda_C}\,\lambda_C \propto \sqrt{\lambda_C/H}$, and fluctuations within $r_d$ are exponentially suppressed \citep[Silk damping,][]{silk:1968}.  Now, while the sound horizon is proportional to $1/H$, the diffusion length is proportional to $1/\sqrt{H}$, due to the random walk nature of the diffusion, thus, $r_d/r_s \propto \sqrt{H}$. As a result, increasing the expansion rate increases the diffusion length relative to the sound horizon, which enhances the Silk damping of the small-scale anisotropy \citep{bashinsky/seljak:2004}. Note that $r_d/r_s$ also depends on the mean free path of the photon, $r_d/r_s \propto \sqrt{H\lambda_C} \propto \sqrt{H/n_e}$, thus one can compensate for the increased expansion rate by increasing the number density of free electrons. One way to achieve this is to reduce the helium abundance, $Y_p$ \citep{bashinsky/seljak:2004,hou/etal:prep}: since helium recombines earlier than the epoch of photon decoupling, the number density of free electrons at the decoupling epoch is given by $n_e = (1-Y_p)\,n_b$, where $n_b$ is the number density of baryons \citep[][see also Section~4.8 of Komatsu et al. 2011]{hu/etal:1995}.  In the bottom-right panel of Figure~\ref{fig:neff}, we show $C_\ell^{TT}$ for the $N_{\rm eff}=7$ model after reducing $Y_p$ from 0.24 to 0.08308, which preserves the ratio $r_d/r_s$. The solid and dashed model curves now agree completely (except for $l \lesssim 130$ where our compensation for anisotropic stress was {\it ad hoc}).
\end{enumerate}

\subsubsection{Measurements of \ensuremath{N_{\rm eff}} and \ensuremath{Y_{\rm He}}: testing Big Bang nucleosynthesis}
\label{sec:bbn}

Using the five-year \wmap\ data alone, \citet{dunkley/etal:2009} measured the effect of anisotropic stress on the power spectrum and set a lower bound on \ensuremath{N_{\rm eff}}.  However, BAO and $H_0$ data were still required to set an upper bound due to a degeneracy with the matter-radiation equality redshift \citep{komatsu/etal:2009}.  This was unchanged for the seven-year analysis \citep{komatsu/etal:2011}.  Now, with much improved measurements of the enhanced damping tail from \spt\ and \act\ (\S\ref{sec:cmb}), CMB data alone are able to determine \ensuremath{N_{\rm eff}} \citep{dunkley/etal:2011,keisler/etal:2011}.  Using the nine-year \wmap\ data combined with \spt\ and \act, we find
$$
\ensuremath{N_{\rm eff} = 3.89\pm 0.67}~(68\%~{\rm CL}) \qquad  
\mbox{\wmap+eCMB; \ensuremath{Y_{\rm He}} fixed}.
$$
The inclusion of lensing in the eCMB likelihood helps this constraint because the primary CMB fluctuations are still relatively insensitive to a combination of \ensuremath{N_{\rm eff}} and \ensuremath{\Omega_mh^2}, as described above.  CMB lensing data help constrain \ensuremath{\Omega_mh^2} by constraining \ensuremath{\sigma_8}.  The measurement is further improved by including the BAO and $H_0$ data, which reduces the degeneracy with the matter-radiation equality redshift. We find
$$
\ensuremath{N_{\rm eff} = 3.84\pm 0.40}~(68\%~{\rm CL}) \qquad 
\mbox{\wmap+eCMB+BAO+$H_0$; \ensuremath{Y_{\rm He}} fixed},
$$
which is consistent with the standard model value of $\ensuremath{N_{\rm eff}}=3.046$. We thus find no evidence for the existence of extra radiation species \citep[see also][]{calabrese/etal:2012}.

\begin{figure}[ht]
\begin{center}
\includegraphics[width=0.6\textwidth]{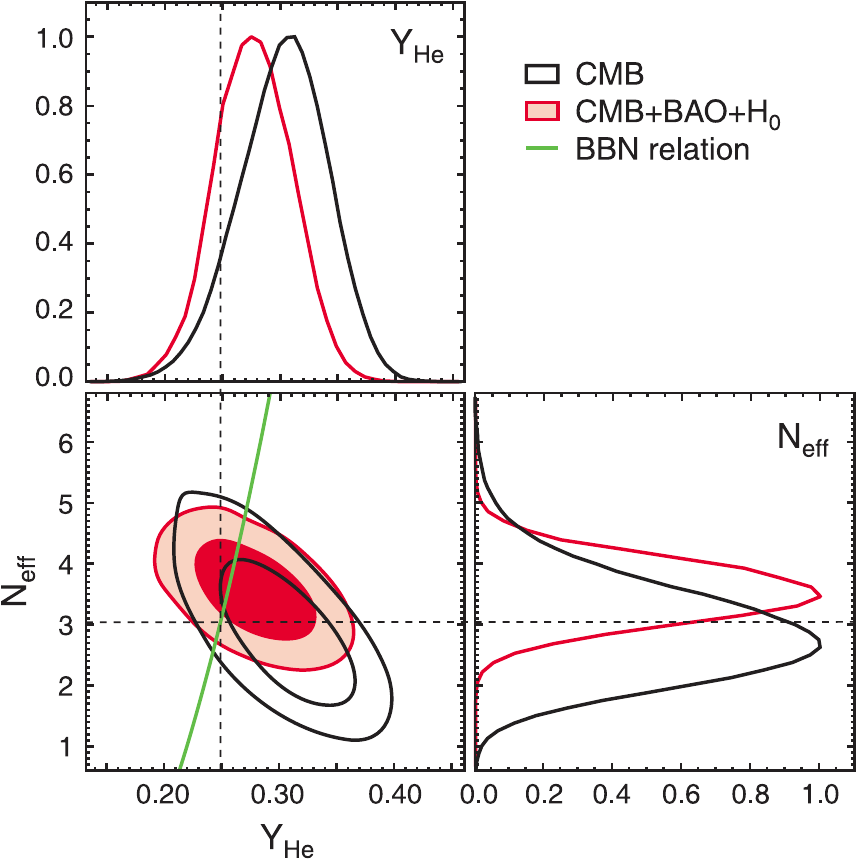}
\caption{Joint, marginalized constraints (68\% and 95\% CL) on the primordial helium abundance, \ensuremath{Y_{\rm He}}, and the energy density of ``extra radiation species,'' parameterized as an effective number of neutrino species, \ensuremath{N_{\rm eff}}.  These constraints are derived from the nine-year \wmap+eCMB data (black), and from \wmap+eCMB+BAO+$H_0$ data (red). The green curve shows the predicted dependence of $Y_{\rm He}$ on $N_{\rm eff}$ from Big Bang Nucleosynthesis; the dashed lines indicate the standard model: $N_{\rm eff} = 3.046$, $Y_{\rm He} = 0.248$.\label{fig:bbn}}
\end{center}
\end{figure}

As noted above, this measurement of \ensuremath{N_{\rm eff}} relies on the damping tail measured by \act\ and \spt, which is also affected by the primordial helium abundance, \ensuremath{Y_{\rm He}}.  Figure~\ref{fig:bbn} shows the joint, marginalized constraints on \ensuremath{N_{\rm eff}} and \ensuremath{Y_{\rm He}} using the above two data combinations.  As expected, these two parameters are anti-correlated when fit to CMB data alone (black contours).  When BAO and $H_0$ measurements are included, we find
$$
\begin{array}{lr}
\ensuremath{N_{\rm eff} = 3.55^{+ 0.49}_{- 0.48}} & \\[2mm]
\ensuremath{Y_{\rm He} = 0.278^{+ 0.034}_{- 0.032}} & \end{array} 
(68\%~{\rm CL})\qquad \mbox{\wmap+eCMB+BAO+}H_0.
$$
When combined with our measurement of the baryon density, both of these values are within the 95\%~CL region of the standard Big Bang nucleosynthesis (BBN) prediction \citep{steigman:2012}, shown by the green curve in Figure~\ref{fig:bbn}.  Our measurement provides strong support for the standard BBN scenario.  Table~\ref{tab:dof} summarizes the nine-year measurements of \ensuremath{N_{\rm eff}} and \ensuremath{Y_{\rm He}}. 

\begin{deluxetable}{lcccc}
\tablecaption{Relativistic degrees of freedom and Big Bang Nucleosynthesis\tablenotemark{a}\label{tab:dof}}
\tablewidth{0pt}
\tabletypesize{\footnotesize}
\tablehead{\colhead{Parameter} & \colhead{\WMAP} & \colhead{+eCMB} & \colhead{+eCMB+BAO} & \colhead{+eCMB+BAO+$H_0$}}
\startdata
\multicolumn{5}{l}{Number of relativistic species\tablenotemark{b}} \\[1mm]
\quad \ensuremath{N_{\rm eff}} 
& \ensuremath{> 1.7\ \mbox{(95\% CL)}}
& \ensuremath{3.89\pm 0.67}
& \ensuremath{3.55\pm 0.60}
& \ensuremath{3.84\pm 0.40} \\[1mm]
\quad \ensuremath{n_s} 
& \ensuremath{0.988\pm 0.027}
& \ensuremath{0.985^{+ 0.018}_{- 0.019}}
& \ensuremath{0.969\pm 0.015}
& \ensuremath{0.975\pm 0.010} \\[1mm]
\multicolumn{5}{l}{Primordial helium abundance\tablenotemark{b}} \\[1mm]
\quad \ensuremath{Y_{\rm He}} 
& \ensuremath{< 0.42\ \mbox{(95\% CL)}}
& \ensuremath{0.299\pm 0.027}
& \ensuremath{0.295\pm 0.027}
& \ensuremath{0.299\pm 0.027} \\[1mm]
\quad \ensuremath{n_s} 
& \ensuremath{0.973\pm 0.016}
& \ensuremath{0.982\pm 0.013}
& \ensuremath{0.973\pm 0.011}
& \ensuremath{0.977\pm 0.011} \\[1mm]
\multicolumn{5}{l}{Big bang nucleosynthesis\tablenotemark{c}} \\[1mm]
\quad \ensuremath{N_{\rm eff}} 
& \nodata
& \ensuremath{2.92\pm 0.79}
& \ensuremath{2.58\pm 0.67}
& \ensuremath{3.55^{+ 0.49}_{- 0.48}} \\[1mm]
\quad \ensuremath{Y_{\rm He}} 
& \nodata
& \ensuremath{0.302^{+ 0.038}_{- 0.039}}
& \ensuremath{0.311^{+ 0.036}_{- 0.037}}
& \ensuremath{0.278^{+ 0.034}_{- 0.032}} \\[1mm]
\quad \ensuremath{n_s} 
& \nodata
& \ensuremath{0.978\pm 0.019}
& \ensuremath{0.965\pm 0.015}
& \ensuremath{0.980\pm 0.011}
\enddata
\tablenotetext{a}{A complete list of parameter values for these models, with additional data combinations, may be found at {\tt http://lambda.gsfc.nasa.gov/}.}
\tablenotetext{b}{The parameters \ensuremath{N_{\rm eff}} and \ensuremath{Y_{\rm He}} comprise one additional parameter each in these table sections.}
\tablenotetext{c}{The parameters \ensuremath{N_{\rm eff}} and \ensuremath{Y_{\rm He}} are fit jointly in this section.}
\end{deluxetable}

\subsection{Neutrino Mass}
\label{sec:mnu}

The mean energy of a relativistic neutrino at the epoch of recombination is $\langle E \rangle = 0.58$ eV.  In order for the CMB power spectrum to be sensitive to a non-zero neutrino mass, at least one species of neutrino must have a mass in excess of this mean energy.  If one assumes that there are $\ensuremath{N_{\rm eff}}=3.046$ neutrino species with degenerate mass eigenstates, this would suggest that the lowest total mass that could be detected with CMB data is $\ensuremath{\sum m_\nu}\sim1.8$ eV.  Using a refined argument, \citet{ichikawa/fukugita/kawasaki:2005} argue that one could reach $\sim$1.5 eV.  When we add \ensuremath{\sum m_\nu} = 93 eV (\ensuremath{\Omega_\nu h^2}) as a parameter to the \lcdm\ model we obtain the fit given in Table~\ref{tab:mnu}, specifically
$$
\ensuremath{\sum m_\nu} \ensuremath{< 1.3\ \mbox{eV}\ \mbox{(95\% CL)}} \qquad \mbox{\wmap\ only},
$$
which is at the basic limit just presented. 

When the mass of individual neutrinos is less than 0.58~eV, the CMB power spectrum alone (excluding CMB lensing) cannot determine \ensuremath{\sum m_\nu}; however, tighter limits can be obtained by combining CMB data with BAO and $H_0$ data.  For a given \ensuremath{\Omega_ch^2} and \ensuremath{H_0}, adding massive neutrinos results in a larger present-day total matter density, \ensuremath{\Omega_m}, giving a smaller dark energy density for a flat universe, and hence a smaller angular diameter distance to the decoupling epoch.  This change in distance can be compensated by lowering $H_0$.  By the same token, for a given \ensuremath{\Omega_ch^2}$+\Omega_\nu h^2$, adding massive neutrinos results in a smaller matter density at the decoupling epoch (since neutrinos are still relativistic then), which produces a larger sound horizon size at that epoch.  Both of these effects cause the angular size of the acoustic scale, $\theta_*$, to be larger, shifting the CMB peaks to larger angular scales.  Furthermore, a reduced matter density at the decoupling epoch produces an earlier matter-radiation equality epoch giving a larger early ISW effect which, in turn, shifts the first peak position to a larger angular scale.  This effect can again be compensated by lowering $H_0$.  Therefore, independent information on $H_0$ obtained from local distance indicators and from BAO data helps tighten the limit on \ensuremath{\sum m_\nu} \citep{ichikawa/fukugita/kawasaki:2005}.  We find
$$
\ensuremath{\sum m_\nu} \ensuremath{< 0.44\ \mbox{eV}\ \mbox{(95\% CL)}}
\qquad \mbox{\wmap+eCMB+BAO+$H_0$},
$$
which is 25\% lower than the bound of 0.58 eV that was set with the seven-year analysis \citep{komatsu/etal:2011}. 

Since massive neutrinos have a large velocity dispersion, they cannot cluster on small scales. This means that a fraction of matter density in a low redshift universe (when neutrinos are non-relativistic) cannot cluster, which yields a shallower gravitational potential well, hence a lower value of \ensuremath{\sigma_8}.  As a result, one sees a clear negative correlation between \ensuremath{\sigma_8} and \ensuremath{\sum m_\nu} \citep[see, e.g., the middle panel of Figure~17 of][]{komatsu/etal:2009}. Therefore, adding independent information on $\sigma_8$ obtained from, e.g., the abundance of galaxy clusters \citep{vikhlinin/etal:2009,mantz/etal:2010b} helps tighten the limit on \ensuremath{\sum m_\nu}.  See \S\ref{sec:matter} for a discussion of recent measurements of \ensuremath{\sigma_8} from various cosmological probes such as cluster abundances, peculiar velocities, and gravitational lensing.

\begin{deluxetable}{lcccc}
\tablecaption{Neutrino mass\tablenotemark{a}\label{tab:mnu}}
\tablewidth{0pt}
\tabletypesize{\footnotesize}
\tablehead{\colhead{Parameter} & \colhead{\WMAP} & \colhead{+eCMB} & \colhead{+eCMB+BAO} & \colhead{+eCMB+BAO+$H_0$}}
\startdata
\multicolumn{5}{l}{New parameter} \\[1mm]
\quad \ensuremath{\sum m_\nu} (eV)\tablenotemark{b}
& \ensuremath{< 1.3\ \mbox{(95\% CL)}}
& \ensuremath{< 1.5\ \mbox{(95\% CL)}}
& \ensuremath{< 0.56\ \mbox{(95\% CL)}}
& \ensuremath{< 0.44\ \mbox{(95\% CL)}} \\[1mm]
\multicolumn{5}{l}{Related parameters} \\[1mm]
\quad \ensuremath{\sigma_8} 
& \ensuremath{0.706^{+ 0.077}_{- 0.076}}
& \ensuremath{0.660^{+ 0.066}_{- 0.061}}
& \ensuremath{0.750^{+ 0.044}_{- 0.042}}
& \ensuremath{0.770\pm 0.038} \\
\quad \ensuremath{\Omega_ch^2} 
& \ensuremath{0.1157^{+ 0.0048}_{- 0.0047}}
& \ensuremath{0.1183\pm 0.0044}
& \ensuremath{0.1133\pm 0.0026}
& \ensuremath{0.1132\pm 0.0025} \\
\quad \ensuremath{\Omega_\Lambda} 
& \ensuremath{0.641^{+ 0.065}_{- 0.068}}
& \ensuremath{0.586^{+ 0.080}_{- 0.076}}
& \ensuremath{0.695\pm 0.013}
& \ensuremath{0.707\pm 0.011} \\
\quad \ensuremath{10^9 \Delta_{\cal R}^2} 
& \ensuremath{2.48\pm 0.12}
& \ensuremath{2.59\pm 0.12}
& \ensuremath{2.452^{+ 0.075}_{- 0.074}}
& \ensuremath{2.438\pm 0.074} \\
\quad \ensuremath{n_s} 
& \ensuremath{0.962\pm 0.016}
& \ensuremath{0.947\pm 0.014}
& \ensuremath{0.9628\pm 0.0086}
& \ensuremath{0.9649^{+ 0.0085}_{- 0.0083}}
\enddata
\tablenotetext{a}{A complete list of parameter values for these models, with additional data combinations, may be found at {\tt http://lambda.gsfc.nasa.gov/}.}
\tablenotetext{b}{In the standard model, \ensuremath{\sum m_\nu} = 93.14 eV (\ensuremath{\Omega_\nu h^2}), when neutrino heating is taken into account.}
\end{deluxetable}

\subsection{Spatial Curvature}
\label{sec:olcdm}

The geometric degeneracy in the angular diameter distance to the surface of last scattering limits  our ability to constrain spatial curvature, \ensuremath{\Omega_k}, with primary CMB fluctuations alone \citep{bond/efstathiou/tegmark:1997, zaldarriaga/spergel/seljak:1997}.  For example, the nine-year \wmap\ data gives a measurement with 4\% uncertainty,
$$
\ensuremath{\Omega_k} = \ensuremath{-0.037^{+ 0.044}_{- 0.042}}
\qquad \mbox{\wmap-only},
$$
(see Table~\ref{tab:olcdm}).  However, with the recent detection of CMB lensing in the high-$l$ power spectrum \citep{das/etal:2011b, vanengelen/etal:2012}, the degeneracy between \ensuremath{\Omega_m} and \ensuremath{\Omega_\Lambda} is now substantially reduced.  This produces a significant detection of dark energy, and tight constraints on spatial curvature using {\em only} CMB data: when the \spt\ and \act\ data, including the lensing constraints, are combined with nine-year \wmap\ data we find
$$
\begin{array}{lclr}
\ensuremath{\Omega_\Lambda} & = & \ensuremath{0.727\pm 0.038} & \\[2mm]
\ensuremath{\Omega_k} & = & \ensuremath{-0.001\pm 0.012} & \end{array} \qquad \mbox{\wmap+eCMB}.
$$
Figure~43 in \citet{bennett/etal:prep} shows the joint constraints on (\ensuremath{\Omega_m},\ensuremath{\Omega_\Lambda}) (and \ensuremath{\Omega_k}, implicitly) from the currently available CMB data.  Combining the CMB data with lower-redshift distance indicators, such $H_0$, BAO, or supernovae further constrains \ensuremath{\Omega_k} \citep{spergel/etal:2007}.  Assuming the dark energy is vacuum energy ($w = -1$), we find
$$
\ensuremath{\Omega_k} = \ensuremath{-0.0027^{+ 0.0039}_{- 0.0038}}
\qquad \mbox{\wmap+eCMB+BAO+$H_0$},
$$
which limits spatial curvature to be no more than 0.4\% (68\% CL) of the critical density.  These (\ensuremath{\Omega_m},\ensuremath{\Omega_\Lambda}) constraints are also shown in Figure~43 of \citet{bennett/etal:prep}.  An independent analysis of non-flat models based on time-delay measurements of two strong gravitational lens systems, combined with seven-year \wmap\ data, give consistent and nearly competitive constraints of \ensuremath{\Omega_k} = $0.003^{+0.005}_{-0.006}$ \citep{suyu/etal:2012}.

The limits on curvature weaken slightly if dark energy is allowed to be dynamical, $w \ne -1$. However, with new distance measurements at somewhat higher redshift, where dynamical dark energy starts to become significant, the degradation factor is substantially less than it was in our previous analyses.  We revisit this topic in \S\ref{sec:de}.

\begin{deluxetable}{lccccc}
\tablecaption{Non-flat \lcdm\ Constraints\tablenotemark{a} \label{tab:olcdm}}
\tablewidth{0pt}
\tabletypesize{\footnotesize}
\tablehead{\colhead{Parameter} & \colhead{\WMAP} & \colhead{+eCMB} & \colhead{+eCMB+BAO} & \colhead{+eCMB+$H_0$} & \colhead{+eCMB+BAO+$H_0$}}
\startdata
\multicolumn{6}{l}{New parameter} \\[1mm]
\quad \ensuremath{\Omega_k} 
& \ensuremath{-0.037^{+ 0.044}_{- 0.042}}
& \ensuremath{-0.001\pm 0.012}
& \ensuremath{-0.0049^{+ 0.0041}_{- 0.0040}}
& \ensuremath{0.0049\pm 0.0047}
& \ensuremath{-0.0027^{+ 0.0039}_{- 0.0038}} \\[1mm]
\multicolumn{6}{l}{Related parameters} \\[1mm]
\quad \ensuremath{\Omega_{\rm tot}}
& \ensuremath{1.037^{+ 0.042}_{- 0.044}}
& \ensuremath{1.001\pm 0.012}
& \ensuremath{1.0049^{+ 0.0040}_{- 0.0041}}
& \ensuremath{0.9951\pm 0.0047}
& \ensuremath{1.0027^{+ 0.0038}_{- 0.0039}} \\[1mm]
\quad \ensuremath{\Omega_m}
& \ensuremath{0.19<\Omega_m<0.95\ \mbox{(95\% CL)}}
& \ensuremath{0.273\pm 0.049}
& \ensuremath{0.292\pm 0.010}
& \ensuremath{0.252\pm 0.017}
& \ensuremath{0.2855^{+ 0.0096}_{- 0.0097}} \\[1mm]
\quad \ensuremath{\Omega_\Lambda}
& \ensuremath{0.22<\Omega_\Lambda<0.79\ \mbox{(95\% CL)}}
& \ensuremath{0.727\pm 0.038}
& \ensuremath{0.713\pm 0.011}
& \ensuremath{0.743\pm 0.015}
& \ensuremath{0.717\pm 0.011} \\[1mm]
\quad \ensuremath{H_0} (km/s/Mpc)
& \ensuremath{38<H_0<84\ \mbox{(95\% CL)}}
& \ensuremath{71.2\pm 6.5}
& \ensuremath{68.0\pm 1.0}
& \ensuremath{73.4^{+ 2.2}_{- 2.3}}
& \ensuremath{68.92^{+ 0.94}_{- 0.95}} \\
\quad \ensuremath{t_0} (Gyr)
& \ensuremath{14.8\pm 1.5}
& \ensuremath{13.71\pm 0.65}
& \ensuremath{13.99\pm 0.17}
& \ensuremath{13.46\pm 0.24}
& \ensuremath{13.88\pm 0.16}
\enddata
\tablenotetext{a}{A complete list of parameter values for these models, with additional data combinations, may be found at {\tt http://lambda.gsfc.nasa.gov/}.}
\end{deluxetable}

\subsection{Dark Energy}
\label{sec:de}

The dark energy equation-of-state parameter, $w\equiv P_{\rm de}/\rho_{\rm de}$, where $P_{\rm de}$ and $\rho_{\rm de}$ are the pressure and density of dark energy, respectively, governs whether $\rho_{\rm de}$ changes with time ($w\ne -1$) or not ($w=-1$).  CMB data alone (excluding the effect of CMB gravitational lensing) are unable to determine $w$ because dark energy only affects the CMB through 1) the comoving angular diameter distance to the decoupling epoch, $d_A(z_*)$, and 2) the late-time ISW effect. The ISW effect has limited ability to constrain dark energy due to its large cosmic variance. The angular diameter distance to $z_*$ depends on several parameters (\ensuremath{\Omega_m},\ensuremath{\Omega_k}, \ensuremath{\Omega_\Lambda}, \ensuremath{w}, and \ensuremath{H_0}), thus a measurement of the angular diameter distance to a single redshift cannot distinguish these parameters.

Distance measurements to multiple redshifts greatly improve the constraint on $w$. These include the Hubble constant, $H_0$, which determines the distance scale in the low-redshift universe; $D_V$'s from BAO measurements; and luminosity distances from high-redshift Type Ia supernovae. Gravitational lensing of the CMB also probes $w$ by measuring the ratio of the angular diameter distance to the source plane (the decoupling epoch) and to the lens planes (matter fluctuations in the range of $z\sim 1-2$).  Current CMB lensing data do not yet provide competitive constraints on $w$, though they do improve the CMB-only measurement.  

In this section we derive dark energy constraints using the full CMB power spectrum information (as opposed to the simplified ``distance posteriors'' given in \S\ref{sec:de_distance}), both alone and in conjunction with the data sets noted above (see Table~\ref{tab:de}).  Linear perturbations in dark energy are treated following the ``parameterized post-Friedmann'' (PPF) approach, implemented in the CAMB code by \citet{fang/hu/lewis:2008}, \citep[see also][]{zhao/etal:2005}.  New measurements of the BAO scale (\S\ref{sec:bao}) and $H_0$ (\S\ref{sec:h0}) significantly tighten the 68\%~CL errors on a constant $w$ for both flat and non-flat models
$$
\ensuremath{w} = \left\{ \begin{array}{lr}
\ensuremath{-1.073^{+ 0.090}_{- 0.089}} & (\mbox{flat}) \\[2mm]
\ensuremath{-1.19\pm 0.12} & (\mbox{non-flat})
\end{array} \right. \qquad \mbox{\wmap+eCMB+BAO+}H_0.
$$
These constraints represent 35\% and 55\% improvements, respectively, over those from the seven-year \wmap+BAO+$H_0$ combination \citep[see the fourth column in Table 4 of][]{komatsu/etal:2011}: $\ensuremath{w} = -1.10 \pm 0.14$ (flat) and $\ensuremath{w} = -1.44 \pm 0.27$ (non-flat).  Adding 472 Type Ia supernovae compiled by \citet{conley/etal:2011} improves these limits to
$$
\ensuremath{w} = \left\{ \begin{array}{lr}
\ensuremath{-1.084\pm 0.063} & (\mbox{flat}) \\[2mm]
\ensuremath{-1.122^{+ 0.068}_{- 0.067}} & (\mbox{non-flat})
\end{array} \right. \qquad \mbox{\wmap+eCMB+BAO+}H_0\mbox{+SNe},
$$
where the errors include systematic uncertainties in the supernova data. Note that these limits are somewhat weaker than those reported in \citet{komatsu/etal:2011}, Table~4, column~6, despite the smaller number of supernovae (397) in the ``Constitution'' sample compiled by \citet{hicken/etal:2009}, as that analysis did not include SNe systematic uncertainties in the seven-year analysis.

\begin{figure}[ht]
\begin{center}
\includegraphics[width=0.6\textwidth]{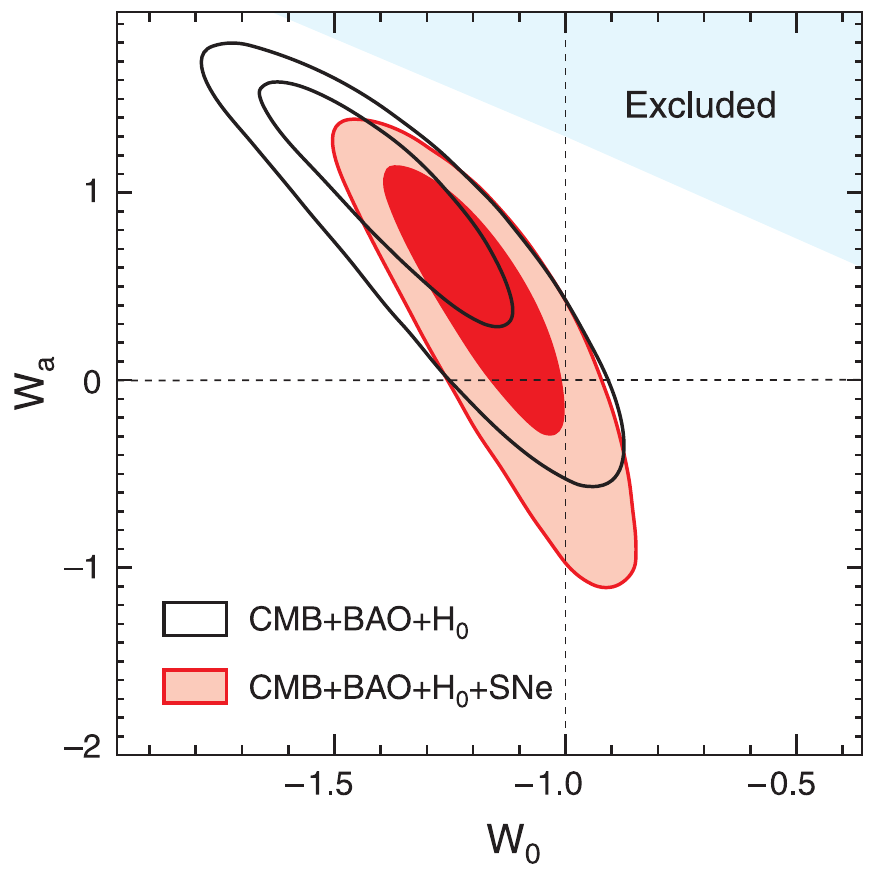}
\caption{The joint, marginalized constraint on $w_0$ and $w_a$, assuming a flat universe.  A cosmological constant ($w_0=-1$, $w_a=0$) is at the boundary of the 68\%~CL region allowed by the\wmap+eCMB+BAO+$H_0$+SNe data, indicating that the current data are consistent with a non-evolving dark energy density. The shaded region is excluded by a hard prior, $w_a < 0.2 - 1.1w_0$, in our fits. \label{fig:w0wa}}
\end{center}
\end{figure}

When $w$ is allowed to vary with the scale factor according to $w(a)=w_0+w_a(1-a)$ \citep{chevallier/polarski:2001,linder:2003}, we find, for a flat universe\footnote{We consider only the flat case here since the non-flat case with $w_a$ is not well-constrained by the present data.}
$$
\begin{array}{lclr}
w_0 & = & \ensuremath{-1.17^{+ 0.13}_{- 0.12}} & \\[2mm]
w_a & = & \ensuremath{0.35^{+ 0.50}_{- 0.49}} & \end{array} \qquad \mbox{\wmap+eCMB+BAO+}H_0\mbox{+SNe}.
$$
Figure~\ref{fig:w0wa} shows the joint, marginalized constraint on $w_0$ and $w_a$.  A cosmological constant ($w_0=-1$ and $w_a=0$) is at the boundary of the 68\%~CL region, indicating that the current data are consistent with a time-independent dark energy density.  Comparing this measurement with the seven-year result in Figure~13 of \citet{komatsu/etal:2011}, we note that adding the new BAO and $H_0$ data significantly reduces the allowed parameter space by eliminating $w_a\lesssim -1$. 

\begin{deluxetable}{lcccc}
\tablecaption{Dark energy constraints\tablenotemark{a} \label{tab:de}}
\tablewidth{0pt}
\tabletypesize{\footnotesize}
\tablehead{\colhead{Parameter} & \colhead{\WMAP} & \colhead{+eCMB} & \colhead{+eCMB+BAO+$H_0$} & \colhead{+eCMB+BAO+$H_0$+SNe}}
\startdata
\multicolumn{5}{l}{Constant equation of state; flat universe} \\[1mm]
\quad \ensuremath{w}
& \ensuremath{-1.71<w<-0.34\ \mbox{(95\% CL)}}
& \ensuremath{-1.07^{+ 0.38}_{- 0.41}}
& \ensuremath{-1.073^{+ 0.090}_{- 0.089}}
& \ensuremath{-1.084\pm 0.063} \\[1mm]
\quad \ensuremath{H_0}
& \ensuremath{> 50\ \mbox{(95\% CL)}}
& \ensuremath{> 55\ \mbox{(95\% CL)}}
& \ensuremath{70.7^{+ 1.8}_{- 1.9}}
& \ensuremath{71.0^{+ 1.4}_{- 1.3}} \\[1mm]
\multicolumn{5}{l}{Constant equation of state; non-flat universe} \\[1mm]
\quad \ensuremath{w}
& \ensuremath{> -2.1\ \mbox{(95\% CL)}}
& \nodata
& \ensuremath{-1.19\pm 0.12}
& \ensuremath{-1.122^{+ 0.068}_{- 0.067}} \\[1mm]
\quad \ensuremath{\Omega_k}
& \ensuremath{-0.052^{+ 0.051}_{- 0.054}}
& \nodata
& \ensuremath{-0.0072^{+ 0.0042}_{- 0.0043}}
& \ensuremath{-0.0059^{+ 0.0038}_{- 0.0039}} \\[1mm]
\quad \ensuremath{H_0}
& \ensuremath{37<H_0<84\ \mbox{(95\% CL)}}
& \nodata
& \ensuremath{71.7\pm 2.0}
& \ensuremath{70.7\pm 1.3} \\[1mm]
\multicolumn{5}{l}{Non-constant equation of state; flat universe} \\[1mm]
\quad $w_0$
& \nodata
& \nodata
& \ensuremath{-1.34\pm 0.18}
& \ensuremath{-1.17^{+ 0.13}_{- 0.12}} \\[1mm]
\quad \ensuremath{w_a}
& \nodata
& \nodata
& \ensuremath{0.85\pm 0.47}\tablenotemark{b}
& \ensuremath{0.35^{+ 0.50}_{- 0.49}} \\[1mm]
\quad \ensuremath{H_0}
& \nodata
& \nodata
& \ensuremath{72.3\pm 2.0}
& \ensuremath{71.0\pm 1.3}
\enddata
\tablenotetext{a}{A complete list of parameter values for these models, with additional data combinations, may be found at {\tt http://lambda.gsfc.nasa.gov/}.}
\tablenotetext{b}{The quoted error on $w_a$ from \wmap+eCMB+BAO+$H_0$ is
 smaller than that from \wmap+eCMB+BAO+$H_0$+SNe.  This is due to the
 imposition of a hard prior, $w_a < 0.2 - 1.1w_0$, depicted in
 Figure~\ref{fig:w0wa}. Without this prior, the upper limit on $w_a$ for
 \wmap+eCMB+BAO+$H_0$ would extend to larger values.}
\end{deluxetable}

\subsubsection{\wmap\ Nine-year Distance Posterior}
\label{sec:de_distance}

The ``\wmap\ distance posterior'' gives the likelihood of three variables: the acoustic scale, $l_A$, the shift parameter, $R$, and the decoupling redshift, $z_*$. This likelihood is based on, and extends, the original idea put forward by several authors \citep{wang/mukherjee:2007, wright:2007, elgaroy/multamaki:2007}.  It allows one to quickly evaluate the likelihood of various dark energy models given the \wmap\ data, without the need to run a full Markov Chain Monte Carlo exploration of the likelihood.

Here, we provide an updated distance posterior based on the nine-year data. For details on how to use this simplified likelihood, the definition of the above variables, and the limitation of this approach, see Section~5.5 of \citet{komatsu/etal:2011} and Section~5.4 of \citet{komatsu/etal:2009}.

The likelihood is given by
\be
-2\ln L = ({\bf x-d})^T{\bf C}^{-1}({\bf x-d}),
\ee
where ${\bf x}=(l_A,R,z_*)$ are the parameter values for the proposed model, and the data vector ${\bf d}$ has components
\bea
&& d_1 = l_A^{\wmap} = 302.40 \\
&& d_2 = R^{\wmap} = 1.7246 \\
&& d_3 = z_*^{\wmap} = 1090.88.
\eea
These are the maximum-likelihood values obtained from the nine-year data assuming a constant dark energy equation of state and non-zero spatial curvature (the `OWCDM' model).  The elements of the inverse covariance matrix, ${\bf C}^{-1}$, are given in Table~\ref{tab:wmap_prior_cov}. 

\begin{deluxetable}{lrrr}
\tablecolumns{4}
\tablewidth{0pt}
\tablecaption{Inverse covariance matrix for the \wmap\ distance posteriors \label{tab:wmap_prior_cov}}
\tablehead{ & \colhead{$l_A$} & \colhead{$R$} & \colhead{$z_*$}}
\startdata
$l_A$  & $3.182$ & $18.253$    & $-1.429$ \\
$R$    &         & $11887.879$ & $-193.808$ \\
$z_*$  &         &             & $4.556$
\enddata
\end{deluxetable}

\subsection{Constraints on Cosmological Birefringence}
\label{sec:TB}

If the polarization direction on the sky were uniformly rotated by an angle $\Delta\alpha$, then some of the $E$-mode polarization would be converted to $B$-mode polarization. This can arise from a mis-calibration of the detector polarization angle, but also from a physical mechanism called ``cosmological birefringence,'' in which global parity symmetry is broken on cosmological scales \citep{lue/wang/kamionkowski:1999,carroll:1998}.  Such an effect yields non-vanishing TB and EB correlations, hence non-vanishing $U_r$. A non-detection of these correlations limits $\Delta\alpha$. 

A rotation of the polarization plane by an angle $\Delta\alpha$ gives the following transformation
\bea
C_l^{\rm TE,obs} &=& C_l^{\rm TE}\cos(2\Delta\alpha),\\
C_l^{\rm TB,obs} &=& C_l^{\rm TE}\sin(2\Delta\alpha),\\
C_l^{\rm EE,obs} &=& C_l^{\rm EE}\cos^2(2\Delta\alpha),\\
C_l^{\rm BB,obs} &=& C_l^{\rm EE}\sin^2(2\Delta\alpha),\\
C_l^{\rm EB,obs} &=& \frac12C_l^{\rm EE}\sin(4\Delta\alpha),
\eea
where the spectra on the right-hand side are the primordial power spectra in the absence of rotation, while the spectra on the left-hand side are what we would observe in the presence of rotation. We assume there is no $B$-mode polarization in the absence of rotation, for the full expressions including $C_l^{\rm BB}$, see \citet{lue/wang/kamionkowski:1999,feng/etal:2005}.

The low-$l$ TB and EB data at $l\le 23$ yield $\Delta\alpha=-0\ddeg07\pm 4\ddeg82$ (68\%~CL), while the high-$l$ TB data yield $\Delta\alpha=-0\ddeg40\pm 1\ddeg30$ (68\%~CL). The high-$l$ EB data are too noisy to yield a significant limit. Combining all the multipoles, we find
$$
\Delta\alpha=-0\ddeg36\pm 1\ddeg24~(\mbox{stat.})\pm 1\ddeg5~(\mbox{syst.})~(\mbox{68\%~CL}).
$$
Here, we have added the systematic uncertainty of $\pm 1\ddeg5$, to account for uncertainty in the \wmap\ detector polarization angle \citep{page/etal:2003,page/etal:2007}.  The \wmap\ limit on $\Delta\alpha$ is now dominated by this systematic uncertainty.  The statistical error has modestly improved from $1\ddeg4$ with the seven-year data \citep{komatsu/etal:2011}, \citep[see also][]{xia/li/zhang:2010}.

\section{OTHER CONSTRAINTS ON MATTER FLUCTUATIONS}
\label{sec:matter}

	In this section, we summarize recent determinations of the matter fluctuation amplitude as traced by various measurements of large-scale structure.  These include: cluster counts from optically-selected, X-ray-selected, and SZ-selected samples; measurements of N-point statistics in SZ maps, measurements of peculiar velocities, measurements of optical shear, and measurements of CMB lensing.  To date, all of these observations are consistent with the \wmap\ nine-year \lcdm\ fits, which give \ensuremath{\sigma_8} = \ensuremath{0.821\pm 0.023} and \ensuremath{\sigma_8 \Omega_m^{0.5}} = \ensuremath{0.434\pm 0.029}.

\subsection{Cluster Observations}

Clusters are rare, high-mass peaks in the density field, hence their number counts provide an important probe of the matter fluctuation amplitude and, in turn, cosmology (see \citet{allen/etal:2011} for a recent review).  For cosmological studies, the main challenge with clusters is relating the astronomical observable (SZ decrement, X-ray flux, optical richness, etc.) to the mass of the cluster.  Since the mass function is so steep, a small error in the zero-point of the mass-to-observable scaling can produce a significant error in the determination of cosmological parameters.  

In the past two years, many new SZ-selected clusters have been reported by Planck \citep{planckearly:VIII}, \act\ \citep{marriage/etal:2011, menanteau/etal:prep}, and \spt\ \citep{williamson/etal:2011, reichardt/etal:2012b}, and they are providing new impetus for cosmological studies.  Clusters are close to virial equilibrium, so they should exhibit a tight relationship between integrated SZ decrement and mass; however, there are significant sources of non-thermal pressure support that need to be modeled \citep{trac/bode/ostriker:2011, battaglia/etal:2012a, battaglia/etal:2012b}.  Estimates of cluster mass based on X-ray data agree well with estimates based on Planck SZ measurements when one has both X-ray and SZ data for the same cluster \citep{planckearly:X, planckearly:XI}, however, there are intriguing discrepancies between some estimates based on optical and SZ data.  This is seen in both the Planck \citep{planckearly/sz_optical} and \act\ data \citep{hand/etal:2011, sehgal/etal:2012}, and several groups are exploring this discrepancy \citep{angulo/etal:2012, rozo/etal:2012, biesiadzinski/etal:2012}.

The abundance of optically-selected clusters provided early, strong evidence that vacuum energy dominates the universe today.  Based on the number density of rich clusters, \citet{fan/bahcall/cen:1997} measured $\sigma_8 = 0.83 \pm 0.15$ and $\Omega_m = 0.3 \pm 0.1$.  A recent analysis by \citet{tinker/etal:2012}, combining two observables from the Sloan Digital Sky Survey: the galaxy two-point correlation function and the mass-to-galaxy number ratio within clusters, found $\sigma_8\Omega_m^{0.5} = 0.465 \pm 0.026$, with $\Omega_m = 0.29 \pm 0.03$ and $\sigma_8 =0.85 \pm 0.06$.   \citet{zu/etal:2012} used weak lensing measurements to calibrate the masses of MaxBCG clusters in the SDSS data; they find $\sigma_8(\Omega_m/0.325)^{0.501} = 0.828 \pm 0.049$.

X-ray-selected cluster samples also provide constraints on the amplitude of matter fluctuations.  X-ray data allow one to both select the sample and calibrate its mass under the assumption of hydrostatic equilibrium.  \citet{vikhlinin/etal:2009} analyze cosmological parameter constraints from their Chandra cluster sample.  Fitting \lcdm\ parameters to the cluster counts, they find $\sigma_8(\Omega_M/0.25)^{0.47} = 0.813$ with a statistical error of $\pm0.012$ and a systematic error of $\pm0.02$, due to absolute mass calibration uncertainty.  

With the new SZ-selected cluster samples, groups are calibrating the SZ-decrement-to-mass scaling using weak-lensing measurements \citep{marrone/etal:2012, miyatake/etal:2012, high/etal:2012, planckintermediate:III}; X-ray measurements \citep{bonomante/etal:2012, benson/etal:2012}; and galaxy velocity dispersions \citep{sifon/etal:2012}.  Remarkably, these different groups are reporting fluctuation amplitudes that are consistent with the \wmap\ \lcdm\ fluctuation amplitude.  For example, \citet{benson/etal:2012} report $\sigma_8(\Omega_m/0.25)^{0.30} = 0.785 \pm 0.037$ based on an analysis of 18 SZ-selected clusters from the \spt\ survey, while \cite{sehgal/etal:2011} report $\sigma_8 = 0.821 \pm 0.044$ and $w = -1.05 \pm 0.20$ based on a joint analysis of \wmap\ seven-year data and 9 optically-confirmed SZ clusters.

The $n$-point correlation function of SZ-selected clusters provides complementary information to cluster counts (the 1-point function).  The 2-point function is a potentially-powerful probe of $\sigma_8$ \citep{komatsu/seljak:2002}; however, it is sensitive to the low-mass end of the $Y(M)$ scaling relation for clusters, which is subject to astrophysical corrections \citep{shaw/etal:2010, battaglia/etal:2010, battaglia/etal:2012a}.  \citet{reichardt/etal:2012a} use simulations and observations to calibrate the SZ power spectrum (2-point function); they apply this to the \spt\ data to find $\sigma_8 = 0.807 \pm 0.016$.  Higher-order correlation functions are less sensitive to low-mass clusters, so these moments are less affected by non-thermal processes and more sensitive to the matter fluctuation amplitude \citep{hill/sherwin:2012, bhattacharya/etal:2012}.  Measurement of the 3-point function in the \act\ SZ data \citep{wilson/etal:2012} yields $\sigma_8 = 0.78^{+0.03}_{-0.04}$.  All of these measurements are consistent with the \wmap\ nine-year measurement of \ensuremath{\sigma_8} = \ensuremath{0.821\pm 0.023}, assuming \lcdm.

\subsection{Peculiar Velocities}

Galaxy peculiar velocities provide another independent probe of gravitational potential fluctuations.  \citet{hudson/etal:2012} combine redshift-space distortion data from BOSS \citep{reid/etal:2012}, 6dFGS \citep{beutler/etal:2012}, and  WiggleZ \citep{blake/etal:2011}, with local measurements of the peculiar velocity field, to find $\Omega_m = 0.259 \pm 0.045$, $\sigma_8 = 0.748 \pm 0.035$, and a growth rate of $\gamma \equiv d\ln D/d\ln a = 0.619 \pm 0.514$.

The kinematic Sunyaev-Zel'dovich (kSZ) effect can probe peculiar velocity fields over a wide range of redshift.  \citet{hand/etal:2012} report the first kSZ measurements of peculiar velocities at $z\sim 0.35$; they detect a signal consistent with predictions from N-body simulations that are based on the \wmap\ seven-year \lcdm\ parameters.  Future kSZ measurements should be able to provide precision tests of cosmology.

\subsection{Gravitational Lensing}

There are a number of complementary gravitational lensing techniques that measure the amplitude of potential fluctuations:

{\bf CMB lensing} - Large-scale structure along the line of sight deflects CMB photons and imparts a non-Gaussian pattern on the CMB fluctuation field.  While this is most easily detected on scales smaller than those probed by \wmap, \citet{smith/zahn/dore:2007} reported the first detection of CMB lensing, by cross-correlating the 3-year \wmap\ data with the NVSS survey \citep[see also][]{feng/etal:2012}.  \citet{das/etal:2011b} reported the first detection of CMB lensing using a measurement of the 4-point correlation function in the \act\ temperature maps.  They parameterize the lensing signal by a dimensionless parameter $A_L$ which scales the lensing power spectrum relative to the prediction of the best-fit \lcdm\ model.  They report an amplitude of $A_L = 1.16 \pm 0.29$, where $A_L = 1$ is the value predicted by the \wmap\ seven-year \lcdm\ model.  (A value of $A_L$ significantly different from 1 would signal a problem with the data and/or the lensing calculation.)  Using a similar technique on \spt\ data, \citet{vanengelen/etal:2012} report $A_L = 0.90 \pm 0.19$. Since $A_L \propto \sigma_8^2$, this corresponds to an 8\% measurement of $\sigma_8$.  When these CMB lensing measurements are combined with \wmap\ seven-year data, they provide strong evidence for Dark Energy based purely on CMB observations \citep{sherwin/etal:2011, vanengelen/etal:2012}.

{\bf Cosmic shear} - Measurements of cosmic shear in large optical surveys directly probe matter fluctuations on small scales.  \citet{huff/etal:2011} analyzed 168 square degrees of co-added equatorial images from the Sloan Digital Sky Survey (SDSS) and found $\sigma_8=0.636^{+0.109}_{-0.154}$ (when other cosmological parameters are fixed to the \wmap\ seven-year \lcdm\ values).  \citet{lin/etal:2011} analyzed 275 square degrees of co-added imaging from SDSS Stripe 82 and found $\Omega_m^{0.7}\sigma_8 = 0.252^{+0.032}_{-0.052}$.  \citet{jee/etal:2012} report $\Omega_m=0.262 \pm 0.051$ and $\sigma_8=0.868 \pm 0.071$ from a cosmic shear study using the Deep Lensing Survey; when their results are combined with the \wmap\ seven-year data, they find $\Omega_m=0.278 \pm 0.018$ and $\sigma_8=0.815 \pm 0.020$.  \citet{semboloni/etal:2012} analyzed both the second and third-order moments of the cosmic shear field in the HST COSMOS data; they found $\sigma_8 (\Omega_M/0.3)^{0.49} = 0.78^{+0.11}_{-0.26}$ using the 3-point statistic, in agreement with their result from the 2-point statistic: $\sigma_8 (\Omega_M/0.3)^{0.67} = 0.70^{+0.11}_{-0.14}$.

{\bf Cross correlation} - Correlating the cosmic shear field with the large-scale galaxy distribution measures galaxy bias: the relationship been galaxies and dark matter.  \citet{mandelbaum/etal:2012} measured this cross-correlation in the SDSS DR7 data and used the inferred bias to determine cosmological parameters.  They report $\sigma_8 (\Omega_m/0.25)^{0.57}=0.80 \pm 0.05$, where the errors include both statistical and systematic effects.  \citet{cacciato/etal:2012} use combined SDSS measurements of galaxy number counts, galaxy clustering, and galaxy-galaxy lensing, together with \wmap\ seven-year priors on the scalar spectral index, the Hubble parameter, and the baryon density, to find $\Omega_m = 0.278_{-0.026}^{+0.023}$ and ${\sigma}_8 = 0.763_{-0.049}^{+0.064}$ (95\% CL). 
 
{\bf Strong lensing} - The statistics of lensed quasars probes the amplitude of fluctuations, the shape of galaxy halos, and the large-scale geometry of the universe.  \citet{oguri/etal:2012} have analyzed the final data from the SDSS Quasar Lens Search (19 lensed quasars selected from 50,836 candidates).   They claim that the number of lensed quasar are consistent with predictions based on \wmap\ seven-year parameters. Assuming  the velocity function of galaxies does not evolve with redshift, they report $\Omega_\Lambda = 0.79^{+0.06}_{-0.07}\,^{+0.06}_{-0.06}$, where the errors are statistical and systematic, respectively. 

\begin{figure}[ht]
\begin{center}
\includegraphics[width=0.8\textwidth]{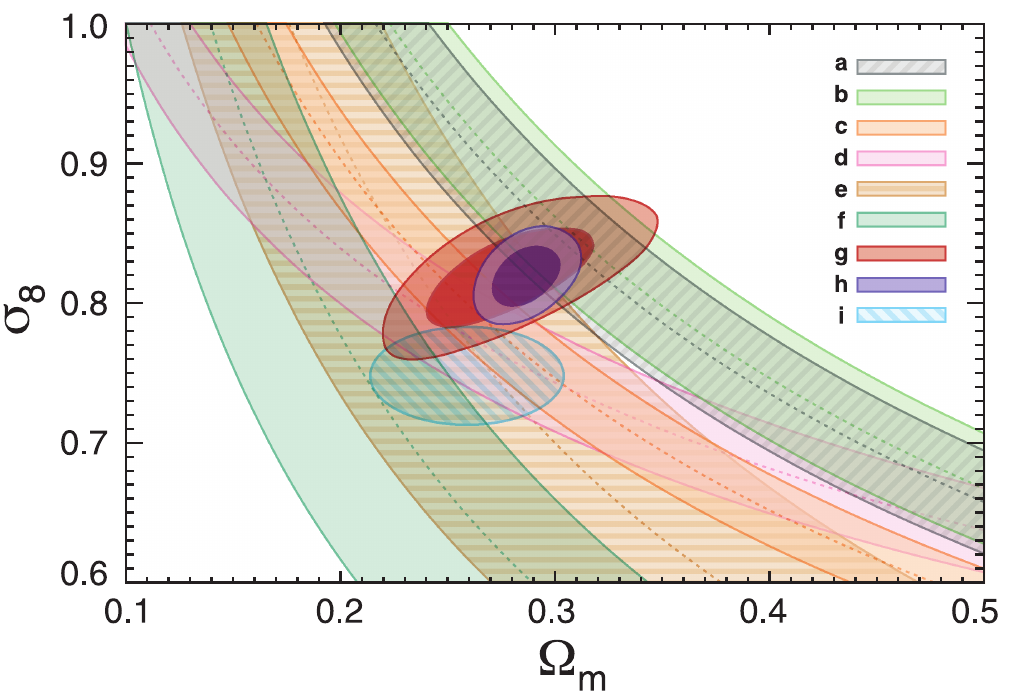}
\end{center}
\caption{A compilation of the ($\Omega_m,\sigma_8$) constraints from large scale structure observations, discussed in \S\ref{sec:matter}, compared to the constraints obtained from CMB, BAO, and $H_0$ data. The various large scale structure probes do not separately constrain the two parameters, and have somewhat different degeneracy slopes among them, but these independent measurements are quite consistent. 
The following 1$\sigma$ regions are plotted:
{\bf (a)} $\sigma_8 \Omega_m^{0.5} = 0.465\pm 0.026$ from \citet{tinker/etal:2012};
{\bf (b)} $\sigma_8 (\Omega_m / 0.325)^{0.501} = 0.828 \pm 0.049$ from \citet{zu/etal:2012};
{\bf (c)} $\sigma_8 (\Omega_m / 0.25)^{0.47} = 0.813 \pm 0.032$ from \citet{vikhlinin/etal:2009};
{\bf (d)} $\sigma_8 (\Omega_m / 0.25)^{0.3} = 0.785 \pm 0.037$ from \citet{benson/etal:2012};
{\bf (e)} $\sigma_8 (\Omega_m / 0.3)^{0.67} = 0.70^{+0.11}_{-0.14}$ from \citet{semboloni/etal:2012};
{\bf (f)} $\sigma_8 \Omega_m^{0.7} = 0.252^{+0.032}_{-0.052}$ from \citet{lin/etal:2011};
{\bf (g)} \wmap\ only;
{\bf (h)} \wmap+eCMB+BAO+$H_0$;
{\bf (i)} ellipse whose major and minor axes are given by $\Omega_m = 0.259 \pm 0.045$ and $\sigma_8 = 0.748 \pm 0.035$ from \citet{hudson/etal:2012}. \label{fig:om_s8}}
\end{figure}

\section{ON THE SZ EFFECT MEASURED BY \wmap\ AND \planck}
\label{sec:sz}

In \citet{komatsu/etal:2011}, we demonstrated that \wmap\ is capable of detecting and characterizing the Sunyaev-Zel'dovich (SZ) effect: the change in CMB temperature due to inverse Compton scattering of CMB photons off hot electrons in clusters of galaxies \citep{zeldovich/sunyaev:1969,sunyaev/zeldovich:1972}.  After our paper was published, the \planck\ collaboration published their first measurements of the SZ effect \citep{planckearly:VIII}. Owing to \planck's higher sensitivity and angular resolution, their measurements improve substantially upon the precision with which the SZ effect is characterized.

In this section, we do not report any new results from the \wmap\ nine-year data, but we compare our seven-year findings with the corresponding \planck\ measurements. In addition, we note that on-going blind SZ surveys at arcminute angular scales by \act\ \citep{marriage/etal:2011,menanteau/etal:prep} and \spt\ \citep{song/etal:prep} provide complementary information on clusters.   

{\bf Coma cluster} - Using the V- and W-band data, we are able to separate the SZ effect and the CMB fluctuation in the direction of the Coma cluster (Abell 1656).  As a result, we find that the Coma cluster is sitting at the bottom of a $\sim -100$ $\mu$K CMB fluctuation, and that all the previous determinations of the SZ effect toward Coma that did not identify the primary CMB overestimated the SZ signal by about 25\%. Our radial profile of the Coma cluster \citep[see Figure~14 of][]{komatsu/etal:2011} is in excellent agreement with the much-improved radial profile measured by \planck\ \citep[see Figure~4 of][]{planckintermediate:X}. 

{\bf Agreement with X-ray-predicted SZ signal} - The seven-year \wmap\ data are sensitive enough to measure the SZ effect toward other nearby clusters.  Among 49 $z<0.1$ clusters with detailed {\sl Chandra} observations \citep{vikhlinin/etal:2009b}, 29 are large enough to be resolved by \wmap\ and are outside the KQ75y7 sky mask. Among these, we detected the SZ effect in 20 clusters whose masses, $M_{500}$, are greater than $2\times 10^{14}~h^{-1}~M_\sun$.  The {\sl Chandra} data allow us to predict the SZ signal in each these clusters {\it without relying on any scaling relations}.  We find very good agreement between the measured and predicted signals \citep[see Figure~15 of][]{komatsu/etal:2011}: when the {\sl Chandra}-based prediction is fit to the SZ data from these 20 clusters, the best-fit amplitude is $0.82 \pm 0.12$ \citep[68\%~CL; see Table~12 of][]{komatsu/etal:2011}.  This agreement has been confirmed by the \planck\ collaboration with striking precision \citep[see the left panel of Figure~4 of][]{planckintermediate:V}.  Their analysis used 62 $z<0.5$ clusters whose masses are greater than $M_{500}=2 \times 10^{14}~h^{-1}~M_\sun$.

{\bf Comparison with the universal pressure profile} - For a given cluster, the measured and predicted SZ signals agree well, if the prediction is derived from the detailed X-ray data on the same cluster. However, the agreement is not as good if the prediction is derived from the so-called ``universal pressure profile,'' (UPP\footnote{Here, the UPP refers to Equation (13) of
\citet{arnaud/etal:2010}. Specifically, the pressure profile with a mass scaling of $M_{500}^{2/3+\alpha_p}$ with $\alpha_p=0.12$, and profile parameters $c_{500}=1.177$, $\alpha=1.051$, $\beta=5.4905$, and $\gamma=0.3081$.}) proposed by \citet{arnaud/etal:2010}: the best-fit amplitude for the 20 clusters above $M_{500}=2\times 10^{14}~h^{-1}~M_\sun$ is $0.660\pm 0.095$ \citep[68\%~CL; see Table~12 of][]{komatsu/etal:2011}.  The \planck\ collaboration observes the same trend with a similar magnitude \citep[see the left panel of Figure~4 of][]{planckintermediate:V}.  This may be caused by sample differences: the UPP is the median of a particular X-ray cluster sample, ``REXCESS,'' derived from the {\sl ROSAT} All-sky Survey \citep{boehringer/etal:2007}.  There is no guarantee that the median of the X-ray sample coincides with the median of our sample or the \planck\ sample.  The disagreement between the UPP-based predictions and the WMAP SZ profiles means that \wmap\ is sensitive to details beyond average cluster properties, provided that the inner structure of the cluster is resolved by the \wmap\ beam.  The same is seen in the \planck\ analysis: see Figure~3 of \citet{planckearly:XI}.

{\bf Cool-core versus non-cool-core clusters} - Motivated by this disagreement, we divided the samples into two sub-samples: (1) cooling-flow (or cool-core) clusters and (2) non-cooling-flow (or non-cool-core) clusters.  Fitting the prediction from the UPP to the measured SZ data, we find best-fit amplitudes of $0.89\pm 0.15$ and $0.48\pm 0.15$ for sub-samples 1 and 2, respectively \citep[68\%~CL; see Table~12 of][]{komatsu/etal:2011}. In other words, there is a statistically significant difference between these two sub-samples. This is not so surprising: the X-ray data \citep{arnaud/etal:2010} indicates that non-cooling-flow clusters have a significantly lower gas pressure in the core.  We argued that this was the first time the same effect has been detected in the SZ data. The \planck\ collaboration has confirmed this \citep[see the right panel of Figure~4 of][]{planckintermediate:V}. 

{\bf Comparing to X-ray surveys} - Combined, resolved measurements of the X-ray emission and SZ effect in a cluster give us a clear picture of the intra-cluster medium.  For example, this has allowed us to detect the difference between cool-core and non-cool-core clusters in the SZ effect.  However, such an investigation is limited to a small sample of clusters.  Many more have been detected in the {\sl ROSAT} All-sky Survey.  How might we best use these clusters? 

In many cases, the only information available for these clusters is a redshift and an X-ray flux measured within a certain aperture.  From this, one can derive an X-ray luminosity measured within a certain physical radius, for given cosmological parameters. We then need to use some scaling relations to relate the measured X-ray luminosity to the size (or the mass). This presents a challenge: while using more clusters increases statistics, using scaling relations introduces systematic errors. We tried three different scaling relations relating the X-ray luminosity, $L_X$, to the size, $r_{500}$:\footnote{Note that the relations 2 and 3 are derived originally for $L_{500}$, which came from the analysis of XMM-Newton observations of the ROSAT-detected clusters \citep{piffaretti/etal:2011}, while we use the published values of $L_X$ from the REFLEX and the extended BCS samples. In \citet{komatsu/etal:2011} we reported there was weak evidence indicating that lower mass clusters tended to be under-represented in their SZ signal. In \citet{melin/etal:2011}, this does not appear to be the case. We suspect the difference is due to the relation between $L_X$ and $L_{500}$ for low-luminosity clusters.} 
\begin{enumerate}
\item $r_{500} = \frac{0.753~h^{-1}~{\rm Mpc}}{E(z)}[L_X/(10^{44}~h^{-2}~{\rm erg~s^{-1}})]^{0.228}$ \citep{boehringer/etal:2007}, derived from the $L_X$-temperature relation of \citet{ikebe/etal:2002} and the size-temperature relation of \citet{arnaud/pointecouteau/pratt:2005},
\item $r_{500} = \frac{0.717~h^{-1}~{\rm Mpc}}{E^{1.19}(z)}[L_X/(10^{44}~h^{-2}~{\rm erg~s^{-1}})]^{0.222}$ \citep[``REXCESS'' scaling relation of][]{melin/etal:2011}, and
\item $r_{500} = \frac{0.745~h^{-1}~{\rm Mpc}}{E^{1.15}(z)}[L_X/(10^{44}~h^{-2}~{\rm erg~s^{-1}})]^{0.207}$ \citep[``intrinsic'' scaling relation of][]{melin/etal:2011}.
\end{enumerate}
The mass (hence $r_{500}$) for the scaling relation 1 is estimated using the hydrostatic equilibrium, while that for the scaling relations 2 and 3 is estimated using the $M_{500}$-$Y_X$ relation of \citet{arnaud/pointecouteau/pratt:2007}.  Scaling relation 1 predicts the largest radius (hence mass) for a given luminosity, and the scaling relation 2 predicts the smallest radius for a given luminosity.  Which relation should we use?  There is no simple answer to this question, as different cluster catalogs have different selection functions. 

For the study presented in \citet{komatsu/etal:2011}, we used 499 clusters at $z<0.2$ with luminosity $L_X > 0.45 \times 10^{44}~h^{-2}~{\rm erg~s^{-1}}$. These clusters were taken from the REFLEX sample \citep{boehringer/etal:2004} in the southern hemisphere, and the extended BCS sample \citep{ebeling/etal:1998,ebeling/etal:2000} in the northern hemisphere. We used the above scaling relations to convert $L_X$ to $r_{500}$, then to mass, $M_{500}$. We used this mass to calculate the predicted SZ signal from the UPP.  We fit these predicted SZ signals to the \wmap\ seven-year data, and found amplitudes of $0.59 \pm 0.07$, $0.78 \pm 0.09$, and $0.69 \pm 0.08$ for scaling relations 1, 2, and 3, respectively. For a high-luminosity sub-sample, with $L_X>4.5 \times 10^{44}~h^{-2}~{\rm erg~s^{-1}}$, we found $0.67 \pm 0.09$, $0.90 \pm 0.12$, and $0.84 \pm 0.11$, respectively.  Note that the quoted errors are 68\% CL statistical errors only. The scaling relation uncertainty increases the errors \citep[see footnote (a) in Table~13
of][]{komatsu/etal:2011}.

Clearly the predicted SZ signal depends significantly on the scaling relation one adopts.  For example, using relation 1, we find that the UPP over-predicts the SZ effect.  However, using relation 2 with the \wmap\ five-year data, \citet{melin/etal:2011} found the UPP-based prediction agreed well with the measured SZ.  Similarly, the \planck\ collaboration used scaling relation 2 and found the UPP-based prediction agreed well with the \planck-measured SZ.  Thus, all analyses to date agree that scaling relation 2 correctly predicts the SZ signal to within $2\sigma$, and that the other scaling relations over-predict SZ signal \citep[see Figure~8 of][]{planckearly:X}. These results indicate that the statistical analysis of many clusters using scaling relations is far less robust than the analysis of single clusters with detailed X-ray data. If statistical analysis yields unexpected results, one should question the scaling relation.   Thus, we do not endorse basing cosmological constraints on \wmap\ data that have been stacked on large numbers of X-ray cluster positions, when none of the clusters have individually detected SZ signals.

\section{ACOUSTIC STRUCTURE IN THE 9-YEAR DATA}
\label{sec:peaks}

\subsection{Motivation}

In the standard model of cosmology based upon adiabatic scalar perturbations, temperature hot spots correspond to potential wells (i.e., over-dense regions) at the surface of last scattering; therefore, matter flows towards these hot spots. A crucial length scale in the CMB is the sound horizon at the epoch of decoupling, $r_s(z_*)$; the angular size of the sound horizon sets the acoustic scale, $\theta_A \equiv r_s/d_A \approx 0\ddeg6$.  At twice the acoustic scale, the flow of matter is accelerating due to gravity, which creates a radial polarization pattern.  At the acoustic scale, the flow is decelerating due to the central photon pressure, which creates a tangential pattern  \citep{baccigalupi:1999,komatsu/etal:2011}.  Around cold spots (potential hills), the polarization follows the opposite pattern, with tangential and radial polarization formed at $1\ddeg2$ and $0\ddeg6$, respectively.

We detected this polarization pattern in the seven-year \wmap\ data with a statistical significance of $8\sigma$. The measured pattern was fully consistent with that predicted by the standard \lcdm\ model \citep[see Section~2.4 of][]{komatsu/etal:2011}. Here we apply the same analysis to the nine-year data and find results that are again fully consistent with the standard model, now with a statistical significance of $10\sigma$.

The small-scale polarization data offer a powerful test of the standard model of cosmology. Once the cosmological parameters are determined by the temperature and large-scale polarization data, one can predict the polarization signal on small angular scales with no free parameters.  This simple description is an important test of the standard cosmological model.

\subsection{Analysis Method}

The nine-year analysis replicates that of the seven-year data \citep[see Section~2.3 and Appendix B of][]{komatsu/etal:2011}.  We first smooth the foreground-reduced temperature maps from differencing assemblies V1 through W4 to a common angular resolution of $0\ddeg5$ (FWHM). We combine these maps with inverse-noise-variance weighting, and remove the monopole from the region outside the KQ85y9 mask \citep{bennett/etal:prep}. The locations of the local maxima and minima are obtained using the software {\tt hotspot} in the HEALPix distribution \citep{gorski/etal:2005}. 

As in the seven-year analysis, we cull the hot spot list by removing all local peaks with $T<0$, and vice versa.  In the 66.34\% of the sky outside the union of the KQ85y9 (temperature) mask and the P06 (polarization) mask, we find 11,536 hot spots and 11,752 cold spots remain in the temperature map. These counts are consistent with the expectation for a Gaussian random field drawn from the best-fit nine-year \wmap\ signal plus noise power spectrum.

The raw polarization maps from differencing assemblies V1 through W4 are combined using inverse-noise-variance weighting. We do not smooth the polarization maps for this analysis.  We extract a $5^\circ \times 5^\circ$ square region in the Stokes $I$, $Q$, and $U$ maps centered on each hot and cold temperature spot.  We combine the extracted temperature images with uniform weighting, while the $Q$ and $U$ images are combined with inverse-noise-variance weighting, excluding pixels masked by either analysis mask.  Afterwards, we remove the monopole from the co-added $Q$ and $U$ images.  There are 625 $0\ddeg2$ pixels in each polarization image, which sets the number of degrees of freedom in the $\chi^2$ analysis.

To make contact with the standard model prediction, we work in a rotated polarization basis, $Q_r$ and $U_r$, introduced by \citet{kamionkowski/kosowsky/stebbins:1997}. These parameters are related to $Q$ and $U$ by 
\bea
\label{eq:Qrdef}
&& Q_r(\bm{\theta}) = -Q(\bm{\theta}) \cos 2\phi - U(\bm{\theta}) \sin 2\phi, \\
\label{eq:Urdef}
&& U_r(\bm{\theta}) = Q(\bm{\theta}) \sin 2\phi - U(\bm{\theta}) \cos 2\phi,
\eea
where $\bm{\theta}=(\theta\cos\phi,\theta\sin\phi)$ is a position vector whose origin is the location of the temperature extremum; see Figure~1 of \citet{komatsu/etal:2011}.  These Stokes parameters offer a simple test of the standard model, which predicts $U_r=0$ everywhere, and $Q_r(\theta)$ alternating between positive (radial polarization) and negative (tangential polarization) values.

\subsection{Results}

\begin{figure}
\begin{center}
\includegraphics[width=0.5\textwidth]{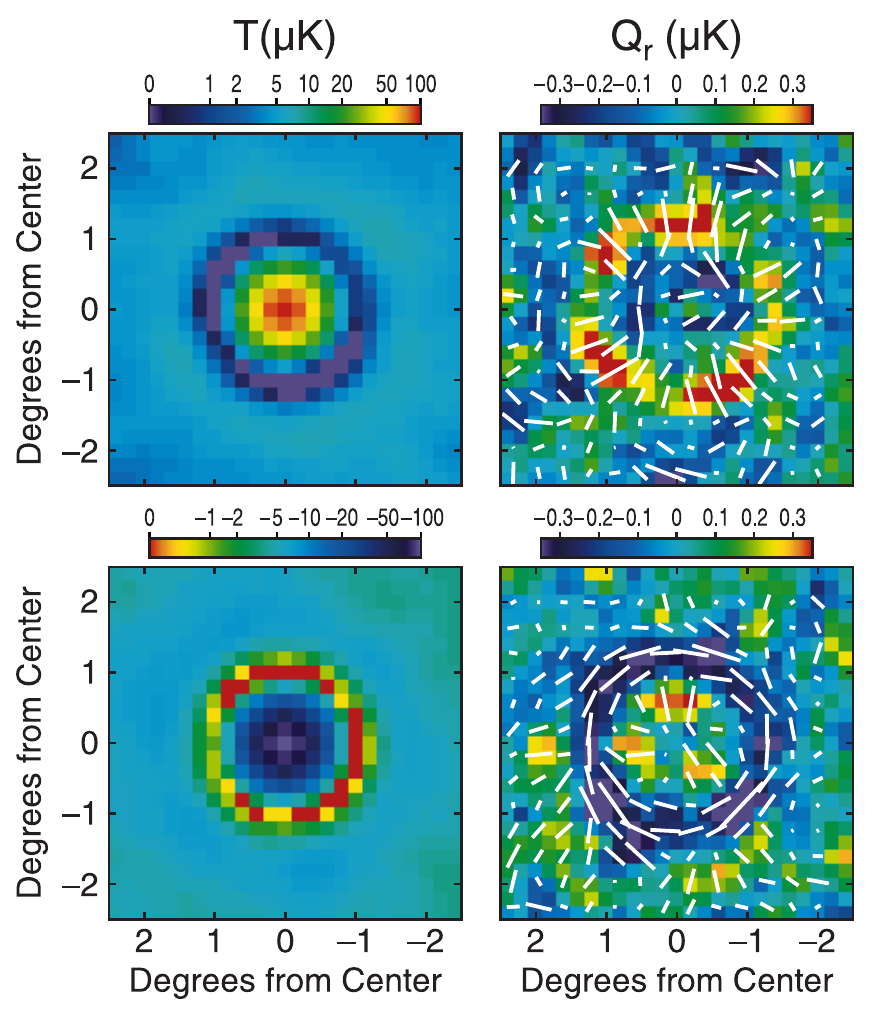}
\end{center}
\figcaption{Co-added maps of temperature, $T$, and polarization, $Q_r$, smoothed to a common resolution of $0\ddeg5$, and stacked by the location of temperature extrema.  (The polarization maps were not smoothed for the analysis, however.) {\it Top-left}: the average temperature hot spot. {\it Top-right}: the rotated polarization map, $Q_r$, stacked around temperature hot spots.  {\it Bottom-left}: the average temperature cold spot. {\it Bottom-right}: the rotated polarization map, $Q_r$, stacked around temperature cold spots. The polarization images are color-coded so that the red ($Q_r>0$) shows the radial polarization pattern, while blue ($Q_r<0$) shows the tangential polarization pattern. The lines indicate polarization direction. These images are a striking illustration of BAO in the early plasma, and phase coherence in their initial conditions. \label{fig:peak_fig}}
\end{figure}

Figure~\ref{fig:peak_fig} shows the co-added $T$ and $Q_r$ images from the nine-year data. We clearly see the alternating radial and tangential polarization pattern around the average hot spot, and vice-versa around the average cold spot.  To test the agreement between data and theory, we fit the $Q_r$ maps to their predicted patterns, and let the amplitude be a free parameter. For the hot-spot $Q_r$, we find a best-fit amplitude of $0.89 \pm 0.14$ (68\%~CL) with $\Delta\chi^2 = -41.3$ relative to zero amplitude. For the cold-spot $Q_r$, we find $1.06 \pm 0.13$ (68\%~CL) with $\Delta\chi^2 = -61.4$. The combined best-fit amplitude is $0.973 \pm 0.096$ (68\%~CL). The data are fully consistent with the standard \lcdm\ prediction, and the combined statistical significance of the detection is $10\sigma$, compared to $8\sigma$ for the seven-year data.

The co-added $U_r$ maps are consistent with zero.  We fit the measured $U_r$ maps to the predicted $Q_r$ patterns and find best-fit amplitudes of $0.02\pm 0.14$ and $0.04\pm 0.13$ (68\%~CL) around the average hot and cold spots, respectively. 

\section{CONCLUSION} 
\label{sec:conclusion}

We have used the final, nine-year \wmap\ temperature and polarization data \citep{bennett/etal:prep} in conjunction with high-$l$ CMB power spectrum data \citep{das/etal:2011a, keisler/etal:2011, reichardt/etal:2012a}, BAO data \citep{beutler/etal:2011, padmanabhan/etal:2012, anderson/etal:2012, blake/etal:2012}, and a new $H_0$ measurement \citep{riess/etal:2011} to place stringent constraints on the six parameters of the minimal \lcdm\ model, and on parameters beyond the minimal set. The six-parameter model continues to describe all the data remarkably well, and we find no convincing evidence for deviations from this model: the geometry of the observable universe is flat and dark energy is consistent with a cosmological constant. The amplitude of matter fluctuations derived from \wmap\ data alone, assuming the minimal model, \ensuremath{\sigma_8 = 0.821\pm 0.023}~(68\%~CL), is consistent with all the existing data on matter fluctuations, including cluster abundances, peculiar velocities, and gravitational lensing. The combined ({\wmap+eCMB+BAO+$H_0$) data set gives \ensuremath{\sigma_8 = 0.820^{+ 0.013}_{- 0.014}}~(68\%~CL).

The basic predictions of single-field inflation models for properties of primordial curvature perturbations are well supported by the data: the temperature fluctuations, which linearly trace primordial curvature perturbations, are Gaussian \citep{bennett/etal:prep} and adiabatic; they exhibit a slight power-law scale dependence, and the limits on primordial gravitational waves are consistent with many inflation models, including one of the oldest, proposed by \citet{starobinsky:1980}.

We find strong support for standard Big Bang nucleosynthesis from the joint constraint on the effective number of relativistic species and the primordial helium abundance, which yields \ensuremath{N_{\rm eff} = 3.55^{+ 0.49}_{- 0.48}} and \ensuremath{Y_{\rm He} = 0.278^{+ 0.034}_{- 0.032}}~(68\%~CL). The total mass of neutrinos is restricted to \ensuremath{\sum m_\nu < 0.44\ \mbox{eV}\ \mbox{(95\% CL)}} without relying on information about the growth of structure.

We compared our seven-year measurements of the SZ effect, presented in \citet{komatsu/etal:2011}, with the recent \planck\ measurements, finding that our results have been confirmed by \planck\ with striking precision.

The improved polarization data around temperature extrema, now detected at $10\sigma$, are in an excellent agreement with the prediction of the standard model based on adiabatic scalar fluctuations, providing a striking illustration of our physical understanding of the formation of acoustic waves in the early universe.  No evidence for rotation of the polarization plane, e.g., by cosmological birefringence, is found: the nine-year \wmap\ bound is $\Delta\alpha=-0\ddeg36 \pm 1\ddeg24~(\mbox{stat.}) \pm 1\ddeg5~(\mbox{syst.})$~(68\%~CL). The error is now dominated by systematic uncertainty.

The nine-year \wmap\ data have reduced the allowable volume of the six-dimensional \lcdm\ parameter space by a factor of 68,000 relative to pre-\wmap\ CMB measurements.  When combined with the high-$l$ CMB, BAO, and $H_0$ data the volume is reduced by an additional factor of 27.  The maximum likelihood values of the \lcdm\ parameters are given in Table~\ref{tab:lcdm_def} and the mean and associated 68\%~CL error bars are given in Table~\ref{tab:lcdm_wmap_ext}.  These results and those presented in the companion paper \citep{bennett/etal:prep} complete the \wmap\ Team's formal analysis and interpretation of the \wmap\ data.

\acknowledgments
The \wmap\ mission was made possible by the support of NASA.  We are grateful to Marian Pospieszalski of the National Radio Astronomy Observatory (NRAO) for his design of the microwave amplifiers that enabled the mission, and to NRAO for the development of the flight amplifiers. We also thank the project managers, Rich Day and Liz Citrin, and system engineers, Mike Bay, and Cliff Jackson, who were both expert and effective in leading the mission to launch, on-schedule and on-budget. It was a special pleasure for the science team to work closely with Cliff Jackson from the earliest times of the proposal development through to the post-launch activities. NASA has never had a finer engineer and we wish him well in his retirement. We also recognize the extraordinary efforts of the engineers, technicians, machinists, data analysts, budget analysts, managers, administrative staff, and reviewers who were all key parts of the team that created the \wmap\ spacecraft. 

G.H. was supported, in part, by the Canadian Institute for Advanced Research.  C.L.B. was supported, in part, by the Johns Hopkins University.  K.M.S. was supported at the Perimeter Institute by the Government of Canada through Industry Canada and by the Province of Ontario through the Ministry of Research \& Innovation. E.K. was supported in part by NASA grants NNX08AL43G and NNX11AD25G and NSF grants AST-0807649 and PHY-0758153.  We acknowledge use of the HEALPix \citep{gorski/etal:2005}, CAMB \citep{lewis/challinor/lasenby:2000}, and CMBFAST \citep{seljak/zaldarriaga:1996} packages. Some computations were performed on the GPC supercomputer at the SciNet HPC Consortium. We thank SciNet, which is funded by the Canada Foundation for Innovation under the auspices of Compute Canada, the Government of Ontario, Ontario Research Fund -- Research Excellence, and the University of Toronto. We acknowledge the use of the Legacy Archive for Microwave Background Data Analysis (LAMBDA). Support for LAMBDA is provided by NASA Headquarters.

\appendix

\section{Predictions of Starobinsky's $R^2$ inflation}
\label{sec:starobinsky}

\subsection{Primordial tilt}

As an example of a model that is consistent with the \wmap\ nine-year data, we examine the predictions of Starobinsky's $R^2$ inflation model.  In 1980, \citet{starobinsky:1980} showed that one-loop quantum corrections to the Einstein-Hilbert action, which generate fourth-order derivative terms of ${\cal O}(R^2)$, where $R$ is the Ricci scalar, lead to a de Sitter-type accelerated expansion of the universe.  Starobinsky's motivation was not to solve the flatness and homogeneity problems of the standard Big Bang model \citep[this was later done by][]{guth:1981}, but to see whether one-loop corrections eliminate the classical singularity at the beginning of the universe.  He was attempting to construct a cosmological model beginning from an initial de Sitter stage (not necessarily expanding) and ending in a radiation-dominated stage with a Friedmann-Robertson-Walker metric, with a mechanism for the graceful exit from inflation and the subsequent reheating phase.  

Starobinsky's work motivated Mukhanov and Chibisov in 1981 to consider quantum fluctuations in this model \citep{mukhanov/chibisov:1981}.  They made the remarkable observation that quantum fluctuations generated during the de Sitter expansion are approximately scale invariant with a logarithmic dependence on wave number, and ``could have lead to formation of galaxies and galactic clusters'' (quoted from the abstract of their paper).  In current notation, \citet{mukhanov/chibisov:1981} show that $\Delta_{\cal R}(k) \propto \left(1+\frac12\ln\frac{aH}{k}\right)$ [see their Equation~(9)], where $H$ is the Hubble rate during inflation and $k$ is the comoving wave number.  In the super-horizon limit, $k \ll aH$, the primordial tilt observed in the wave number range accessible to \wmap, $k_{\rm WMAP}$, is given by
\be
n_s - 1 = \left.\frac{d\ln\Delta^2_{\cal R}(k)}{d\ln k}\right|_{k=k_{\rm WMAP}}
 = -\frac{2}{\ln\frac{k_{\rm WMAP}}{aH}} = -\frac{2}{N},
\ee
where $N$ is the number of inflationary $e$-folds between the epoch when $k_{\rm WMAP}$ left the horizon and the end of inflation.  With $N=50$, for example, one obtains $n_s=0.96$, which is in agreement with our measurement.

Among the one-loop corrections considered by \citet{starobinsky:1980}, a term proportional to $R^2$ is found to be sufficient for driving inflation and creating curvature perturbations with the above spectrum; the other terms vanish when the metric is conformally flat (e.g., a flat Friedmann-Robertson-Walker metric).  One can obtain this result from the standard slow-roll calculation. As first shown by \citet{whitt:1984}, an action containing $R$ and $R^2$ is equivalent to an action containing $R$ and a scalar field.  To see this, start with
\be
I = \frac{1}{2}\int d^4x \sqrt{-g} \left(R+\alpha R^2\right),
\ee
where we have set $8\pi G=1$.  Perform the conformal transformation, $g_{\mu\nu} \to \hat{g}_{\mu\nu} =(1+2\alpha R)g_{\mu\nu}$, and introduce a canonically normalized scalar field, $\Psi = \sqrt{3/2} \ln(1+2\alpha R)$.  \citet{maeda:1988} showed that this system is described by a scalar field $\Psi$ governed by a potential
\be
 V(\Psi)=\frac1{8\alpha}\left(1-e^{-\sqrt{2/3}\Psi}\right)^2,
 \label{V_psi}
\ee
which is quite flat for large $\Psi$ \citep[see also][]{barrow/costakis:1988, salopek/bond/bardeen:1989}.

Once $V(\Psi)$ is specified, it is straightforward to compute the slow-roll parameters,
\be
 \epsilon\equiv \frac12\left(\frac{V'}{V}\right)^2,\qquad \eta\equiv \frac{V''}{V},
\ee
where prime denotes derivative with respect to $\Psi$.  The number of inflationary $e$-folds that occur between the epoch when a given perturbation scale leaves the horizon, $t$, and the end of inflation, $t_e$, is, to the leading order in $\epsilon$, given by 
\be
 N\equiv \int_{t}^{t_e} Hdt = \int_{\Psi_e}^{\Psi}\frac{d\Psi}{\sqrt{2\epsilon}}.
\ee
For $V(\Psi)$ as given in equation~(\ref{V_psi}), with $\Psi \gg \Psi_e$,
\be
 \epsilon = \frac{4}{3}e^{-2\sqrt{2/3}\Psi}, \qquad
 \eta = -\frac{4}{3}e^{-\sqrt{2/3}\Psi}, \qquad
 N = \frac{3}{4}e^{\sqrt{2/3}\Psi},
\ee
or
\be
 \epsilon = \frac{3}{4N^2}, \qquad \eta = -\frac{1}{N},
\ee
i.e., $\epsilon \ll |\eta|$.  In terms of the slow-roll parameters, the primordial tilt is given by \citep{liddle/lyth:CIALSS}
\be
 n_s-1 = -6\epsilon + 2\eta \simeq -\frac2{N},
\ee
in agreement with the original result of \citet{mukhanov/chibisov:1981}. 

\subsection{Tensor-to-scalar ratio}

Prior to inventing $R^2$ inflation, \citet{starobinsky:1979} calculated the energy spectrum of long-wavelength gravitational waves produced during de Sitter expansion, assuming the Einstein-Hilbert action \citep[also see][]{grishchuk:1975}.  Such long-wavelength gravitational waves induce temperature and polarization anisotropy in the CMB \citep{rubakov/sazhin/veryaskin:1982,fabbri/pollock:1983,abbott/wise:1984,starobinsky:1985}.  Later, \citet{starobinsky:1983} calculated the specific energy spectrum of gravitational waves from $R^2$ inflation \citep[also see][]{mijic/morris/suen:1986}.  This can most easily be expressed in terms of the slow-roll results, $r = 16\epsilon$ \citep{liddle/lyth:CIALSS}, giving
\be
r = \frac{12}{N^2}.
\ee
Thus, while $R^2$ inflation predicts a tilted power spectrum, the tensor-to-scalar ratio is much smaller than ${\cal O}(1-n_s) = {\cal O}(1/N)$, as it is of order ${\cal O}(1/N^2)$. 

\subsection{Predictions of non-minimally-coupled inflation}

The $R^2$ inflation predictions for $n_s$ and $r$ are exactly the same as those of a model based on a scalar field non-minimally coupled to the Ricci scalar \citep{spokoiny:1984,accetta/zoller/turner:1985,futamase/maeda:1989,salopek/bond/bardeen:1989,fakir/unruh:1990},
\be
I = \frac1{2}\int d^4x\sqrt{-g}\left(1+\xi\phi^2\right)R,
\ee
where $\xi = -1/6$ for conformal coupling.  For inflation occurring in a large-field regime, $\xi\phi^2\gg 1$, with a scalar potential $V(\phi)\propto \phi^4$, the tilt is given by $n_s-1=-2/N$ \citep{salopek:1992,kaiser:1995}, and the tensor-to-scalar ratio is given by \citep[see Equation~(5.1) of][]{komatsu/futamase:1999}
\be
 r=\frac{12}{N^2}\frac{1+6\xi}{6\xi},
\ee
which has $r \to 12/N^2$ for $\xi \gg 1$ \citep[also see][]{hwang/noh:1998}.

This model has attracted renewed attention since \citet{bezrukov/shaposhnikov:2008} showed that the standard-model Higgs field can drive inflation if the Higgs field is non-minimally coupled to the Ricci scalar with $\xi \gg 1$.  The observable predictions of the original Higgs inflation are therefore $n_s-1 = -2/N$ and $r = 12/N^2$ \citep[e.g.,][]{bezrukov/gorbunov/shaposhnikov:2009}.

\bibliography{wmap}

\begin{thebibliography}{200}
\expandafter\ifx\csname natexlab\endcsname\relax\def\natexlab#1{#1}\fi

\bibitem[{Abbott \& Wise(1984)}]{abbott/wise:1984}
Abbott, L.~F. \& Wise, M.~B. 1984, Nucl. Phys., B244, 541

\bibitem[{{Accetta} et~al.(1985){Accetta}, {Zoller}, \&
  {Turner}}]{accetta/zoller/turner:1985}
{Accetta}, F.~S., {Zoller}, D.~J., \& {Turner}, M.~S. 1985, \prd, 31, 3046

\bibitem[{{Addison} et~al.(2012)}]{addison/etal:2012}
{Addison}, G.~E., et~al. 2012, \apj, 752, 120

\bibitem[{{Allen} et~al.(2011){Allen}, {Evrard}, \& {Mantz}}]{allen/etal:2011}
{Allen}, S.~W., {Evrard}, A.~E., \& {Mantz}, A.~B. 2011, \araa, 49, 409

\bibitem[{{Anderson} et~al.(2012)}]{anderson/etal:2012}
{Anderson}, L., et~al. 2012, ArXiv e-prints, arXiv:1203.6594

\bibitem[{{Angulo} et~al.(2012){Angulo}, {Springel}, {White}, {Jenkins},
  {Baugh}, \& {Frenk}}]{angulo/etal:2012}
{Angulo}, R.~E., {Springel}, V., {White}, S.~D.~M., {Jenkins}, A., {Baugh},
  C.~M., \& {Frenk}, C.~S. 2012, \mnras, 426, 2046

\bibitem[{{Archidiacono} et~al.(2011){Archidiacono}, {Calabrese}, \&
  {Melchiorri}}]{archidiacono/calabrese/melchiorri:2011}
{Archidiacono}, M., {Calabrese}, E., \& {Melchiorri}, A. 2011, \prd, 84, 123008

\bibitem[{{Archidiacono} et~al.(2012){Archidiacono}, {Giusarma}, {Melchiorri},
  \& {Mena}}]{archidiacono/etal:2012}
{Archidiacono}, M., {Giusarma}, E., {Melchiorri}, A., \& {Mena}, O. 2012, \prd,
  86, 043509

\bibitem[{{Arnaud} et~al.(2005){Arnaud}, {Pointecouteau}, \&
  {Pratt}}]{arnaud/pointecouteau/pratt:2005}
{Arnaud}, M., {Pointecouteau}, E., \& {Pratt}, G.~W. 2005, \aap, 441, 893

\bibitem[{{Arnaud} et~al.(2007){Arnaud}, {Pointecouteau}, \&
  {Pratt}}]{arnaud/pointecouteau/pratt:2007}
---. 2007, \aap, 474, L37

\bibitem[{{Arnaud} et~al.(2010){Arnaud}, {Pratt}, {Piffaretti},
  {B{\"o}hringer}, {Croston}, \& {Pointecouteau}}]{arnaud/etal:2010}
{Arnaud}, M., {Pratt}, G.~W., {Piffaretti}, R., {B{\"o}hringer}, H., {Croston},
  J.~H., \& {Pointecouteau}, E. 2010, \aap, 517, A92+

\bibitem[{{Astier} et~al.(2006)}]{astier/etal:2006}
{Astier}, P., et~al. 2006, \aap, 447, 31

\bibitem[{Baccigalupi(1999)}]{baccigalupi:1999}
Baccigalupi, C. 1999, Phys. Rev. D, 59, 123004

\bibitem[{{Barrow} \& {Cotsakis}(1988)}]{barrow/costakis:1988}
{Barrow}, J.~D. \& {Cotsakis}, S. 1988, Physics Letters B, 214, 515

\bibitem[{{Bashinsky} \& {Seljak}(2004)}]{bashinsky/seljak:2004}
{Bashinsky}, S. \& {Seljak}, U. 2004, \prd, 69, 083002

\bibitem[{{Battaglia} et~al.(2012{\natexlab{a}}){Battaglia}, {Bond},
  {Pfrommer}, \& {Sievers}}]{battaglia/etal:2012a}
{Battaglia}, N., {Bond}, J.~R., {Pfrommer}, C., \& {Sievers}, J.~L.
  2012{\natexlab{a}}, \apj, 758, 74

\bibitem[{{Battaglia} et~al.(2012{\natexlab{b}}){Battaglia}, {Bond},
  {Pfrommer}, \& {Sievers}}]{battaglia/etal:2011}
---. 2012{\natexlab{b}}, \apj, 758, 75

\bibitem[{{Battaglia} et~al.(2012{\natexlab{c}}){Battaglia}, {Bond},
  {Pfrommer}, \& {Sievers}}]{battaglia/etal:2012b}
---. 2012{\natexlab{c}}, ArXiv e-prints, arXiv:1209.4082

\bibitem[{{Battaglia} et~al.(2010){Battaglia}, {Bond}, {Pfrommer}, {Sievers},
  \& {Sijacki}}]{battaglia/etal:2010}
{Battaglia}, N., {Bond}, J.~R., {Pfrommer}, C., {Sievers}, J.~L., \& {Sijacki},
  D. 2010, \apj, 725, 91

\bibitem[{Bean et~al.(2006)Bean, Dunkley, \&
  Pierpaoli}]{bean/dunkley/pierpaoli:2006}
Bean, R., Dunkley, J., \& Pierpaoli, E. 2006, Phys. Rev., D74, 063503

\bibitem[{Bennett et~al.(2012)}]{bennett/etal:prep}
Bennett, C.~L., et~al. 2012, Submitted to \apjs, --

\bibitem[{{Benson} et~al.(2011)}]{benson/etal:2012}
{Benson}, B.~A., et~al. 2011, ArXiv e-prints, arXiv:1112.5435

\bibitem[{{Beutler} et~al.(2011)}]{beutler/etal:2011}
{Beutler}, F., et~al. 2011, \mnras, 416, 3017

\bibitem[{{Beutler} et~al.(2012)}]{beutler/etal:2012}
---. 2012, \mnras, 423, 3430

\bibitem[{{Bezrukov} et~al.(2009){Bezrukov}, {Gorbunov}, \&
  {Shaposhnikov}}]{bezrukov/gorbunov/shaposhnikov:2009}
{Bezrukov}, F., {Gorbunov}, D., \& {Shaposhnikov}, M. 2009, Journal of
  Cosmology and Astro-Particle Physics, 6, 29

\bibitem[{{Bezrukov} \& {Shaposhnikov}(2008)}]{bezrukov/shaposhnikov:2008}
{Bezrukov}, F. \& {Shaposhnikov}, M. 2008, Physics Letters B, 659, 703

\bibitem[{{Bhattacharya} et~al.(2012){Bhattacharya}, {Nagai}, {Shaw},
  {Crawford}, \& {Holder}}]{bhattacharya/etal:2012}
{Bhattacharya}, S., {Nagai}, D., {Shaw}, L., {Crawford}, T., \& {Holder}, G.~P.
  2012, \apj, 760, 5

\bibitem[{{Biesiadzinski} et~al.(2012){Biesiadzinski}, {McMahon}, {Miller},
  {Nord}, \& {Shaw}}]{biesiadzinski/etal:2012}
{Biesiadzinski}, T., {McMahon}, J., {Miller}, C.~J., {Nord}, B., \& {Shaw}, L.
  2012, \apj, 757, 1

\bibitem[{{Blake} et~al.(2011)}]{blake/etal:2011}
{Blake}, C., et~al. 2011, \mnras, 418, 1725

\bibitem[{{Blake} et~al.(2012)}]{blake/etal:2012}
---. 2012, \mnras, 425, 405

\bibitem[{{B{\"o}hringer} et~al.(2004)}]{boehringer/etal:2004}
{B{\"o}hringer}, H., et~al. 2004, \aap, 425, 367

\bibitem[{{B{\"o}hringer} et~al.(2007)}]{boehringer/etal:2007}
---. 2007, \aap, 469, 363

\bibitem[{{Bonamente} et~al.(2012)}]{bonomante/etal:2012}
{Bonamente}, M., et~al. 2012, New Journal of Physics, 14, 025010

\bibitem[{{Bond} et~al.(1997){Bond}, {Efstathiou}, \&
  {Tegmark}}]{bond/efstathiou/tegmark:1997}
{Bond}, J.~R., {Efstathiou}, G., \& {Tegmark}, M. 1997, \mnras, 291, L33

\bibitem[{{Bond} et~al.(1998){Bond}, {Jaffe}, \& {Knox}}]{bond/jaffe/knox:1998}
{Bond}, J.~R., {Jaffe}, A.~H., \& {Knox}, L. 1998, \prd, 57, 2117

\bibitem[{{Bowen} et~al.(2002){Bowen}, {Hansen}, {Melchiorri}, {Silk}, \&
  {Trotta}}]{bowen/etal:2002}
{Bowen}, R., {Hansen}, S.~H., {Melchiorri}, A., {Silk}, J., \& {Trotta}, R.
  2002, \mnras, 334, 760

\bibitem[{{Brown} et~al.(2009)}]{brown/etal:2009}
{Brown}, M.~L., et~al. 2009, \apj, 705, 978

\bibitem[{{Cacciato} et~al.(2012){Cacciato}, {van den Bosch}, {More}, {Mo}, \&
  {Yang}}]{cacciato/etal:2012}
{Cacciato}, M., {van den Bosch}, F.~C., {More}, S., {Mo}, H., \& {Yang}, X.
  2012, ArXiv e-prints, 1207.0503

\bibitem[{{Calabrese} et~al.(2012){Calabrese}, {Archidiacono}, {Melchiorri}, \&
  {Ratra}}]{calabrese/etal:2012}
{Calabrese}, E., {Archidiacono}, M., {Melchiorri}, A., \& {Ratra}, B. 2012,
  \prd, 86, 043520

\bibitem[{{Carroll}(1998)}]{carroll:1998}
{Carroll}, S.~M. 1998, \prl, 81, 3067

\bibitem[{Chevallier \& Polarski(2001)}]{chevallier/polarski:2001}
Chevallier, M. \& Polarski, D. 2001, Int. J. Mod. Phys., D10, 213

\bibitem[{{Chiang} et~al.(2010)}]{chiang/etal:2010}
{Chiang}, H.~C., et~al. 2010, \apj, 711, 1123

\bibitem[{{Conley} et~al.(2011)}]{conley/etal:2011}
{Conley}, A., et~al. 2011, \apjs, 192, 1

\bibitem[{{Contreras} et~al.(2010)}]{contreras/etal:2010}
{Contreras}, C., et~al. 2010, \aj, 139, 519

\bibitem[{{Crocce} \& {Scoccimarro}(2008)}]{crocce/scoccimarro:2008}
{Crocce}, M. \& {Scoccimarro}, R. 2008, \prd, 77, 023533

\bibitem[{{Das} et~al.(2011{\natexlab{a}})}]{das/etal:2011b}
{Das}, S., et~al. 2011{\natexlab{a}}, Physical Review Letters, 107, 021301

\bibitem[{{Das} et~al.(2011{\natexlab{b}})}]{das/etal:2011a}
---. 2011{\natexlab{b}}, \apj, 729, 62

\bibitem[{Dicus et~al.(1982)}]{dicus/etal:1982}
Dicus, D.~A. et~al. 1982, Phys. Rev., D26, 2694

\bibitem[{{Dunkley} et~al.(2009)}]{dunkley/etal:2009}
{Dunkley}, J., et~al. 2009, \apjs, 180, 306

\bibitem[{{Dunkley} et~al.(2011)}]{dunkley/etal:2011}
---. 2011, \apj, 739, 52

\bibitem[{{Ebeling} et~al.(2000){Ebeling}, {Edge}, {Allen}, {Crawford},
  {Fabian}, \& {Huchra}}]{ebeling/etal:2000}
{Ebeling}, H., {Edge}, A.~C., {Allen}, S.~W., {Crawford}, C.~S., {Fabian},
  A.~C., \& {Huchra}, J.~P. 2000, \mnras, 318, 333

\bibitem[{{Ebeling} et~al.(1998){Ebeling}, {Edge}, {Bohringer}, {Allen},
  {Crawford}, {Fabian}, {Voges}, \& {Huchra}}]{ebeling/etal:1998}
{Ebeling}, H., {Edge}, A.~C., {Bohringer}, H., {Allen}, S.~W., {Crawford},
  C.~S., {Fabian}, A.~C., {Voges}, W., \& {Huchra}, J.~P. 1998, \mnras, 301,
  881

\bibitem[{{Eisenstein} et~al.(2007){Eisenstein}, {Seo}, {Sirko}, \&
  {Spergel}}]{eisenstein/etal:2007}
{Eisenstein}, D.~J., {Seo}, H.-J., {Sirko}, E., \& {Spergel}, D.~N. 2007, \apj,
  664, 675

\bibitem[{{Elgar{\o}y} \& {Multam{\"a}ki}(2007)}]{elgaroy/multamaki:2007}
{Elgar{\o}y}, O. \& {Multam{\"a}ki}, T. 2007, \aap, 471, 65

\bibitem[{Fabbri \& Pollock(1983)}]{fabbri/pollock:1983}
Fabbri, R. \& Pollock, M.~d. 1983, Phys. Lett., B125, 445

\bibitem[{{Fakir} \& {Unruh}(1990)}]{fakir/unruh:1990}
{Fakir}, R. \& {Unruh}, W.~G. 1990, \prd, 41, 1783

\bibitem[{{Fan} et~al.(1997){Fan}, {Bahcall}, \& {Cen}}]{fan/bahcall/cen:1997}
{Fan}, X., {Bahcall}, N.~A., \& {Cen}, R. 1997, \apjl, 490, L123+

\bibitem[{{Fang} et~al.(2008){Fang}, {Hu}, \& {Lewis}}]{fang/hu/lewis:2008}
{Fang}, W., {Hu}, W., \& {Lewis}, A. 2008, \prd, 78, 087303

\bibitem[{Feng et~al.(2005)Feng, Li, Li, \& Zhang}]{feng/etal:2005}
Feng, B., Li, H., Li, M.-z., \& Zhang, X.-m. 2005, Phys. Lett., B620, 27

\bibitem[{{Feng} et~al.(2012){Feng}, {Aslanyan}, {Manohar}, {Keating}, {Paar},
  \& {Zahn}}]{feng/etal:2012}
{Feng}, C., {Aslanyan}, G., {Manohar}, A.~V., {Keating}, B., {Paar}, H.~P., \&
  {Zahn}, O. 2012, \prd, 86, 063519

\bibitem[{{Fixsen}(2009)}]{fixsen:2009}
{Fixsen}, D.~J. 2009, \apj, 707, 916

\bibitem[{{Fowler} et~al.(2010)}]{fowler/etal:2010}
{Fowler}, J.~W., et~al. 2010, \apj, 722, 1148

\bibitem[{{Freedman} et~al.(2012){Freedman}, {Madore}, {Scowcroft}, {Burns},
  {Monson}, {Persson}, {Seibert}, \& {Rigby}}]{freedman/etal:2012}
{Freedman}, W.~L., {Madore}, B.~F., {Scowcroft}, V., {Burns}, C., {Monson}, A.,
  {Persson}, S.~E., {Seibert}, M., \& {Rigby}, J. 2012, \apj, 758, 24

\bibitem[{{Freedman} et~al.(2001)}]{freedman/etal:2001}
{Freedman}, W.~L., et~al. 2001, \apj, 553, 47

\bibitem[{Futamase \& Maeda(1989)}]{futamase/maeda:1989}
Futamase, T. \& Maeda, K.-i. 1989, Phys. Rev., D39, 399

\bibitem[{Gorski et~al.(2005)Gorski, Hivon, Banday, Wandelt, Hansen, Reinecke,
  \& Bartlemann}]{gorski/etal:2005}
Gorski, K.~M., Hivon, E., Banday, A.~J., Wandelt, B.~D., Hansen, F.~K.,
  Reinecke, M., \& Bartlemann, M. 2005, \apj, 622, 759

\bibitem[{Grishchuk(1975)}]{grishchuk:1975}
Grishchuk, L.~P. 1975, Sov. Phys. JETP, 40, 409

\bibitem[{Guth(1981)}]{guth:1981}
Guth, A.~H. 1981, \prd, 23, 347

\bibitem[{{Guy} et~al.(2010)}]{guy/etal:2010}
{Guy}, J., et~al. 2010, \aap, 523, A7

\bibitem[{{Hamuy} et~al.(1996){Hamuy}, {Phillips}, {Suntzeff}, {Schommer},
  {Maza}, \& {Aviles}}]{hamuy/etal:1996}
{Hamuy}, M., {Phillips}, M.~M., {Suntzeff}, N.~B., {Schommer}, R.~A., {Maza},
  J., \& {Aviles}, R. 1996, \aj, 112, 2391

\bibitem[{{Hand} et~al.(2011)}]{hand/etal:2011}
{Hand}, N., et~al. 2011, \apj, 736, 39

\bibitem[{{Hand} et~al.(2012)}]{hand/etal:2012}
---. 2012, Physical Review Letters, 109, 041101

\bibitem[{{Hicken} et~al.(2009){Hicken}, {Wood-Vasey}, {Blondin}, {Challis},
  {Jha}, {Kelly}, {Rest}, \& {Kirshner}}]{hicken/etal:2009}
{Hicken}, M., {Wood-Vasey}, W.~M., {Blondin}, S., {Challis}, P., {Jha}, S.,
  {Kelly}, P.~L., {Rest}, A., \& {Kirshner}, R.~P. 2009, \apj, 700, 1097

\bibitem[{{High} et~al.(2012)}]{high/etal:2012}
{High}, F.~W., et~al. 2012, \apj, 758, 68

\bibitem[{{Hill} \& {Sherwin}(2012)}]{hill/sherwin:2012}
{Hill}, J.~C. \& {Sherwin}, B.~D. 2012, ArXiv e-prints, arXiv:1205.5794

\bibitem[{{Hivon} et~al.(2002){Hivon}, {G{\' o}rski}, {Netterfield}, {Crill},
  {Prunet}, \& {Hansen}}]{hivon/etal:2002}
{Hivon}, E., {G{\' o}rski}, K.~M., {Netterfield}, C.~B., {Crill}, B.~P.,
  {Prunet}, S., \& {Hansen}, F. 2002, \apj, 567, 2

\bibitem[{{Holtzman} et~al.(2008)}]{holtzman/etal:2008}
{Holtzman}, J.~A., et~al. 2008, \aj, 136, 2306

\bibitem[{{Hou} et~al.(2011){Hou}, {Keisler}, {Knox}, {Millea}, \&
  {Reichardt}}]{hou/etal:prep}
{Hou}, Z., {Keisler}, R., {Knox}, L., {Millea}, M., \& {Reichardt}, C. 2011,
  arXiv:1104.2333

\bibitem[{{Hu} et~al.(1999){Hu}, {Eisenstein}, {Tegmark}, \&
  {White}}]{hu/etal:1999}
{Hu}, W., {Eisenstein}, D.~J., {Tegmark}, M., \& {White}, M. 1999, \prd, 59,
  023512

\bibitem[{Hu et~al.(1995)Hu, Scott, Sugiyama, \& White}]{hu/etal:1995}
Hu, W., Scott, D., Sugiyama, N., \& White, M. 1995, \prd, 52, 5498

\bibitem[{{Hu} \& {Sugiyama}(1995)}]{hu/sugiyama:1995}
{Hu}, W. \& {Sugiyama}, N. 1995, \apj, 444, 489

\bibitem[{{Hu} \& {Sugiyama}(1996)}]{hu/sugiyama:1996}
---. 1996, \apj, 471, 542

\bibitem[{{Hudson} \& {Turnbull}(2012)}]{hudson/etal:2012}
{Hudson}, M.~J. \& {Turnbull}, S.~J. 2012, \apjl, 751, L30

\bibitem[{{Huff} et~al.(2011){Huff}, {Eifler}, {Hirata}, {Mandelbaum},
  {Schlegel}, \& {Seljak}}]{huff/etal:2011}
{Huff}, E.~M., {Eifler}, T., {Hirata}, C.~M., {Mandelbaum}, R., {Schlegel}, D.,
  \& {Seljak}, U. 2011, ArXiv e-prints, arXiv:1112.3143

\bibitem[{{Hwang} \& {Noh}(1998)}]{hwang/noh:1998}
{Hwang}, J. \& {Noh}, H. 1998, Physical Review Letters, Volume 81, Issue 24,
  December 14, 1998, pp.5274-5277, 81, 5274

\bibitem[{{Ichikawa} et~al.(2005){Ichikawa}, {Fukugita}, \&
  {Kawasaki}}]{ichikawa/fukugita/kawasaki:2005}
{Ichikawa}, K., {Fukugita}, M., \& {Kawasaki}, M. 2005, \prd, 71, 043001

\bibitem[{{Ikebe} et~al.(2002){Ikebe}, {Reiprich}, {B{\" o}hringer}, {Tanaka},
  \& {Kitayama}}]{ikebe/etal:2002}
{Ikebe}, Y., {Reiprich}, T.~H., {B{\" o}hringer}, H., {Tanaka}, Y., \&
  {Kitayama}, T. 2002, \aap, 383, 773

\bibitem[{{Jee} et~al.(2012){Jee}, {Tyson}, {Schneider}, {Wittman}, {Schmidt},
  \& {Hilbert}}]{jee/etal:2012}
{Jee}, M.~J., {Tyson}, J.~A., {Schneider}, M.~D., {Wittman}, D., {Schmidt}, S.,
  \& {Hilbert}, S. 2012, ArXiv e-prints, arXiv:1210.2732

\bibitem[{{Jeong} \& {Komatsu}(2006)}]{jeong/komatsu:2006}
{Jeong}, D. \& {Komatsu}, E. 2006, \apj, 651, 619

\bibitem[{{Jeong} \& {Komatsu}(2009)}]{jeong/komatsu:2009}
---. 2009, \apj, 691, 569

\bibitem[{{Jha} et~al.(2006)}]{jha/etal:2006}
{Jha}, S., et~al. 2006, \aj, 131, 527

\bibitem[{{Kaiser}(1995)}]{kaiser:1995}
{Kaiser}, D.~I. 1995, \prd, 52, 4295

\bibitem[{{Kamionkowski} et~al.(1997){Kamionkowski}, {Kosowsky}, \&
  {Stebbins}}]{kamionkowski/kosowsky/stebbins:1997}
{Kamionkowski}, M., {Kosowsky}, A., \& {Stebbins}, A. 1997, \prd, 55, 7368

\bibitem[{{Kazin} et~al.(2010)}]{kazin/etal:2010}
{Kazin}, E.~A., et~al. 2010, \apj, 710, 1444

\bibitem[{{Keisler} et~al.(2011)}]{keisler/etal:2011}
{Keisler}, R., et~al. 2011, \apj, 743, 28

\bibitem[{{Knox}(1995)}]{knox:1995}
{Knox}, L. 1995, \prd, 52, 4307

\bibitem[{Komatsu \& Futamase(1999)}]{komatsu/futamase:1999}
Komatsu, E. \& Futamase, T. 1999, Phys. Rev., D59, 064029

\bibitem[{{Komatsu} \& {Seljak}(2002)}]{komatsu/seljak:2002}
{Komatsu}, E. \& {Seljak}, U. 2002, \mnras, 336, 1256

\bibitem[{{Komatsu} et~al.(2009)}]{komatsu/etal:2009}
{Komatsu}, E., et~al. 2009, \apjs, 180, 330

\bibitem[{{Komatsu} et~al.(2011)}]{komatsu/etal:2011}
---. 2011, \apjs, 192, 18

\bibitem[{{Kosowsky} \& {Turner}(1995)}]{kosowsky/turner:1995}
{Kosowsky}, A. \& {Turner}, M.~S. 1995, \prd, 52, 1739

\bibitem[{{Larson} et~al.(2011)}]{larson/etal:2011}
{Larson}, D., et~al. 2011, \apjs, 192, 16

\bibitem[{{Lewis} \& {Bridle}(2002)}]{lewis/bridle:2002}
{Lewis}, A. \& {Bridle}, S. 2002, \prd, 66, 103511

\bibitem[{{Lewis} et~al.(2000){Lewis}, {Challinor}, \&
  {Lasenby}}]{lewis/challinor/lasenby:2000}
{Lewis}, A., {Challinor}, A., \& {Lasenby}, A. 2000, \apj, 538, 473

\bibitem[{Liddle \& Lyth(2000)}]{liddle/lyth:CIALSS}
Liddle, A.~R. \& Lyth, D.~H. 2000, Cosmological inflation and large-scale
  structure (Cambridge University Press)

\bibitem[{{Lin} et~al.(2012)}]{lin/etal:2011}
{Lin}, H., et~al. 2012, \apj, 761, 15

\bibitem[{Linde(1983)}]{linde:1983}
Linde, A.~D. 1983, Phys. Lett., B129, 177

\bibitem[{Linder(2003)}]{linder:2003}
Linder, E.~V. 2003, Phys. Rev. Lett., 90, 091301

\bibitem[{{Lue} et~al.(1999){Lue}, {Wang}, \&
  {Kamionkowski}}]{lue/wang/kamionkowski:1999}
{Lue}, A., {Wang}, L., \& {Kamionkowski}, M. 1999, Physical Review Letters, 83,
  1506

\bibitem[{{Maeda}(1988)}]{maeda:1988}
{Maeda}, K.-I. 1988, \prd, 37, 858

\bibitem[{{Mandelbaum} et~al.(2012){Mandelbaum}, {Slosar}, {Baldauf}, {Seljak},
  {Hirata}, {Nakajima}, {Reyes}, \& {Smith}}]{mandelbaum/etal:2012}
{Mandelbaum}, R., {Slosar}, A., {Baldauf}, T., {Seljak}, U., {Hirata}, C.~M.,
  {Nakajima}, R., {Reyes}, R., \& {Smith}, R.~E. 2012, ArXiv e-prints,
  arXiv:1207.1120

\bibitem[{Mangano et~al.(2002)Mangano, Miele, Pastor, \&
  Peloso}]{mangano/etal:2002}
Mangano, G., Miele, G., Pastor, S., \& Peloso, M. 2002, Phys. Lett., B534, 8

\bibitem[{{Mantz} et~al.(2010){Mantz}, {Allen}, \&
  {Rapetti}}]{mantz/etal:2010b}
{Mantz}, A., {Allen}, S.~W., \& {Rapetti}, D. 2010, \mnras, 406, 1805

\bibitem[{{Marriage} et~al.(2011)}]{marriage/etal:2011}
{Marriage}, T.~A., et~al. 2011, \apj, 737, 61

\bibitem[{{Marrone} et~al.(2012)}]{marrone/etal:2012}
{Marrone}, D.~P., et~al. 2012, \apj, 754, 119

\bibitem[{{Matsubara}(2008)}]{matsubara:2008}
{Matsubara}, T. 2008, \prd, 77, 063530

\bibitem[{{McAllister} et~al.(2010){McAllister}, {Silverstein}, \&
  {Westphal}}]{mcallister/silverstein/westphal:2010}
{McAllister}, L., {Silverstein}, E., \& {Westphal}, A. 2010, \prd, 82, 046003

\bibitem[{{Melin} et~al.(2011){Melin}, {Bartlett}, {Delabrouille}, {Arnaud},
  {Piffaretti}, \& {Pratt}}]{melin/etal:2011}
{Melin}, J.-B., {Bartlett}, J.~G., {Delabrouille}, J., {Arnaud}, M.,
  {Piffaretti}, R., \& {Pratt}, G.~W. 2011, \aap, 525, A139

\bibitem[{{Menanteau} et~al.(2012)}]{menanteau/etal:prep}
{Menanteau}, F., et~al. 2012, arXiv:1210.4048

\bibitem[{{Miji{\'c}} et~al.(1986){Miji{\'c}}, {Morris}, \&
  {Suen}}]{mijic/morris/suen:1986}
{Miji{\'c}}, M.~B., {Morris}, M.~S., \& {Suen}, W.-M. 1986, \prd, 34, 2934

\bibitem[{{Miyatake} et~al.(2012)}]{miyatake/etal:2012}
{Miyatake}, H., et~al. 2012, ArXiv e-prints, arXiv:1209.4643

\bibitem[{{Mukhanov} \& {Chibisov}(1981)}]{mukhanov/chibisov:1981}
{Mukhanov}, V.~F. \& {Chibisov}, G.~V. 1981, JETP Letters, 33, 532

\bibitem[{{Nolta} et~al.(2009)}]{nolta/etal:2009}
{Nolta}, M.~R., et~al. 2009, \apjs, 180, 296

\bibitem[{{Oguri} et~al.(2012)}]{oguri/etal:2012}
{Oguri}, M., et~al. 2012, \aj, 143, 120

\bibitem[{{Padmanabhan} \& {White}(2009)}]{padmanabhan/white:2009}
{Padmanabhan}, N. \& {White}, M. 2009, \prd, 80, 063508

\bibitem[{{Padmanabhan} et~al.(2012){Padmanabhan}, {Xu}, {Eisenstein},
  {Scalzo}, {Cuesta}, {Mehta}, \& {Kazin}}]{padmanabhan/etal:2012}
{Padmanabhan}, N., {Xu}, X., {Eisenstein}, D.~J., {Scalzo}, R., {Cuesta},
  A.~J., {Mehta}, K.~T., \& {Kazin}, E. 2012, ArXiv e-prints, arXiv:1202.0090

\bibitem[{{Page} et~al.(2003)}]{page/etal:2003}
{Page}, L., et~al. 2003, \apj, 585, 566

\bibitem[{{Page} et~al.(2007)}]{page/etal:2007}
---. 2007, \apjs, 170, 335

\bibitem[{{Percival} et~al.(2010)}]{percival/etal:2010}
{Percival}, W.~J., et~al. 2010, \mnras, 401, 2148

\bibitem[{{Perlmutter} et~al.(1999)}]{perlmutter/etal:1999}
{Perlmutter}, S., et~al. 1999, \apj, 517, 565

\bibitem[{{Piffaretti} et~al.(2011){Piffaretti}, {Arnaud}, {Pratt},
  {Pointecouteau}, \& {Melin}}]{piffaretti/etal:2011}
{Piffaretti}, R., {Arnaud}, M., {Pratt}, G.~W., {Pointecouteau}, E., \&
  {Melin}, J.-B. 2011, \aap, 534, A109

\bibitem[{{Planck Collaboration III}(2012)}]{planckintermediate:III}
{Planck Collaboration III}. 2012, arXiv:1204.2743

\bibitem[{{Planck Collaboration V}(2012)}]{planckintermediate:V}
{Planck Collaboration V}. 2012, arXiv:1207.4061

\bibitem[{{Planck Collaboration VIII}(2011)}]{planckearly:VIII}
{Planck Collaboration VIII}. 2011, \aap, 536, A8

\bibitem[{{Planck Collaboration X}(2011)}]{planckearly:X}
{Planck Collaboration X}. 2011, \aap, 536, A10

\bibitem[{{Planck Collaboration X}(2012)}]{planckintermediate:X}
---. 2012, arXiv:1208.3611

\bibitem[{{Planck Collaboration XI}(2011)}]{planckearly:XI}
{Planck Collaboration XI}. 2011, \aap, 536, A11

\bibitem[{{Planck Collaboration XII}(2011)}]{planckearly/sz_optical}
{Planck Collaboration XII}. 2011, \aap, 536, A12

\bibitem[{{QUIET Collaboration}(2011)}]{quiet:2011}
{QUIET Collaboration}. 2011, \apj, 741, 111

\bibitem[{{QUIET Collaboration}(2012)}]{quiet2:2012}
---. 2012, \apj, 760, 145

\bibitem[{{Reichardt} et~al.(2012{\natexlab{a}})}]{reichardt/etal:2012a}
{Reichardt}, C.~L., et~al. 2012{\natexlab{a}}, \apj, 755, 70

\bibitem[{{Reichardt} et~al.(2012{\natexlab{b}})}]{reichardt/etal:2012b}
---. 2012{\natexlab{b}}, ArXiv e-prints, arXiv:1203.5775

\bibitem[{{Reid} et~al.(2012)}]{reid/etal:2012}
{Reid}, B.~A., et~al. 2012, ArXiv e-prints, arXiv:1203.6641

\bibitem[{{Riess} et~al.(1998)}]{riess/etal:1998}
{Riess}, A.~G., et~al. 1998, \aj, 116, 1009

\bibitem[{{Riess} et~al.(1999)}]{riess/etal:1999}
---. 1999, \aj, 117, 707

\bibitem[{{Riess} et~al.(2007)}]{riess/etal:2007}
---. 2007, \apj, 659, 98

\bibitem[{{Riess} et~al.(2009)}]{riess/etal:2009}
---. 2009, \apj, 699, 539

\bibitem[{{Riess} et~al.(2011)}]{riess/etal:2011}
---. 2011, \apj, 730, 119

\bibitem[{{Rozo} et~al.(2012){Rozo}, {Bartlett}, {Evrard}, \&
  {Rykoff}}]{rozo/etal:2012}
{Rozo}, E., {Bartlett}, J.~G., {Evrard}, A.~E., \& {Rykoff}, E.~S. 2012, ArXiv
  e-prints, arXiv:1204.6305

\bibitem[{Rubakov et~al.(1982)Rubakov, Sazhin, \&
  Veryaskin}]{rubakov/sazhin/veryaskin:1982}
Rubakov, V.~A., Sazhin, M.~V., \& Veryaskin, A.~V. 1982, Phys. Lett., B115, 189

\bibitem[{{Salopek}(1992)}]{salopek:1992}
{Salopek}, D.~S. 1992, Physical Review Letters, 69, 3602

\bibitem[{{Salopek} et~al.(1989){Salopek}, {Bond}, \&
  {Bardeen}}]{salopek/bond/bardeen:1989}
{Salopek}, D.~S., {Bond}, J.~R., \& {Bardeen}, J.~M. 1989, \prd, 40, 1753

\bibitem[{{S{\'a}nchez} et~al.(2012)}]{sanchez/etal:2012}
{S{\'a}nchez}, A.~G., et~al. 2012, \mnras, 425, 415

\bibitem[{{Sandage} et~al.(2006){Sandage}, {Tammann}, {Saha}, {Reindl},
  {Macchetto}, \& {Panagia}}]{sandage/etal:2006}
{Sandage}, A., {Tammann}, G.~A., {Saha}, A., {Reindl}, B., {Macchetto}, F.~D.,
  \& {Panagia}, N. 2006, \apj, 653, 843

\bibitem[{{Schmidt} et~al.(1998)}]{schmidt/etal:1998}
{Schmidt}, B.~P., et~al. 1998, \apj, 507, 46

\bibitem[{{Sehgal} et~al.(2011)}]{sehgal/etal:2011}
{Sehgal}, N., et~al. 2011, \apj, 732, 44

\bibitem[{{Sehgal} et~al.(2012)}]{sehgal/etal:2012}
---. 2012, ArXiv e-prints, arXiv:1205.2369

\bibitem[{{Seljak} \& {Zaldarriaga}(1996)}]{seljak/zaldarriaga:1996}
{Seljak}, U. \& {Zaldarriaga}, M. 1996, \apj, 469, 437

\bibitem[{{Semboloni} et~al.(2011){Semboloni}, {Schrabback}, {van Waerbeke},
  {Vafaei}, {Hartlap}, \& {Hilbert}}]{semboloni/etal:2012}
{Semboloni}, E., {Schrabback}, T., {van Waerbeke}, L., {Vafaei}, S., {Hartlap},
  J., \& {Hilbert}, S. 2011, \mnras, 410, 143

\bibitem[{Seo \& Eisenstein(2005)}]{seo/eisenstein:2005}
Seo, H.-J. \& Eisenstein, D.~J. 2005, Astrophys. J., 633, 575

\bibitem[{{Seo} \& {Eisenstein}(2007)}]{seo/eisenstein:2007}
{Seo}, H.-J. \& {Eisenstein}, D.~J. 2007, \apj, 665, 14

\bibitem[{{Shaw} et~al.(2010){Shaw}, {Nagai}, {Bhattacharya}, \&
  {Lau}}]{shaw/etal:2010}
{Shaw}, L.~D., {Nagai}, D., {Bhattacharya}, S., \& {Lau}, E.~T. 2010, \apj,
  725, 1452

\bibitem[{{Sherwin} et~al.(2011)}]{sherwin/etal:2011}
{Sherwin}, B.~D., et~al. 2011, Physical Review Letters, 107, 021302

\bibitem[{{Sifon} et~al.(2012)}]{sifon/etal:2012}
{Sifon}, C., et~al. 2012, ArXiv e-prints, arXiv:1201.0991

\bibitem[{{Silk}(1968)}]{silk:1968}
{Silk}, J. 1968, \apj, 151, 459

\bibitem[{{Smith} et~al.(2007){Smith}, {Zahn}, \&
  {Dor{\'e}}}]{smith/zahn/dore:2007}
{Smith}, K.~M., {Zahn}, O., \& {Dor{\'e}}, O. 2007, \prd, 76, 043510

\bibitem[{{Smith} et~al.(2012){Smith}, {Das}, \& {Zahn}}]{smith/das/zahn:2012}
{Smith}, T.~L., {Das}, S., \& {Zahn}, O. 2012, \prd, 85, 023001

\bibitem[{{Song} et~al.(2012)}]{song/etal:prep}
{Song}, J., et~al. 2012, arXiv:1207.4369

\bibitem[{{Spergel} et~al.(2003)}]{spergel/etal:2003}
{Spergel}, D.~N., et~al. 2003, \apjs, 148, 175

\bibitem[{{Spergel} et~al.(2007)}]{spergel/etal:2007}
---. 2007, \apjs, 170, 377

\bibitem[{{Spokoiny}(1984)}]{spokoiny:1984}
{Spokoiny}, B.~L. 1984, Physics Letters B, 147, 39

\bibitem[{{Starobinsky}(1979)}]{starobinsky:1979}
{Starobinsky}, A.~A. 1979, Soviet Journal of Experimental and Theoretical
  Physics Letters, 30, 682

\bibitem[{{Starobinsky}(1980)}]{starobinsky:1980}
---. 1980, Physics Letters B, 91, 99

\bibitem[{{Starobinsky}(1983)}]{starobinsky:1983}
---. 1983, Soviet Astronomy Letters, 9, 302

\bibitem[{{Starobinsky}(1985)}]{starobinsky:1985}
---. 1985, Sov. Astron. Lett., 11, 133

\bibitem[{{Steigman}(2012)}]{steigman:2012}
{Steigman}, G. 2012, ArXiv e-prints

\bibitem[{{Story} et~al.(2012)}]{story/etal:2012}
{Story}, K.~T., et~al. 2012, ArXiv e-prints, arXiv:1210.7231

\bibitem[{{Sullivan} et~al.(2011)}]{sullivan/etal:2011}
{Sullivan}, M., et~al. 2011, \apj, 737, 102

\bibitem[{{Sunyaev} \& {Zel'dovich}(1972)}]{sunyaev/zeldovich:1972}
{Sunyaev}, R.~A. \& {Zel'dovich}, Y.~B. 1972, Comments on Astrophysics and
  Space Physics, 4, 173

\bibitem[{{Suyu} et~al.(2013)}]{suyu/etal:2012}
{Suyu}, S.~H., et~al. 2013, \apj, 766, 70

\bibitem[{{Taruya} \& {Hiramatsu}(2008)}]{taruya/hiramatsu:2008}
{Taruya}, A. \& {Hiramatsu}, T. 2008, \apj, 674, 617

\bibitem[{{Tegmark}(1997)}]{tegmark/etal:1997}
{Tegmark}, M., e.~a. 1997, \apj, 474, 1

\bibitem[{{Tinker} et~al.(2012)}]{tinker/etal:2012}
{Tinker}, J.~L., et~al. 2012, \apj, 745, 16

\bibitem[{{Trac} et~al.(2011){Trac}, {Bode}, \&
  {Ostriker}}]{trac/bode/ostriker:2011}
{Trac}, H., {Bode}, P., \& {Ostriker}, J.~P. 2011, \apj, 727, 94

\bibitem[{{van Engelen} et~al.(2012)}]{vanengelen/etal:2012}
{van Engelen}, A., et~al. 2012, \apj, 756, 142

\bibitem[{{Verde} et~al.(2003)}]{verde/etal:2003}
{Verde}, L., et~al. 2003, \apjs, 148, 195

\bibitem[{{Vikhlinin} et~al.(2009{\natexlab{a}})}]{vikhlinin/etal:2009b}
{Vikhlinin}, A., et~al. 2009{\natexlab{a}}, \apj, 692, 1033

\bibitem[{{Vikhlinin} et~al.(2009{\natexlab{b}})}]{vikhlinin/etal:2009}
---. 2009{\natexlab{b}}, \apj, 692, 1060

\bibitem[{Wang et~al.(2003)Wang, Tegmark, Jain, \&
  Zaldarriaga}]{wang/etal:2002}
Wang, X., Tegmark, M., Jain, B., \& Zaldarriaga, M. 2003, Phys. Rev., D68,
  123001

\bibitem[{{Wang} \& {Mukherjee}(2007)}]{wang/mukherjee:2007}
{Wang}, Y. \& {Mukherjee}, P. 2007, \prd, 76, 103533

\bibitem[{Weinberg(1972)}]{weinberg:GAC}
Weinberg, S. 1972, Gravitation and Cosmology (New York, NY: John Wiley)

\bibitem[{{Whitt}(1984)}]{whitt:1984}
{Whitt}, B. 1984, Physics Letters B, 145, 176

\bibitem[{{Williamson} et~al.(2011)}]{williamson/etal:2011}
{Williamson}, R., et~al. 2011, \apj, 738, 139

\bibitem[{{Wilson} et~al.(2012)}]{wilson/etal:2012}
{Wilson}, M.~J., et~al. 2012, ArXiv e-prints, arXiv:1203.6633

\bibitem[{{Wright}(2007)}]{wright:2007}
{Wright}, E.~L. 2007, ArXiv Astrophysics e-prints, astro-ph/0703640

\bibitem[{{Xia} et~al.(2010){Xia}, {Li}, \& {Zhang}}]{xia/li/zhang:2010}
{Xia}, J., {Li}, H., \& {Zhang}, X. 2010, Physics Letters B, 687, 129

\bibitem[{Zaldarriaga et~al.(1997)Zaldarriaga, Spergel, \&
  Seljak}]{zaldarriaga/spergel/seljak:1997}
Zaldarriaga, M., Spergel, D.~N., \& Seljak, U. 1997, \apj, 488, 1

\bibitem[{{Zel'dovich} \& {Sunyaev}(1969)}]{zeldovich/sunyaev:1969}
{Zel'dovich}, Y.~B. \& {Sunyaev}, R.~A. 1969, \apss, 4, 301

\bibitem[{Zhao et~al.(2005)Zhao, Xia, Li, Feng, \& Zhang}]{zhao/etal:2005}
Zhao, G.-B., Xia, J.-Q., Li, M., Feng, B., \& Zhang, X. 2005, Phys. Rev., D72,
  123515

\bibitem[{{Zu} et~al.(2012){Zu}, {Weinberg}, {Rozo}, {Sheldon}, {Tinker}, \&
  {Becker}}]{zu/etal:2012}
{Zu}, Y., {Weinberg}, D.~H., {Rozo}, E., {Sheldon}, E.~S., {Tinker}, J.~L., \&
  {Becker}, M.~R. 2012, ArXiv e-prints

\end{thebibliography}

\end{document}